\documentclass[12pt,preprint]{aastex61}

\shortauthors{Bower et al.}
\shorttitle{Polarimetry of Sgr A*}
\begin{document}

\newcommand\degd{\ifmmode^{\circ}\!\!\!.\,\else$^{\circ}\!\!\!.\,$\fi}
\newcommand{\etal}{{\it et al.\ }}
\newcommand{\uv}{(u,v)}
\newcommand{\rdm}{{\rm\ rad\ m^{-2}}}
\newcommand{\msun}{{\rm\ M_{\sun}}}
\newcommand{\msuny}{{\rm\ M_{\sun}\ y^{-1}}}
\newcommand{\mylesssim}{\stackrel{\scriptstyle <}{\scriptstyle \sim}}
\newcommand{\lsim}{\stackrel{\scriptstyle <}{\scriptstyle \sim}}
\newcommand{\gsim}{\stackrel{\scriptstyle >}{\scriptstyle \sim}}
\newcommand{\sci}{Science}
\newcommand{\sgr}{PSR J1745-2900}
\newcommand{\sgra}{Sgr~A*}
\newcommand{\kms}{\ensuremath{{\rm km\,s}^{-1}}}
\newcommand{\kkms}{\ensuremath{{\rm K\,km\,s}^{-1}}}
\newcommand{\masy}{\ensuremath{{\rm mas\,yr}^{-1}}}
\newcommand{\frb}{FRB 121102}

\def\kbar{{\mathchar'26\mkern-9mu k}}
\def\totd{{\mathrm{d}}}


\title{ALMA Polarimetry of Sgr A*:  Probing the Accretion Flow from the Event Horizon
to the Bondi Radius}

\author[0000-0003-4056-9982]{Geoffrey C.\ Bower}
\affiliation{Academia Sinica Institute of Astronomy and Astrophysics, 645 N. A'ohoku Place, Hilo, HI 96720, USA}
\email{gbower@asiaa.sinica.edu.tw}

\author[0000-0002-3351-760X]{Avery Broderick}
\affiliation{Department of Physics and Astronomy, University of Waterloo, 200 University Avenue West, Waterloo, ON, N2L 3G1, Canada ; Perimeter Institute for Theoretical Physics, 31 Caroline Street North, Waterloo, ON, N2L 2Y5, Canada}

\author[0000-0003-3903-0373]{Jason Dexter}
\affiliation{Max Planck Institute for Extraterrestrial Physics, Giessenbachstr. 1, 85748 Garching, Germany}

\author{Shepherd Doeleman}
\affiliation{Harvard-Smithsonian Center for Astrophysics, 60 Garden Street, Cambridge, MA 02138, USA}

\author[0000-0002-2526-6724]{Heino Falcke}
\affiliation{
Department of Astrophysics, Institute for Mathematics, Astrophysics and Particle Physics (IMAPP), Radboud University, PO Box 9010, 6500 GL Nijmegen, The Netherlands
}
\affiliation{
ASTRON, P.O. Box 2, 7990 AA Dwingeloo, The Netherlands 
}
\affiliation{
Max-Planck-Institut f\"{u}r Radioastronomie, Auf dem H\"{u}gel 69, D-53121 Bonn, Germany}

\author{Vincent Fish}
\affiliation{Massachusetts Institute of Technology, Haystack Observatory, Route 40, Westford, MA 01886, USA}

\author[0000-0002-4120-3029]{Michael D. Johnson}
\affiliation{Harvard-Smithsonian Center for Astrophysics, 60 Garden Street, Cambridge, MA 02138, USA}

\author[0000-0002-2367-1080]{Daniel P. Marrone}
\affiliation{Steward Observatory, University of Arizona, 933 North Cherry Avenue, Tucson, AZ 85721, USA}

\author[0000-0002-3882-4414]{James M. Moran}
\affiliation{Harvard-Smithsonian Center for Astrophysics, 60 Garden Street, Cambridge, MA 02138, USA}

\author[0000-0002-4661-6332]{Monika Moscibrodzka}
\affiliation{
Department of Astrophysics, Institute for Mathematics, Astrophysics and Particle Physics (IMAPP), Radboud University, PO Box 9010, 6500 GL Nijmegen, The Netherlands
}

\author[0000-0001-8276-0000]{Alison Peck}
\affiliation{Gemini Observatory, 670 N. A'ohoku Pl., Hilo, HI 96720, USA}

\author[0000-0001-6765-9609]{Richard L. Plambeck}
\affiliation{Radio Astronomy Laboratory, University of California, Berkeley, CA 94720-3411, USA}

\author[0000-0002-1407-7944]{Ramprasad Rao}
\affiliation{Academia Sinica Institute of Astronomy and Astrophysics, 645 N. A'ohoku Place, Hilo, HI 96720, USA}

\begin{abstract}
Millimeter polarimetry of \sgra\ probes the linearly polarized emission 
region on a scale of $\sim 10$ Schwarzschild radii ($R_S$) as well as the
dense, magnetized accretion flow on scales out to the Bondi radius 
($\sim 10^5 R_S$) through Faraday rotation.  We present here 
multi-epoch ALMA Band 6 (230 GHz) polarimetry of \sgra.  The results
confirm a mean rotation measure, ${\rm RM} \approx -5 \times 10^5 \rdm$, 
consistent with measurements over the past 20 years and
support an interpretation of the RM as originating from a radiatively
inefficient accretion flow (RIAF) with $\dot{M} \approx 10^{-8} \msuny$.
Variability is observed for the first time in the RM on time scales
that range from hours to months.  The long-term variations may be
the result of changes in the line of sight properties
in a turbulent accretion flow.  Short-term variations in the apparent
RM are not necessarily the result of Faraday rotation and may be the
result of complex emission and propagatation effects close 
to the black hole, some of which have been predicted in numerical
modeling.  We also confirm the detection of circular
polarization at a mean value of $-1.1 \pm 0.2 \%$.  It is variable
in amplitude on time scales from hours to months but 
the handedness remains unchanged from that 
observed in past centimeter- and millimeter-wavelength
detections. 
These results provide critical constraints
for the analysis and interpretation of Event Horizon Telescope 
data of \sgra, M87, and similar sources.
\end{abstract}


\section{Introduction}

\sgra\ is the $\sim 4.1 \times 10^6 \msun$ black hole in the Galactic Center
\citep{2013CQGra..30x4003F,2016ApJ...830...17B,2018A&A...615L..15G}.
As the nearest supermassive black hole, it serves as a powerful laboratory
for the understanding of accretion, outflow, and jet physics, as well
as detailed physics associated with particle acceleration and
magnetic fields \citep{2017MNRAS.467.3604R,2018A&A...612A..34D}. \sgra\ is
also an important target for tests of general relativity and 
measurement of intrinsic black hole parameters such as spin and
the presence of the event horizon
through a variety of approaches including measurement of stellar orbits 
\citep{2018MNRAS.tmp..464W} and imaging of event
horizon scale structure \citep{2000ApJ...528L..13F,2014ApJ...784....7B}.  Convincing tests of GR and 
characterization of black hole properties through imaging rely on a thorough and
detailed understanding of accretion, outflow, and particle acceleration
physics.

Imaging of \sgra\ at millimeter and submillimeter wavelengths is a major
goal of the Event Horizon Telescope \citep{2009astro2010S..68D,2013arXiv1309.3519F}.  EHT observations
will have an angular resolution comparable to the Schwarzschild radius
($1 R_S \approx 10 \mu as$ for a distance of 8.1 kpc) and are sensitive to
structures on scales as large as a few $\times 10 R_S$.  
Imaging has established 
the dominance of angular broadening due to scattering by interstellar 
electrons along the line of sight \citep{2006ApJ...648L.127B} and an intrinsic source size
that is $\sim 10 R_S$ at 3mm wavelength \citep{2004Sci...304..704,2005Natur.438...62S,2014ApJ...790....1B,2016ApJ...824...40O,2016MNRAS.462.1382B}
and $\sim 4 R_S$ at 1.3 mm wavelength \citep{2008Natur.455...78D,2011ApJ...727L..36F,2015Sci...350.1242J,2016ApJ...820...90F,2018ApJ...859...60L}.  These 
radio and mm/submm results cannot be conclusively
interpreted in terms of either accretion disk or jet models, leaving
the question of whether a jet is present unanswered.
Images obtained with the EHT will be sensitive to the 
accretion flow and/or jet launching region on scales of a few $R_S$
\citep{2014A&A...570A...7M}.

The region imaged by the EHT is embedded within the larger accretion
flow of \sgra.  Chandra X-ray imaging shows an extended structure
with a scale comparable to the Bondi radius $\sim 10^5 R_S$ \citep{2013Sci...341..981W}.
The accretion rate at the Bondi radius is estimated to be $\sim 10^{-4}  - 10^{-5} \msuny$ and fed by stellar winds of massive stars outside the accretion
flow \citep{1999ApJ...517L.101Q}.  This accretion rate appears to be inconsistent
with Bondi accretion onto the black hole producing
the very low bolometric luminosity $L_{bol} \sim 10^{35} {\rm\, erg\, s^{-1}}$ of \sgra.  This has driven the development of a number of theoretical
models that fall under the umbrella of radiatively inefficient accretion
flows \citep[RIAFS; ][]{2014ARA&A..52..529Y}.  Broadly, these models produce
the low luminosity of \sgra\ through two mechanisms:  stalled accretion
at large radii which reduces the accretion rate; and, two-temperature
plasmas in which the lower-temperature 
electrons do not equilibrate with the full gravitational
potential energy of infall and, therefore, radiate a small fraction of the
total available energy.

Millimeter and submillimeter wavelength polarimetry of \sgra\ has been
a powerful tool for characterization of the accretion flow on scales
inside the Bondi radius that are inaccessible to other techniques.
\sgra\ shows linear (LP) and circular (CP) polarization properties that are not
common in higher power active galactic nuclei.  
\sgra\ shows no LP at centimeter wavelengths \citep{1999ApJ...521..582B,1999ApJ...527..851B},
while at mm/submm wavelengths the polarization fraction rises to $\sim 10\%$
\citep{2000ApJ...534L.173A,2003ApJ...588..331B,2006ApJ...646L.111M,2007ApJ...654L..57M,2016A&A...593A..44L,2016A&A...593A.107L}.  The LP has been
shown to undergo significant Faraday rotation, with one of the largest
rotation measures observed in any source, ${\rm RM} \approx -5 \times 10^5 
\rdm$.  The RM is proportional to the line-of-sight integrated electron 
density and parallel magnetic field strength:
\begin{equation}
{\rm RM} = 0.81 \int n_e {\bf B \cdot ds} \, \rdm,
\end{equation}  
where $n_e$ is in units of cm$^{-3}$, $\bf B$ is in $\mu$G, and the length
scale is in pc.
The LP 
properties have been interpreted as intrinsic polarization arising
within $\sim 10 R_S$ of the event horizon and propagating through the
dense, magnetized accretion flow.  The RM$=-7 \times 10^4 \rdm$ found for the GC
pulsar J1745-2900, which is separated by $\sim 0.1$ pc from \sgra\,
supports the hypothesis that the majority of the \sgra\ RM originates in the accretion
flow \citep{2013Natur.501..391E,2015ApJ...798..120B}.   Monitoring
of the pulsar RM also demonstrates that ISM changes in the RM
are $\sim 10^4 \rdm$, significant for a pulsar, but small relative
to the \sgra\ RM \citep{2018ApJ...852L..12D}.  In the accretion
flow interpretation for \sgra, the RM
demonstrates a profile for the electron density as a function of
radius (e.g., $n_e \propto r^{-1}$) that is flatter than required by 
advection dominated accretion flow (ADAF) models
and sets an accretion rate onto \sgra\ at the event horizon of
$\sim 10^{-8} \msuny$.  

1.3 mm wavelength VLBI supports the conclusion that
the LP originates within $\sim 10 R_S$ \citep{2015Sci...350.1242J}.
These observations show that the LP does not originate from
a simple, homogeneous source but from a more complex source with 
structure in the magnetic field (and, hence, polarization angle) on scales
of $\lsim 10 R_S$.  Complex polarization features are predicted in
general-relativistic
magneto-hydrodynamical (GRMHD) models of \sgra\ accretion disks and 
jets \citep{2012ApJ...755..133S,2017MNRAS.468.2214M,
2017ApJ...837..180G}.

Circular polarization is present in \sgra\ 
from cm to submm wavelengths, a factor of more than 200 in 
wavelength \citep{1999ApJ...523L..29B,1999ApJ...526L..85S,2002ApJ...571..843B,2012ApJ...745..115M}. The CP has constant handedness across all wavelengths and a magnitude
$\lsim 1\%$.  The origin of the CP is not well understood but it is
unlikely to be produced through the synchrotron mechanism.  It is more
likely that the process of Faraday conversion \citep{1977OISNP..89.....P} 
transforms
LP into CP via thermal electrons that are mixed with the relativistic
electrons responsible for the linearly-polarized synchrotron emission
\citep{2002A&A...388.1106B,2002ApJ...573..485R,2008ApJ...676L.119H}.
The stability of the handedness of the CP over decades suggests a
stable magnetic field configuration in the emission and conversion region.

Time variablity of the polarization properties has long been recognized
as an important diagnostic of the accretion flow properties \citep{2005ApJ...618L..29B,2007ApJ...654L..57M}.
The timescale of variability for the RM 
can be translated into a radius at which the RM originates, a kind
of Faraday tomography.  The orbital period at the innermost stable circular
orbit is $\sim 30$ minutes, while the orbital period at the Bondi radius
is $\sim 10^3$ years.  Characterization of the RM variability over
timescales up to 10 years provides sensitivity to radii as large 
as $10^3 R_S$, much larger than can be probed through submm VLBI imaging
and much smaller than can be probed through direct X-ray imaging of
the accretion flow.
A variety of models of turbulence and magnetic field structure in
the accretion flow predict different degrees of RM variability
\citep{2007ApJ...671.1696S,2011MNRAS.415.1228P}.  GRMHD models are now explicitly modeling variations
in polarization properties including Faraday effects on scales as small
as the emission region \citep{2017MNRAS.468.2214M}.

\sgra\ is not alone in showing these unusual polarization properties.
Three other low luminosity AGN (LLAGN) have now been demonstrated to
have large RMs, M87 \citep{2014ApJ...783L..33K},  3C 84 \citep{2014ApJ...797...66P}, and 3C 273 \citep{2018arXiv180309982H}.
A number of other LLAGN appear to have suppressed LP at cm and mm
wavelengths, possibly as the result of extreme RMs \citep{2001ApJ...560L.123B,2002ApJ...578L.103B,2017ApJ...843L..31B}.

The sensitivity and systematic control of the Atacama Large Millimeter Array
(ALMA) provides a powerful tool for detailed characterization
of the LP, CP, and Faraday properties of \sgra.  We present here
new full-Stokes observations obtained via ALMA Cycle 2.  In
Section 2, we present the observations and data reduction.  In Section 3,
we present our results.  Given the novelty of ALMA polarimetry, we 
place significant emphasis on validation of the results through examination
of calibrator sources in Appendix A.  In Section 4, we discuss
these results and their implications for accretion and outflow models
of \sgra.  We give our conclusions in Section 5.

\section{Observations}

ALMA observed \sgra\ in Band 6 (1.3 mm wavelength) in full polarimetric continuum
mode on three epochs (Table~\ref{tab:obs}).  The correlator was configured with
four spectral windows (SPWs), each with 2 GHz bandwidth in 64 channels.  The 
SPW center frequencies are 223.96 GHz, 225.96 GHz, 239.96 GHz, and 241.96 GHz.
The Band 6 receivers are sensitive to linear polarization (X and Y) and the correlator 
produces XX, YY, XY, and YX correlations.  The array was in a relatively
compact configuration with naturally-weighted 
synthesized beam sizes $\sim 1$ arcsec.

Sources were observed 
for amplitude, bandpass, and polarization calibration
(J1751+0939) and for phase calibration (J1733-3722).  Absolute flux
calibration was set by observations of the moon Titan in epochs 1 and 2.
In epoch 3, absolute flux calibration was set by
ALMA monitoring of J1751+0939 at 90 and 345 GHz; a
power-law extrapolation to 230 GHz was used to provide the estimated flux
density of 2.4 Jy, constant across all SPWs, with
an accuracy of $\sim 20\%$.  A check source (J1713-3418)
was interleaved with phase calibration and \sgra\ observations.  In epoch 3,
additional short observations were obtained on J1517-2422, J1924-2914, and J1733-1304.
A typical observation cycle included 30 seconds on the phase calibrator, 20 seconds
on the check source, and 7 minutes on \sgra.  Approximately three hours of 
observations were obtained in epochs 1 and 2.  Approximately five hours of
observations were obtained over the 7-hour duration of epoch 3 due to failure
to observe \added{two hour-long observing sequences known as}
execution blocks.

Data reduction was performed in CASA using pipelines that applied standard
{\em a priori} and calculated calibrations.  This produced calibrated 
measurement sets with the full time and frequency resolution of the observations.
We extracted source data by imaging with all of the data averaged over
all channels and various subsets of the data, sliced in frequency and in time.
We rejected all baselines shorter than $50 k\lambda$ in order to eliminate any 
extended structure around \sgra.  For consistency, we applied the same baseline cut to all calibrator data.  Results were obtained by fitting point sources 
in the image domain.  Fits in the Stokes Q, U, and V domains were  obtained
with the point-source position fixed at the fitted peak of the Stokes I 
image.  These image-domain results are consistent with point-source
fits obtained in the visibility domain.

\begin{deluxetable}{ll}
\tablecaption{ALMA Observations \label{tab:obs}}
\tablehead{
\colhead{Epoch} & \colhead{UT} 
}
\startdata
03 Mar 2016 & 0936 -- 1314  \\
03 May 2016 & 0537 -- 0913  \\
13 Aug 2016 & 2057 -- 0357  \\
\enddata
\end{deluxetable}

\begin{deluxetable}{llrrrrr}
\tablecaption{Average Polarization Properties \label{tab:avg}}
\tablehead{
\colhead{Source} & \colhead{Epoch} & \colhead{SPW} & \colhead{$I$} & \colhead{$Q$} & \colhead{$U$} & \colhead{$V$} \\
                 &                 &               & \colhead{(mJy)} & \colhead{(mJy)} & \colhead{(mJy)} & \colhead{(mJy)} 
} 
\startdata
  SgrA* &     1 & 0 &  $  4062 \pm    130 $ & $  140.533 \pm     4.370 $ & $   79.584 \pm    2.486 $ & $  -50.173 \pm    1.618 $\\ 
     \dots & \dots & 1 &  $  3962 \pm    126 $ & $  132.145 \pm     4.038 $ & $   84.040 \pm    2.619 $ & $  -49.793 \pm    1.568 $\\ 
     \dots & \dots & 2 &  $  4073 \pm    125 $ & $   88.882 \pm     2.590 $ & $  106.037 \pm    3.239 $ & $  -54.265 \pm    1.649 $\\ 
     \dots & \dots & 3 &  $  4110 \pm    126 $ & $   81.717 \pm     2.385 $ & $  107.850 \pm    3.195 $ & $  -55.053 \pm    1.730 $\\ 
\hline
  SgrA* &     2 & 0 &  $  3435 \pm     68 $ & $ -176.558 \pm     3.425 $ & $ -190.601 \pm    3.795 $ & $  -30.207 \pm    0.667 $\\ 
     \dots & \dots & 1 &  $  3350 \pm     64 $ & $ -167.737 \pm     3.173 $ & $ -191.118 \pm    3.724 $ & $  -29.792 \pm    0.633 $\\ 
     \dots & \dots & 2 &  $  3432 \pm     64 $ & $ -132.402 \pm     2.511 $ & $ -228.727 \pm    4.218 $ & $  -34.542 \pm    0.707 $\\ 
     \dots & \dots & 3 &  $  3448 \pm     64 $ & $ -129.888 \pm     2.492 $ & $ -234.454 \pm    4.307 $ & $  -31.992 \pm    0.669 $\\ 
\hline
  SgrA* &     3 & 0 &  $  2657 \pm     82 $ & $   82.909 \pm     2.512 $ & $ -121.412 \pm    3.909 $ & $  -39.496 \pm    1.222 $\\ 
     \dots & \dots & 1 &  $  2582 \pm     80 $ & $   85.942 \pm     2.617 $ & $ -120.428 \pm    3.894 $ & $  -36.879 \pm    1.134 $\\ 
     \dots & \dots & 2 &  $  2735 \pm     85 $ & $  126.549 \pm     3.873 $ & $ -138.611 \pm    4.565 $ & $  -33.426 \pm    1.009 $\\ 
     \dots & \dots & 3 &  $  2741 \pm     86 $ & $  131.360 \pm     4.055 $ & $ -142.502 \pm    4.719 $ & $  -30.809 \pm    0.920 $\\ 
\enddata
\end{deluxetable}

\movetabledown=1in
\begin{rotatetable}
\begin{deluxetable}{llrrrrrrrr}
\tabletypesize{\scriptsize}
\tablecaption{Frequency-Averaged Polarization and Rotation Measures \label{tab:avgrm}}
\tablehead{
\colhead{Source} & \colhead{Epoch} & 
\colhead{$I$} & \colhead{$\delta I$} &
\colhead{$p$} & \colhead{$\delta p$} &
\colhead{$V$} & \colhead{$\delta V$} &
\colhead{RM} & \colhead{$\chi_0$} 
\\
                 &                 & 
\colhead{(mJy)} & \colhead{(mJy GHz$^{-1}$)} &
\colhead{(mJy)} & \colhead{(mJy GHz$^{-1}$)} &
\colhead{(mJy)} & \colhead{(mJy GHz$^{-1}$)} &
\colhead{($10^5 \rdm$)} & \colhead{(deg)} 
}
\startdata
  SgrA* & 1 &  $   4051 \pm     28 $ & $   4.8 \pm  28.1 $ & $ 147.92 \pm   0.48 $ & $  -1.39 \pm   0.06 $ & $  -52.309 \pm    0.253 $ & $ -0.290 \pm 0.032 $ & $ -7.83 \pm  0.10 $ & $  95.3 \pm   0.9 $ \\ 
     \dots & 2 &  $   3415 \pm     23 $ & $   2.8 \pm  22.5 $ & $ 261.51 \pm   1.85 $ & $   0.56 \pm   0.23 $ & $  -31.576 \pm    0.742 $ & $ -0.186 \pm 0.092 $ & $ -4.77 \pm  0.07 $ & $ -17.4 \pm   0.7 $ \\ 
     \dots & 3 &  $   2678 \pm     23 $ & $   7.1 \pm  23.2 $ & $ 169.07 \pm   1.26 $ & $   2.68 \pm   0.16 $ & $  -35.100 \pm    0.663 $ & $ 0.395 \pm 0.082 $ & $ -2.90 \pm  0.11 $ & $   2.0 \pm   1.1 $ \\ 
\enddata
\end{deluxetable}
\end{rotatetable}

\section{Results}

In Appendix~\ref{sec:appendix}, we present results for calibrators with the
goal of demonstrating the stability of the ALMA polarization measurements
and determination of systematic limits on polarization quantities.  We
find that results are most stable and accurate for sources observed
at multiple parallactic angles.  Inter-epoch results have limits for
fractional LP and CP 
near $\sim 0.1$ -- 0.2 \%.  Position angles are measured to an accuracy $\sim 1$ deg and rotation measures are determined to an accuracy $\lsim 10^5 \rdm$.
Intra-epoch measurements have similar accuracy.
We caution that calibration errors will play a larger role the smaller the
polarization fraction; all calibrators have a polarization fraction larger
than 1\%. 
In this
Section, we present results for \sgra\ in inter-epoch and intra-epoch
measurements.

\subsection{Inter-Epoch Polarization Properties of \sgra}

\begin{figure}[t!]
\includegraphics[]{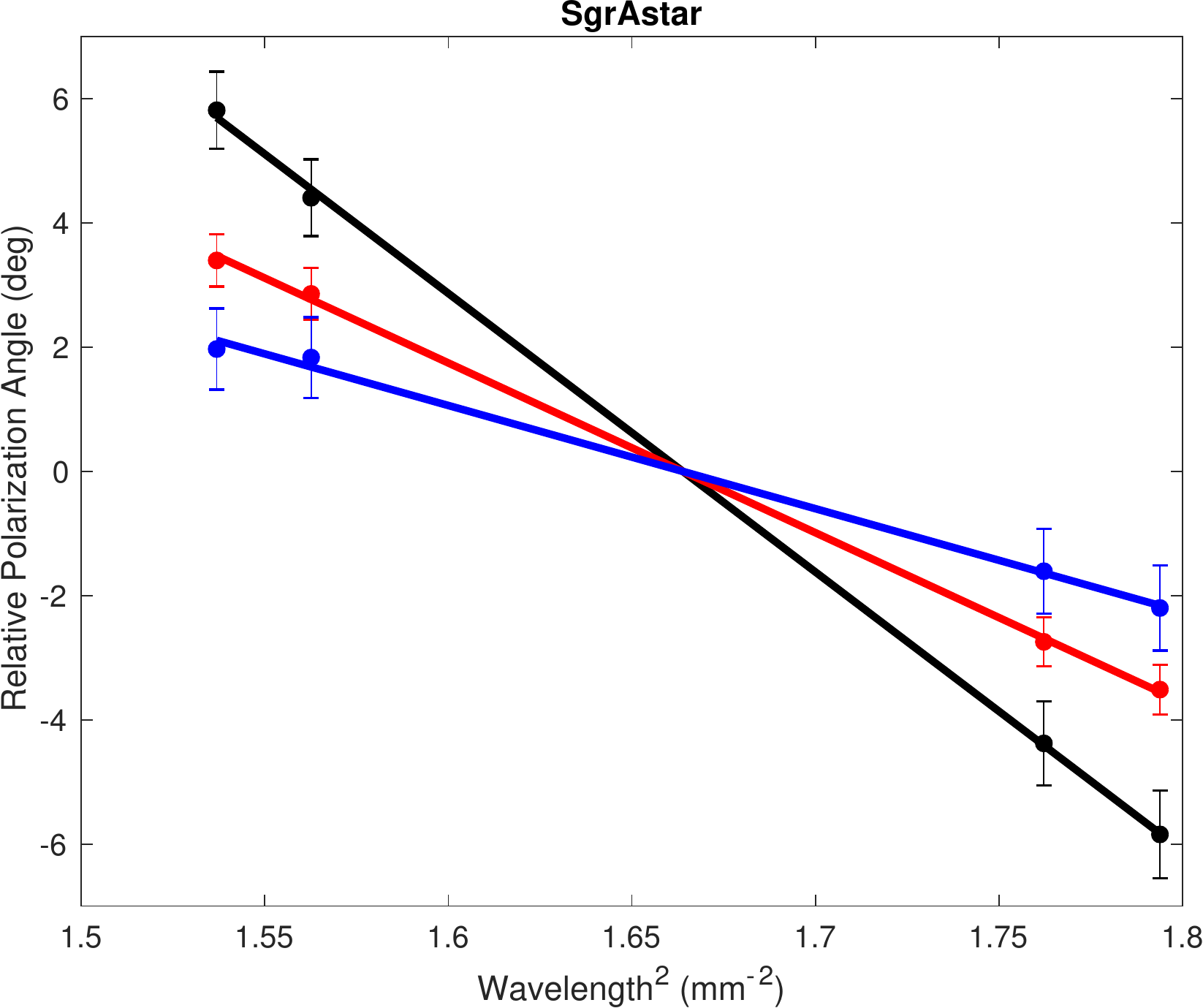}
\caption{Average residual polarization position angle as a function of wavelength-squared for Sgr A* and each of the calibrators in epochs 1 (black), 2 (red), and 3 (blue).  We have removed the mean position angle in each epoch to enable clear comparison.  
\label{fig:avg}
}
\end{figure}

\begin{figure}[t!]
\includegraphics[width=0.33\textwidth]{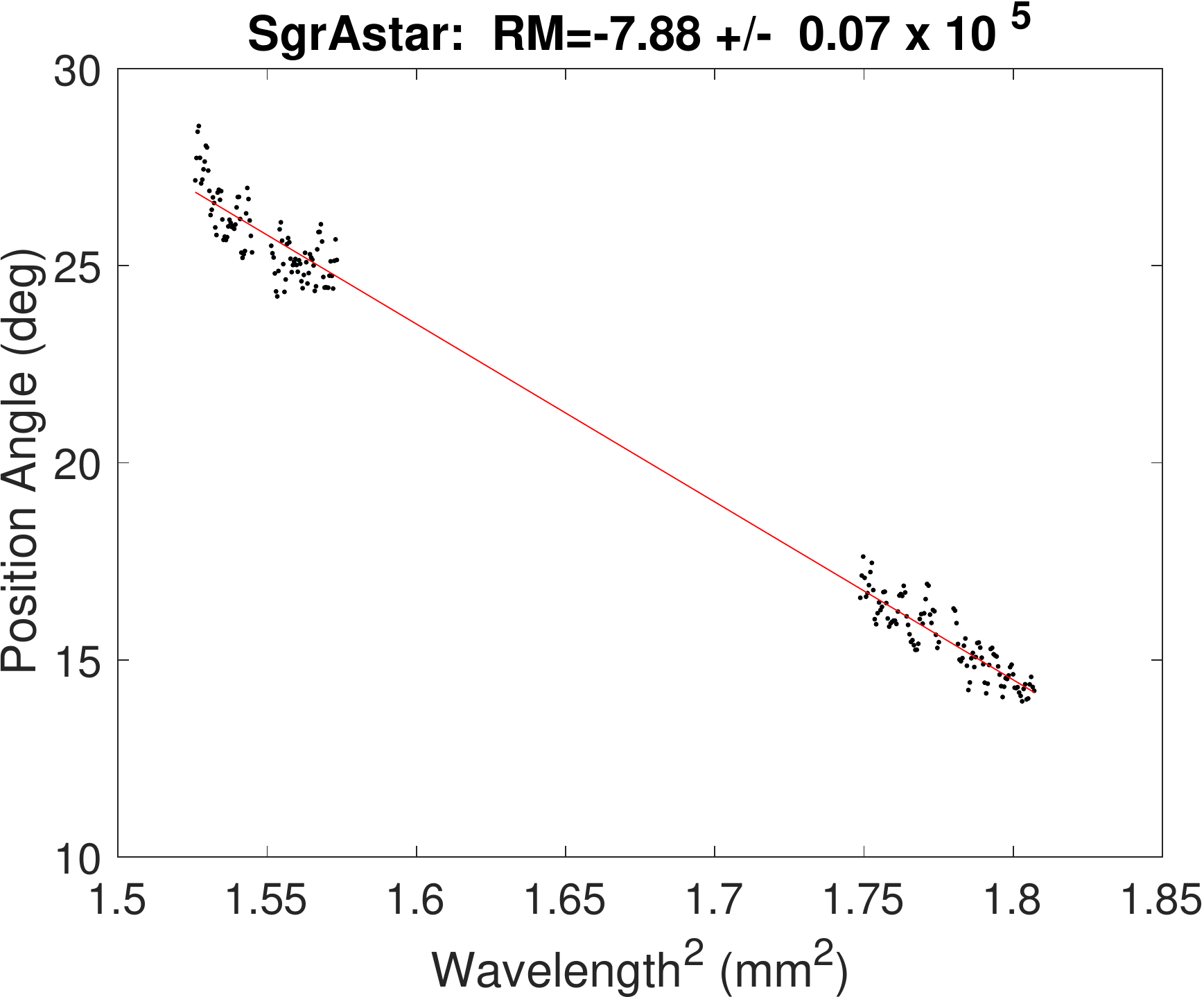}\includegraphics[width=0.33\textwidth]{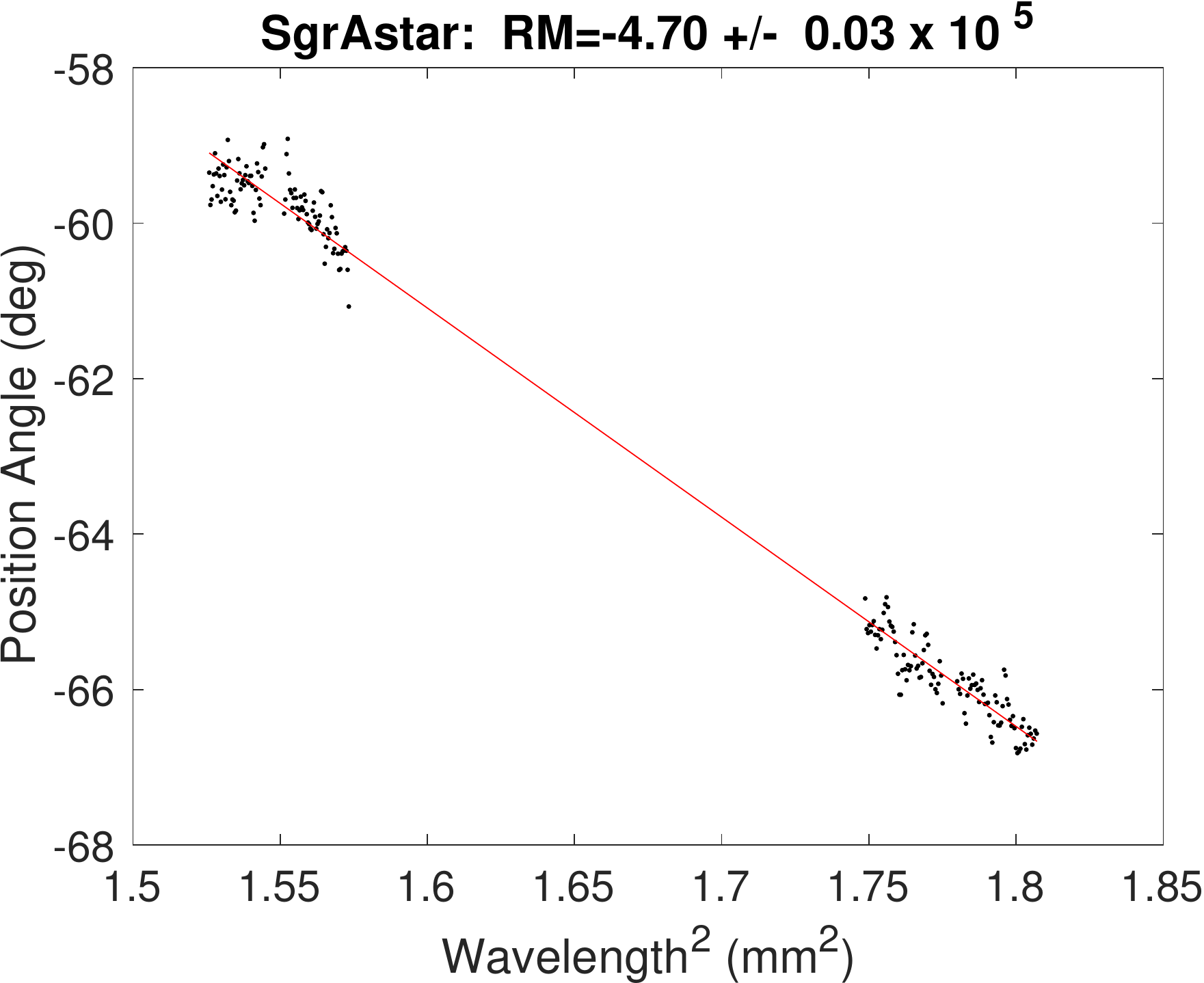}\includegraphics[width=0.33\textwidth]{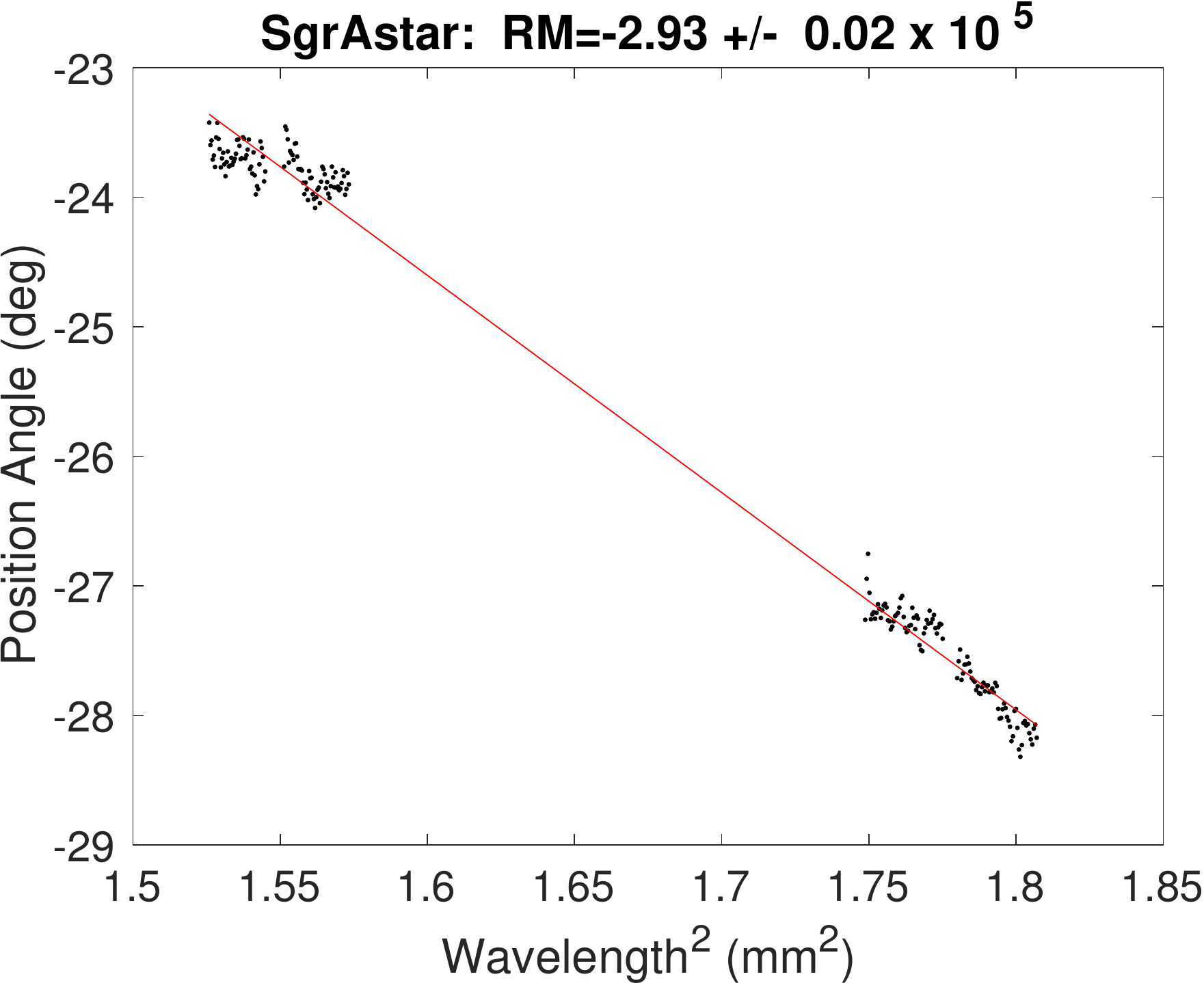}
\caption{Polarization angle as a function of wavelength-squared, presented for each channel and for each epoch.  Epochs 1, 2, and 3
are in left, middle, and right columns.  Note the different scales for position angle.  The title gives the RM in units of $\rdm$.
\label{fig:rmchannel}
}
\end{figure}

In Table~\ref{tab:avg} we summarize the time-averaged polarization properties of
\sgra\ in each epoch and for each SPW.  \sgra\ is detected with high significance
in each epoch, each SPW, and each Stokes parameter.  We show SPW-averaged polarization position angle as a function of wavelength squared in the
three epochs in Figure~\ref{fig:avg}, revealing a clear variation in RM between
these epochs.  In Figure~\ref{fig:rmchannel}, we show the polarization position angle as a function of wavelength-squared for each individual channel, with separate plots for each epoch.  The results 
are consistent in a comparison between the SPW-averaged and channel-averaged 
presentations.

We fit the rotation measure (RM) and the mean-wavelength ($\bar{\lambda}$) position angle ($\bar{\chi}$) following the relation
\begin{equation}
\chi = \chi_0 + RM \lambda^2,
\end{equation}
where $\chi$ is the observed position angle at  wavelength $\lambda$ and
$\chi_0$ is the position angle at zero-wavelength.
An RM of $1.1 \times 10^7 
\rdm$ corresponds to a full rotation of the position angle over the
full frequency range.  Thus, observations with RM $>$ few $\times 10^6 \rdm$
will have $\sim 1$ rad of phase wrap between the highest and lowest frequencies,
which could lead to a phase-wrap ambiguity.
\added{All RM fitting in this paper is done using a weighted least-squares
method of the position angle against $\lambda^2$.  This method is suitable
for the limited range of position angles typically present and the uniformity
of errors in the data.}
In Table~\ref{tab:avgrm}, we summarize the average RM fits to the data.  These fits are also plotted in Figures~\ref{fig:avg} and~\ref{fig:rmchannel}.
The quality of these fits are consistent with no deviation from 
a $\lambda^2$ law for the position angle in the average polarization
properties.  
The significance of the RM detection for \sgra\ is $\gsim 100\sigma$
in each epoch.

\added{In fact, we find the goodness of fit 
$\chi^2_\nu \sim 10^{-3}$ for the fits to the SPW-averaged
data (Fig.~\ref{fig:avg}), suggesting that we are significantly overestimating
the errors in $Q$ and $U$ in our fitting. This is not surprising given
the extremely high dynamic range of these images and the limited number 
of data points (four) contributing to an individual RM calculation.  We 
calculate errors in the RM based on the scatter of the residual phases after 
fitting.
For the individual channel results (Fig.~\ref{fig:rmchannel}), we find
$\chi^2_\nu = 31, 2.5,$ and 0.4 for the three epochs, respectively.
These results suggest their could be some additional systematic
contributions to the RM residual.}

We also fit slopes $\delta$ to $I$, $V$, and $p$ as a function of frequency and summarize these in Table~\ref{tab:avgrm}.  These slopes test whether a non-zero
spectral index is a reasonable fit to these data.  For all of the sources
we see marginal or no evidence for a Stokes $I$ spectral index change over the 
18-GHz range of these observations.  In LP and CP, $\delta p$ 
and $\delta V$ for
\sgra\ are significant, change sign between epochs,  and are an order of magnitude
larger than for other calibrators observed for a full track.

\subsection{Intra-Epoch Properties}

\begin{figure}
\includegraphics[width=0.95\textwidth]{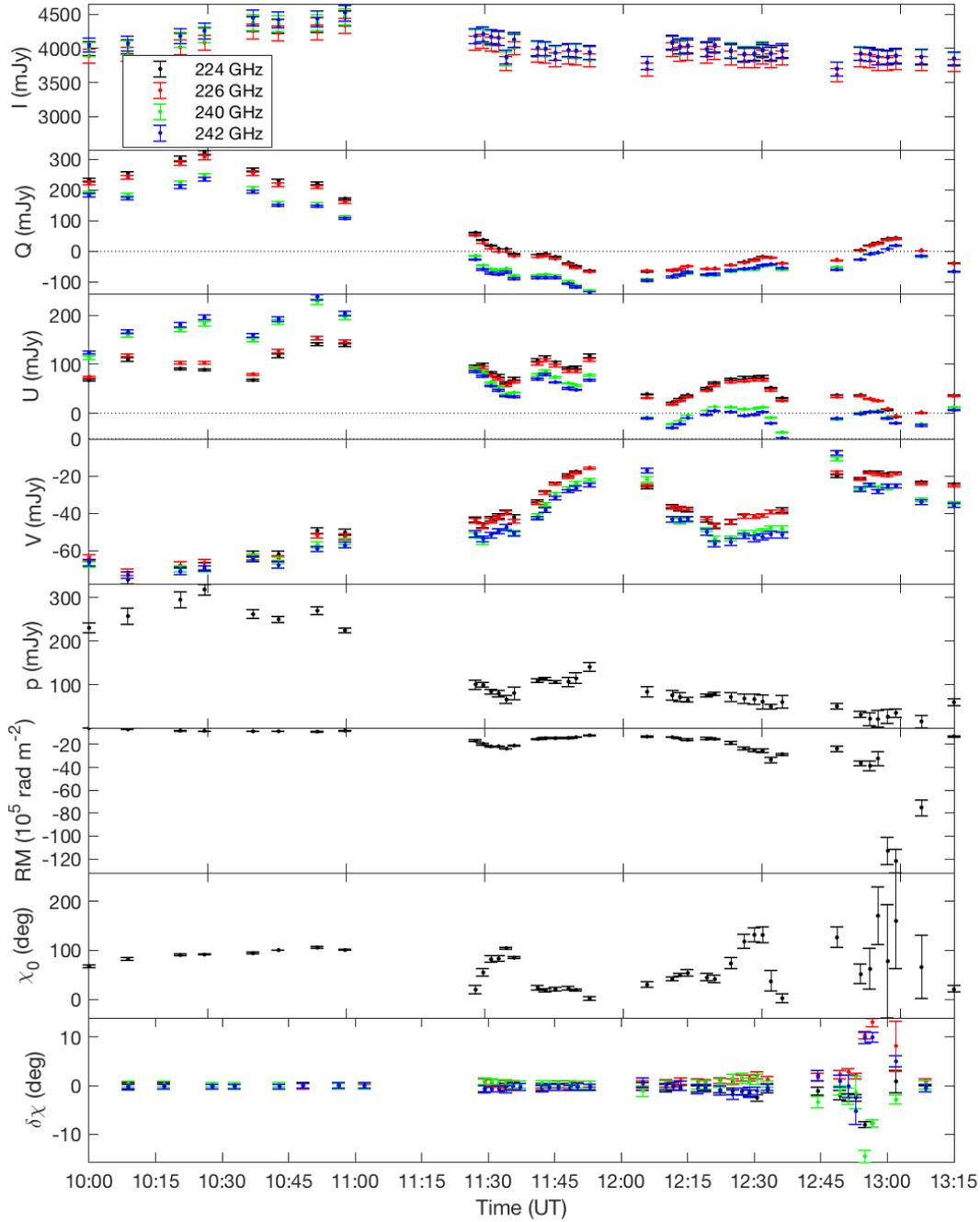}
\caption{Time series of \sgra\ polarimetric data in epoch 1.  The top four 
panels show Stokes I, Q, U, and V for each SPW.  The fifth panel 
shows total linearly polarized intensity ($p$).  The sixth panel
shows RM in units of $10^5 \rdm$.  The seventh panel shows the
zero-wavelength position angle ($\chi_0$).  The eighth panel shows
residual position angle after fitting the RM and position angle.
Where multi SPWs are shown: 224 GHz (black), 226 GHz (red), 240 GHz (green), and 242 GHz (blue).  
\label{fig:sgratime1}}
\end{figure}

\begin{figure}
\includegraphics[width=0.95\textwidth]{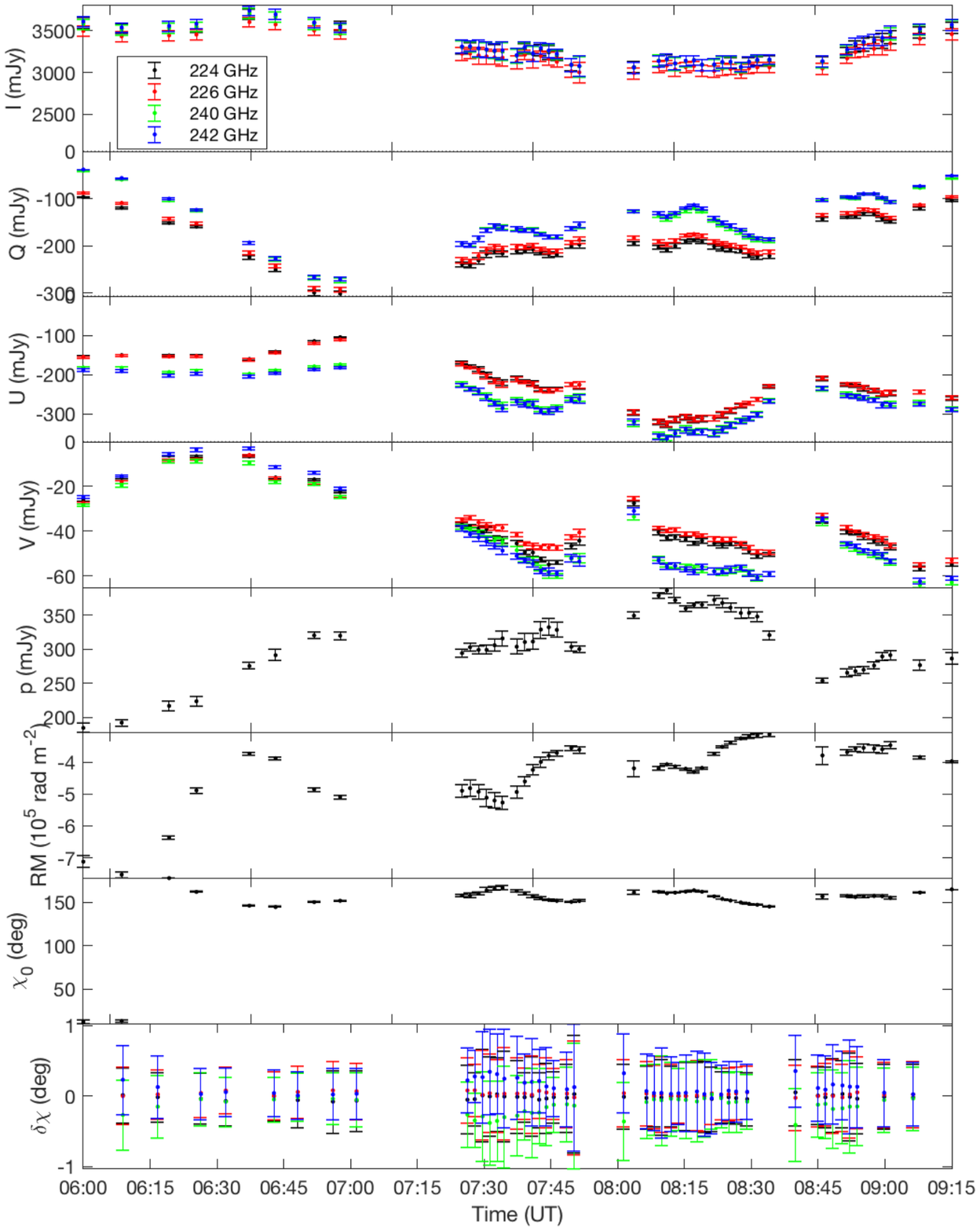}
\caption{Same as Figure~\ref{fig:sgratime1} but for epoch 2.
\label{fig:sgratime2}}
\end{figure}

\begin{figure}
\includegraphics[width=0.95\textwidth]{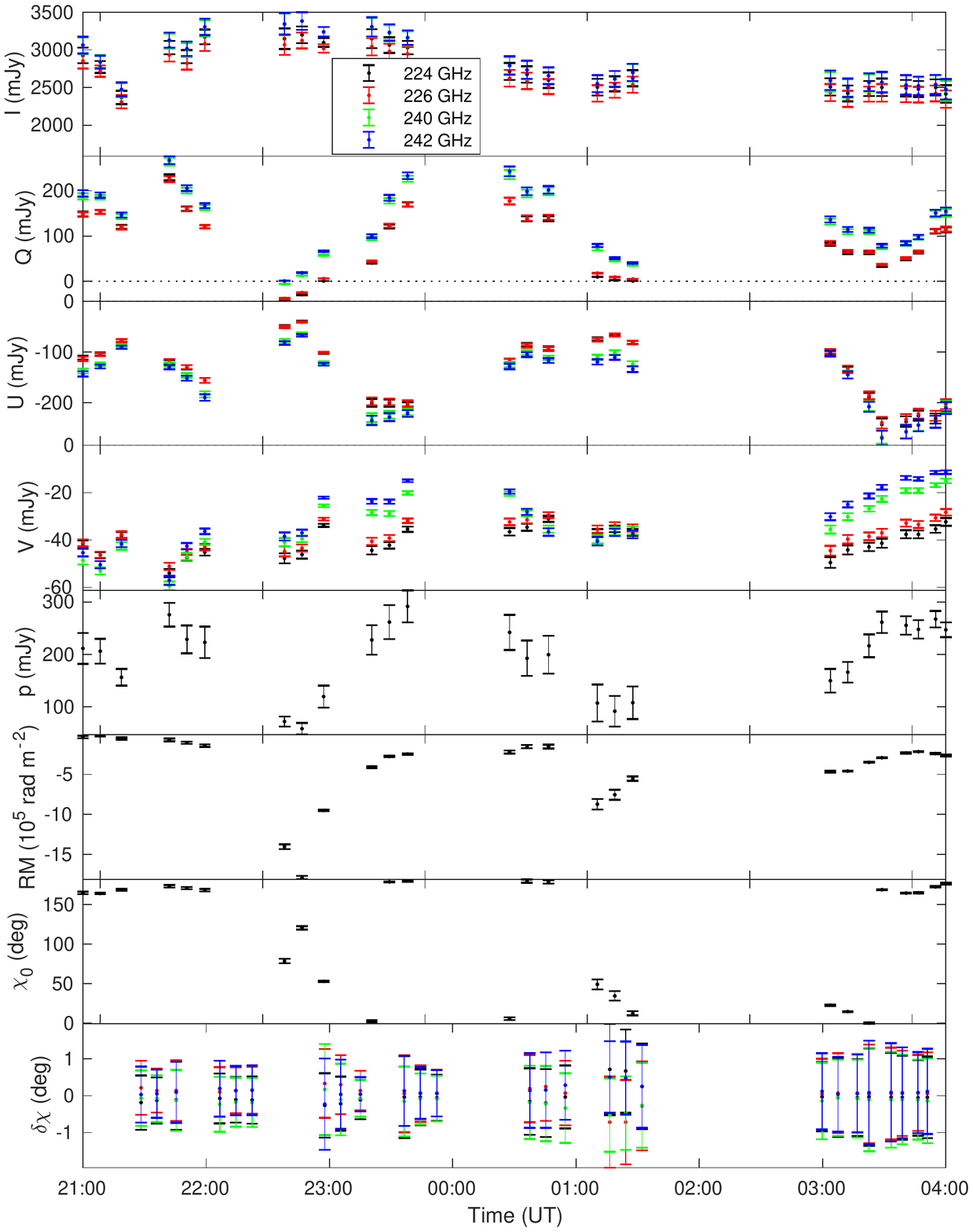}
\caption{Same as Figure~\ref{fig:sgratime1} but for epoch 3.
\label{fig:sgratime3}}
\end{figure}

In Figures~\ref{fig:sgratime1} through ~\ref{fig:sgratime3}, we present intra-epoch light curves in Stokes I, Q, U, and V for Sgr A*.  Data are averaged over individual scans, which range from tens of seconds to 7 minutes in duration.  
We also show time-dependent fits of RM, $\chi_0$, and $p$ to the data for these same sources, as well as the residual position angle after the fits.

We see significant variability in all four Stokes parameters for Sgr A* within each epoch.  We also see SPW-dependent variations in the Stokes parameters.  The variations in RM and $\chi_0$ for \sgra\ are an order of magnitude larger than those seen for the calibrator and check source (Appendix~\ref{sec:appendix}).  Position angle changes in the three epochs are 90 deg, 25 deg, and 50 deg, respectively.  Large intra-epoch polarization angle changes have been previously
seen \citep{2006JPhCS..54..354M,2015Sci...350.1242J}.  Apparent
RM variations are several times $10^5 \rdm$ per epoch and as large as $10^7 \rdm$ in epoch 1.  The largest fitted values of the RM are suspect for three
reasons:  one, the phase-wrap ambiguity at $1.1 \times 10^7 \rdm$; two, 
the large error bars in the estimates and the large residual phases,
indicative of poor quality fits;
and three, the low polarization fraction at the time of these measurements, 
placing our analysis in a regime where calibration errors can have a 
larger effect.

\added{We calculate the goodness of fit statistic $\chi^2_\nu$ for the
results presented in Figures~\ref{fig:sgratime1} through ~\ref{fig:sgratime3}.
For Epochs 2 and 3, $\chi^2_\nu \approx 0.1$ at all times.  For epoch 1,
$\chi^2_\nu \lsim 1$ up to approximately 12:15 UT.  After this time,
$\chi^2_\nu >> 1$.  The small values of $\chi^2_\nu$ in epochs 2 and 3
and the beginning of epoch 1 are consistent with 
excellent fits to standard Faraday rotation with 
a modest overestimate of errors at some of the time.  
The large scatter in $\delta\chi_0$ in the latter half of epoch 1
reflect the time in which the $\chi^2_\nu >> 1$.  These poor fits
come at times when the polarization fraction is the lowest.}

We do not explore in detail the time variable properties of the RM determined
through fitting of the highest-frequency resolution (64-channel per SPW) data.  
In Figures~\ref{fig:polchantime1},~\ref{fig:polchantime2}, and~\ref{fig:polchantime3}, we show selected scans from each epoch
at high frequency resolution.  
These results
are consistent with the SPW-averaged results.
That is, we find similar values of RM and $\chi_0$ from these data.
Further, the slopes within each SPW are consistent with the slope between
SPWs.
In principle, these data can be searched for multiple RM components and/or
non-$\lambda^2$ effects.  
However, 
the possibility for uncalibrated and time-variable 
systematic error in the 
\replaced{highest-frequency}{highest frequency resolution}
polarization data prevents
us from making further use of these results.  
\citet{2016ApJ...824..132N},
for example, find frequency-dependent polarization leakage terms that,
if not properly calibrated, could be misinterpreted as complex Faraday
signatures.

\added{These high-frequency resolution plots give a clear indication of
where non-$\lambda^2$ effects are most prominent, because the slopes within and between SPWs can be more directly compared.  For instance, the
fit in epoch 1 at 12:54 UT gives ${\bf |RM|} > 10^7 \rdm$ but clearly
the slope between adjacent SPWs is consistent with a much smaller value
of the RM.  The fit at 12:56 UT shows similar properties but derives
a ${\bf |RM|} \approx 10^6 \rdm$ in part because of the phase wrap.
In epoch 3, the fit at 01:24 UT also shows a broken slope, although 
the SPW-averaged fit at this time returns $\chi^2_\nu < 1$.}

\section{Analysis}

The results of these observations are qualitatively similar to previous 
measurements of Sgr A*.  But they are significantly more accurate, which 
enables a more detailed picture of the polarization properties as a function
of time and frequency.  We obtain several significant results, which we explore
in greater detail below:  1) polarized intensity variability on a time scale
of months, with a variable spectral index; 2) confirmation of the presence
of circular polarization and detection of variability on a time scale 
of months; 3) variability in the linear and circular polarization on 
a time scale of hours; 4) variability in the rotation measure on a time 
scale of months, while remaining consistent with the long-term average
and sign; and, 5) short-term variability in the rotation measure, that is
coupled with changes in the polarized intensity and position angle.  

The agreement in mean RM with the historical value and the small variations
in the epoch-averaged properties strongly support the interpretation
of Faraday rotation arising in the accretion flow with contributions
dominating at radii $10^3$ - $10^5 R_S$.  Under the RIAF interpretation, 
we find consistency with a constraint of $\dot{M} \sim 10^{-8} \msuny$
\citep{2007ApJ...654L..57M}.  
The uncertainty in $\dot{M}$ arise from assumptions about the accretion flow
model, rather than measurement uncertainty.

\subsection{Average Polarized Intensity}

\begin{figure}
\includegraphics{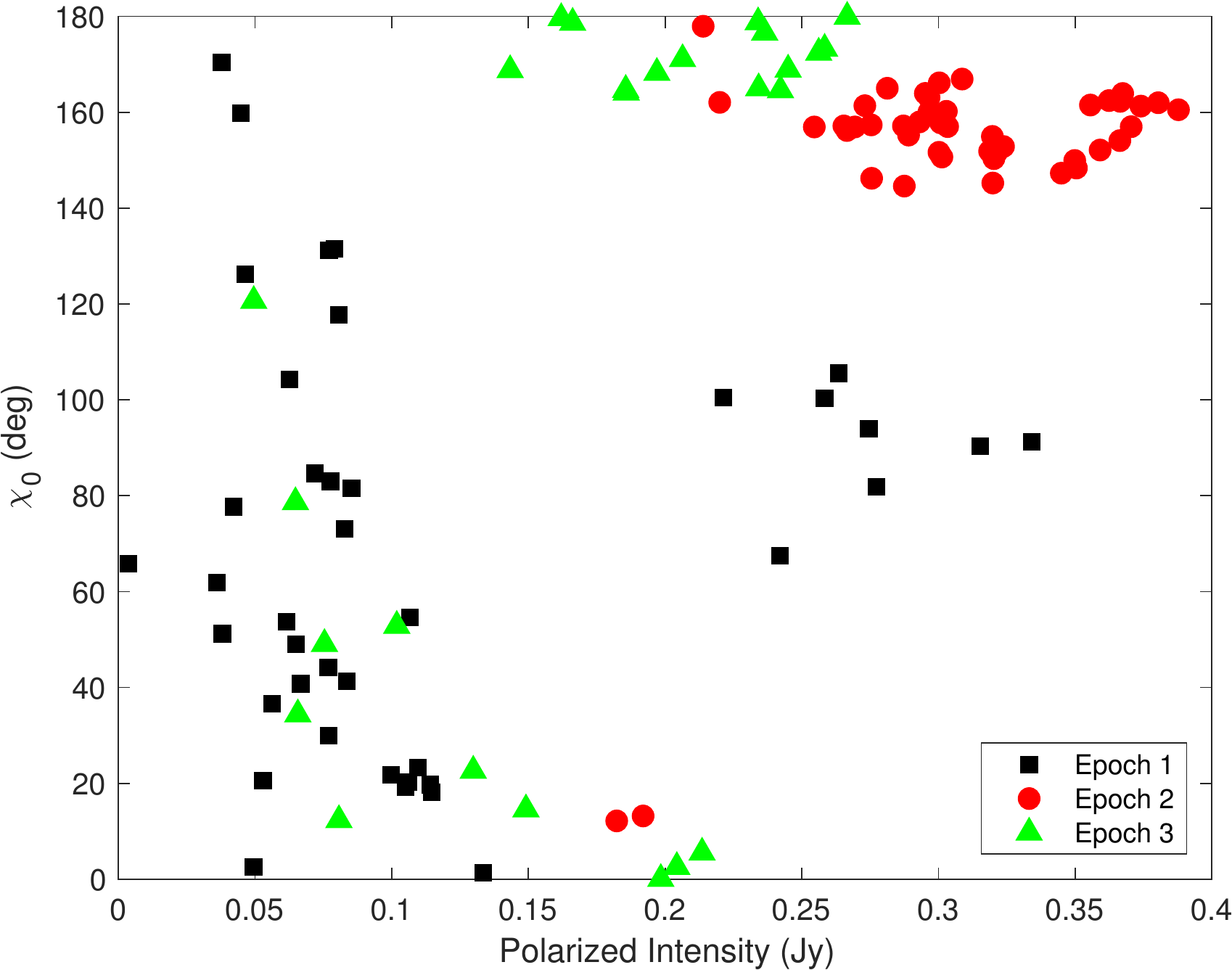}
\caption{Zero-wavelength position angle as a function of polarized intensity
for Sgr A* in 3 epochs.
\label{fig:polangle}
}
\end{figure}

The mean LP between epochs ranges from 3.6 to 7.8\%, consistent with previous measurements \citep{2000ApJ...534L.173A,2003ApJ...588..331B,2006ApJ...646L.111M,2007ApJ...654L..57M}.  The degree of LP appears to be independent of the total intensity per epoch.  The total intensity has a flat spectrum across our frequency range.  The polarized intensity, on the other hand, shows evidence for a slope across the band, corresponding to a change of $\lsim 20$\% with a time variable sign.
Analyses of 
unpublished SMA and CARMA data suggest that there may be a preferred
position angle at zero wavelength, $\chi_0 \sim 180$ deg.  
Epochs 2 and 3 show similar values for $\chi_0$ but epoch 1 has a
mean value $\chi_0=95.3 \pm 0.9$ deg (Table~\ref{tab:avgrm} and
Figure~\ref{fig:polangle}).  We 
conclude that there is 
preferred intrinsic position angle, near 180 deg, but there is 
substantial variability in the intrinsic position angle.

We find a mean CP of $\approx -1.1 \pm 0.2$\%.  We find a change in the CP across the band that is as large as 25\% in the third epoch.  The sign of the frequency-dependent slope is time-variable.  We note that for the three epochs, there is a linear correlation between $\delta P$ and $\delta V$, but there is no clear connection
between $P$ and $V$. There is also no apparent relationship between either of the polarized and total intensity quantities.

We confirm the SMA detection of CP with a value of $-1.2 \pm 0.3$\% at 1.3 mm and $-1.6 \pm 0.4$\% at 0.86 mm \citep{2012ApJ...745..115M}.  The results also suggest that the handedness of the millimeter-wavelength CP is stable on time scales greater than the 11 year span between the earliest SMA observation and the latest ALMA observation.  This stability mirrors that of the centimeter-wavelength CP, which has been shown to be stable for greater than 20 years \citep{2002ApJ...571..843B}. If the centimeter and millimeter wavelength CP originate via the same mechanism, then the handedness of that mechanism is apparently stable over almost 40 years.

The LP is best explained through an
origin in synchrotron emission close to the event horizon.
\citet{2012ApJ...745..115M} provide a detailed discussion of potential
origins for the CP emission.  Faraday conversion
is the favored mechanism for the production of CP in which thermal
electrons that are co-spatial with the relativistic synchrotron-emitting
electrons.
A coherent magnetic field on the scale of the $\tau=1$ surface is required
to produce Faraday conversion and the stable sign of the CP seen at all
wavelengths.  In a uniform medium, the sign of CP is expected to alternate
as the phase shift between two linear polarizations is $\propto n_e B^2 \lambda^3$.  This leads to frequent reversals at long wavelengths.  Alternatively,
for a stratified synchrotron source an appropriate scaling of
the electron density and magnetic field with radius can 
counter the wavelength dependence and lead to an apparently
flat spectrum.  We then require a stable
magnetic field geometry on scales of the source size,
which ranges from a few $R_S$ to hundreds or thousands of $R_S$.
One specific model for achieving the proper stratification is through
magnetic field shear in the accretion flow
(Broderick et al., in prep).

\begin{deluxetable}{lrr}
\tablecaption{Historical Measurements of Rotation Measure \label{tab:rmhist}}
\tablehead{
\colhead{Source} & \colhead{Date} & \colhead{RM} \\
                 &                & \colhead{($10^5 \rdm$)} \\
}
\startdata
   BIMA+JCMT & 2002 Apr 01 & $ -4.3 \pm  0.1 $ \\ 
       \dots & 2004 Apr 01 & $ -4.4 \pm  0.3 $ \\ 
         SMA & 2005 Jun 04 & $ -6.7 \pm 2.9 $ \\ 
       \dots & 2005 Jun 06 & $ -23.1 \pm 12.6 $ \\ 
       \dots & 2005 Jun 09 & $ -5.0 \pm 1.7 $ \\ 
       \dots & 2005 Jun 15 & $ -11.7 \pm 13.6 $ \\ 
       \dots & 2005 Jun 16 & $ -5.4 \pm 1.8 $ \\ 
       \dots & 2005 Jun 17 & $ -22.3 \pm 7.4 $ \\ 
       \dots & 2005 Jul 20 & $ -7.5 \pm -.6 $ \\ 
       \dots & 2005 Jul 21 & $  1.1 \pm 8.2 $ \\ 
       \dots & 2005 Jul 22 & $ -3.7 \pm 1.8 $ \\ 
       \dots & 2005 Jul 30 & $ -4.8 \pm 1.4 $ \\ 
       \dots & 2006 Jul 17 & $ -5.6 \pm 1.6 $ \\ 
       \dots & 2007 Mar 31 & $ -3.7 \pm 0.6 $ \\ 
\enddata
\end{deluxetable}

\subsection{Rotation Measure in Multiple Epochs}

We show clearly that the RM has varied across the three epochs of our observations.  Even with a conservative estimate of a random systematic error per epoch of $10^5 \rdm$, the change $\delta {\rm RM} = 4.93 \times 10^5 \rdm$ across the three epochs is detected at a $5 \sigma$ threshold.  Using only the thermal error, the change has a significance of $>30\sigma$.  

\begin{figure}[t!]
\includegraphics[]{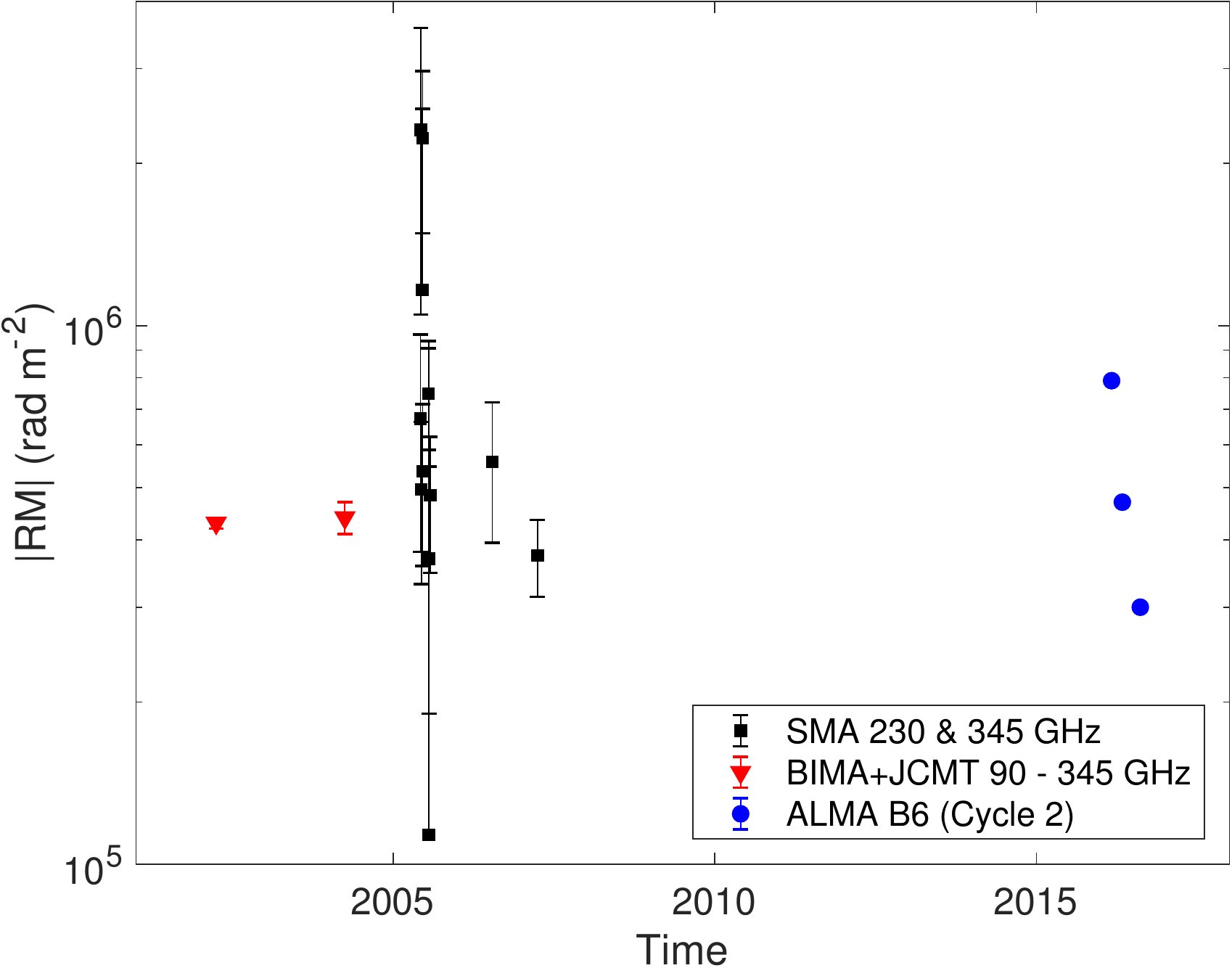}
\caption{RM absolute value versus time from these new and previously published measurements.
BIMA+JCMT measurements \citep{2003ApJ...588..331B,2006ApJ...646L.111M} were widely separated in time but covered a broad
frequency range; thus, they have a small statistical error but a large systematic 
uncertainty due to variability.  SMA measurements were obtained simultaneously
but with a small frequency range \citep{2007ApJ...654L..57M}.
ALMA measurements are averaged over multi-hour epochs, which show significant
variability in the RM on shorter time scales.  ALMA error bars are smaller 
than the displayed points.
\label{fig:rmtime}
}
\end{figure}

In Table~\ref{fig:rmtime} and Figure~\ref{fig:rmtime}, we summarize published historical measurements of the RM.  The RM values that we find fall within the bounds of previous variations.  But none of these previous variations could be determined to be significant given the lower SNR of detection and small lever arms.  Additional unpublished RMs obtained with CARMA and SMA in intervening years fall within the same range.

\begin{figure}[t!]
\includegraphics{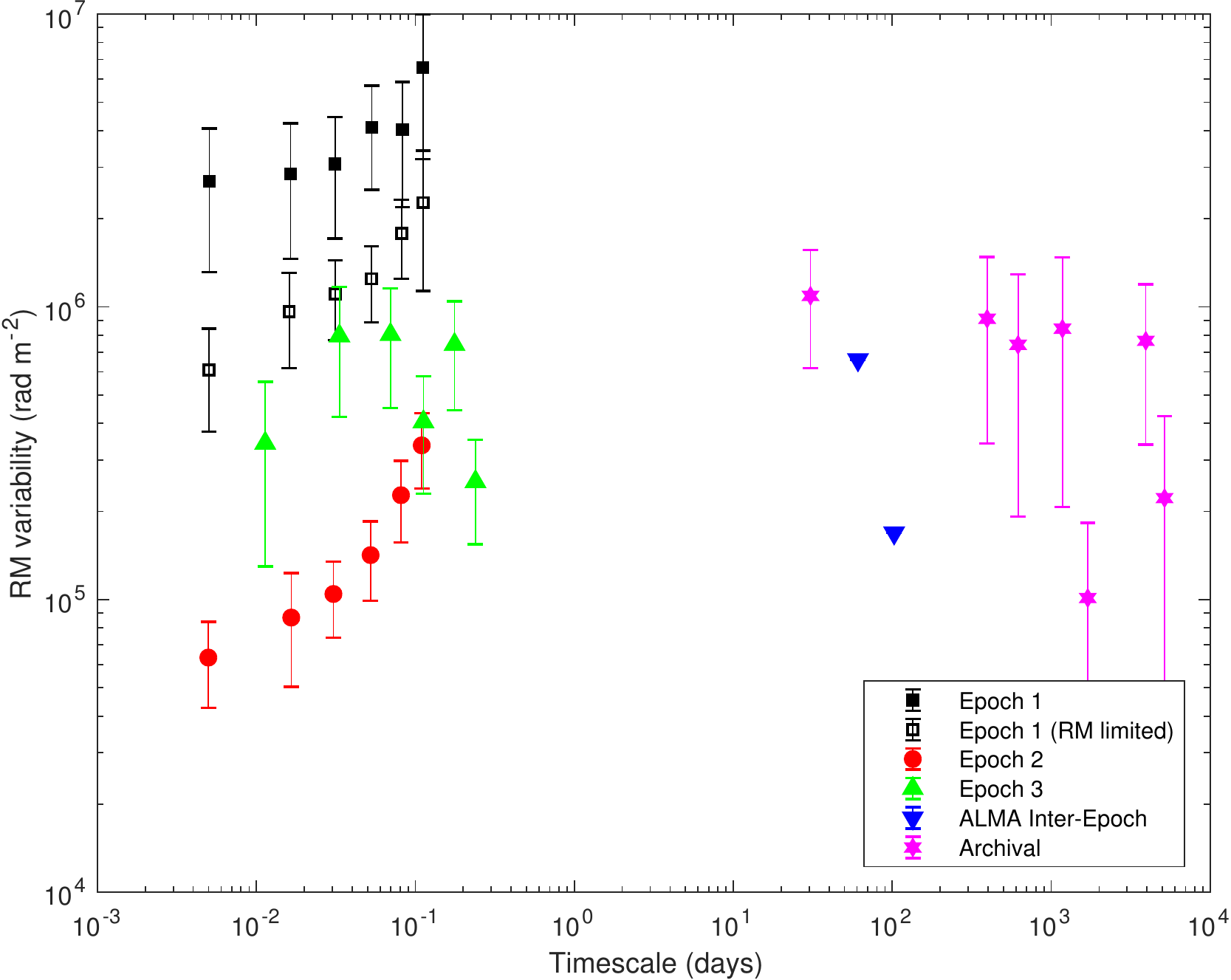}
\caption{RM variability as a function of time scale. We show
the square root of the structure function calculated for individual
epochs and in between epochs.  Open symbols for epoch 1
are calculated excluding the largest and most ambiguous RMs with
$|RM| > 5 \times 10^6 \rdm/$. \label{fig:rmvar}}
\end{figure}

We analyze variability in the RM using the structure function 
\begin{equation}
{\rm SF^2 (\tau)} = \langle ({\rm RM}(t+\tau) - {\rm RM}(t))^2 \rangle .
\label{eq:sf}
\end{equation}
The structure function calculates the characteristic variability on
a timescale $\tau$.  While the structure function has some known
limitations in accurately determining saturation time scales \citep{2010MNRAS.404..931E},
it is suitable for our purposes.  In Figure~\ref{fig:rmvar}, we
show the SF calculated for each epoch individually and for all of the
epoch-averaged results, archival and current.  
\added{Our data are too sparsely sampled to achieve the statistical
ensemble average implied by Equation~\ref{eq:sf}.  
Errors are determined from the scatter of RM differences measured within
a time bin.  Hence, the error is not defined when we calculate the
SF for the ALMA inter-epoch time scales; these points should be
treated as instances to be included in the larger ensemble.  Further,
we cannot properly address uncertainty in the SF
on the longest time scales since we are only capturing a small number
of instances from the ensemble.  We would, for instance, not clearly detect
a red-noise spectrum with a characteristic time scale of $\gsim 10$ years
in this data.}

\added{Nevertheless, we are able to draw some conclusions from this analysis.}
We observe more than
an order of magnitude range in the SF on hour timescales.  This range
is somewhat reduced but still large when we excluded the largest RMs in Epoch 1.
The dominant
result from this structure function analysis is that we do not identify
any characteristic time scale between hours and decades for variability
of the RM.  
That is, RM variability is driven on
a wide range of scales, from as close in as $10 R_S$ all the way out
to the Bondi radius.  

\citet{2011MNRAS.415.1228P} perform 
numerical simulations of magnetized accretion flows with weak convection
to determine the time
scale and magnitude of RM variations.  Variability saturates 
on a Bondi time, $t_B \approx 100$ yr, but can be significant on
shorter time scales.
The characteristic timescale for variability is determined as
\begin{equation}
\tau = 20 \left( R_{rel} / R_B \right)^2 \left( R_{in} / R_B \right)^{-1/2} t_B,\end{equation}
where $R_{rel}\sim 10 R_S$ is the radius at which electrons become relativistic, $R_B$ is the Bondi radius, and $R_{in}$ is the reconnection scale in the simulation.  For $R_{in}/R_B=10^{-5}$ as recommended by
\citet{2011MNRAS.415.1228P},  we calculate $\tau \sim 1$ yr, with order of magnitude
uncertainty.  It is difficult to reconcile the flat structure function that
we observe with a distribution that has a characteristic time scale.  The
results suggest a more complex accretion flow structure than currently modeled
or the existence of multiple processes contributing to RM variability.

\citet{2007ApJ...671.1696S} also performed 
MHD simulations and study of accretion flows with emphasis on
geometry.  Their conclusions focus on orientation effects.  
The equatorial plane of the accretion disc is modeled to be highly turbulent.
Thus, a viewing angle through the plane would lead to RM sign reversals, which
are not seen in our data.
Note that the GC pulsar RM$=-7 \times 10^4 \rdm$ sets a threshold
for the external RM that needs to be removed
but there are still no reversals that would be seen.
On the other hand, polar viewing angles will produce a significant 
variability on a time scale of hours but with a consistent sign over
long periods of time.

These ALMA observations were originally obtained with the goal of searching for RM
changes due to disruption of the G2 cloud \citep{2012Natur.481...51G}.  Numerical
models suggested that tidal forces could lead to a change in the accretion
rate that would change the RM \citep[e.g.,][]{2012ApJ...752L...1M}.  
The cloud reached pericenter in 2014, approximately 2 years before these
observations were obtained. The cloud appeared to remain 
intact after close passage although there may be large scale diffuse
features \citep{2017ApJ...840...50P,2017ApJ...847...80W}.  There is
no evidence for enhanced accretion onto \sgra \citep{2014HEAD...1410004H,2015ApJ...802...69B}.
The epoch-averaged RM changes are linear with time,
potentially the result of a secular change in RM but also 
within the bounds of archival variability and intra-epoch variability.  
Longer term monitoring could detect a continuation of any secular trend
and test the hypothesis of an enhanced accretion rate as the result of
tidal streamers from G2 and other features.

\subsection{Polarization and Rotation Measure Variations within Epochs}

We see clear variations in the polarization properties within epochs as well as
between epochs.  Total intensity variations within epochs 
are relatively small ($\sim 20\%$)
and there are no well-defined flaring events in total intensity.
The spectral index of the total intensity remains essentially flat
during each epoch, as well.  Stokes Q, U, and V flux densities, however,
are seen to vary by as much as 100\%.

\begin{figure}
\includegraphics{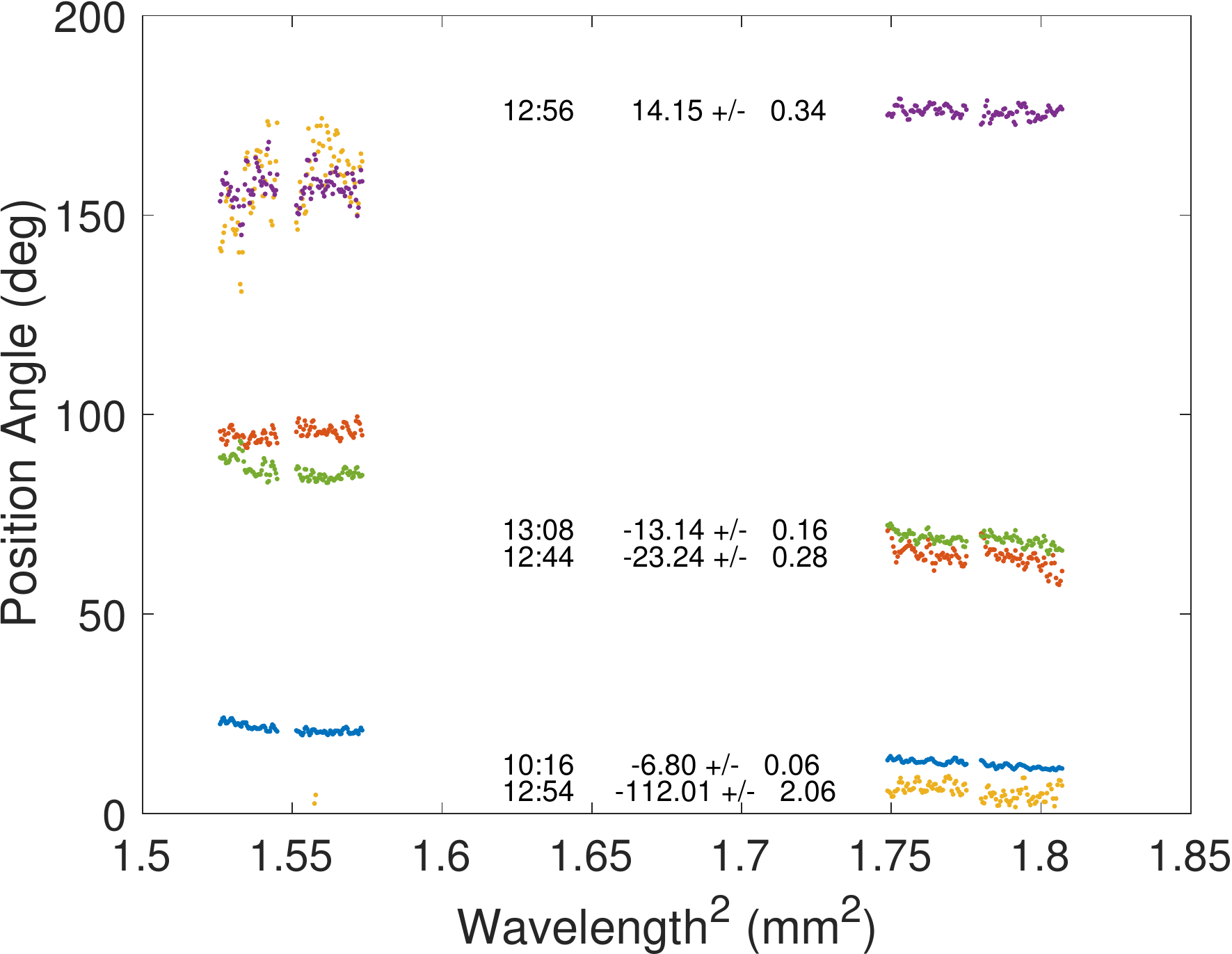}
\caption{Polarization angle versus wavelength squared for a selected
set of scans from epoch 1.  The labels give the time of the scan and
the fitted RM in units of $10^5 \rdm$ and are aligned with the longer
wavelength points to their right.
\label{fig:polchantime1}
}
\end{figure}

\begin{figure}
\includegraphics{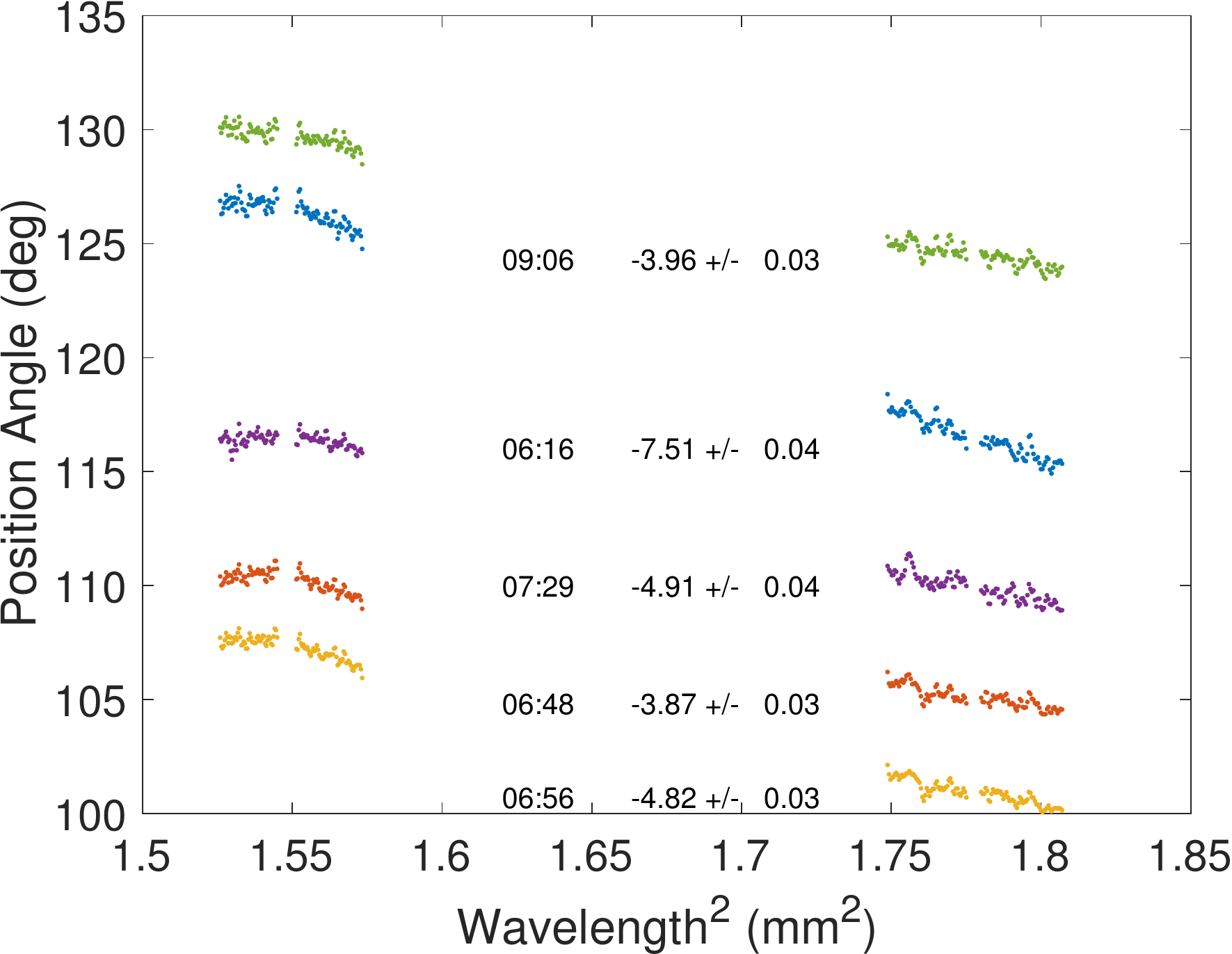}
\caption{Polarization angle versus wavelength squared for a selected
set of scans from epoch 2.  The labels give the time of the scan and
the fitted RM in units of $10^5 \rdm$ and are aligned with the longer
wavelength points to their right.
\label{fig:polchantime2}
}
\end{figure}

\begin{figure}
\includegraphics{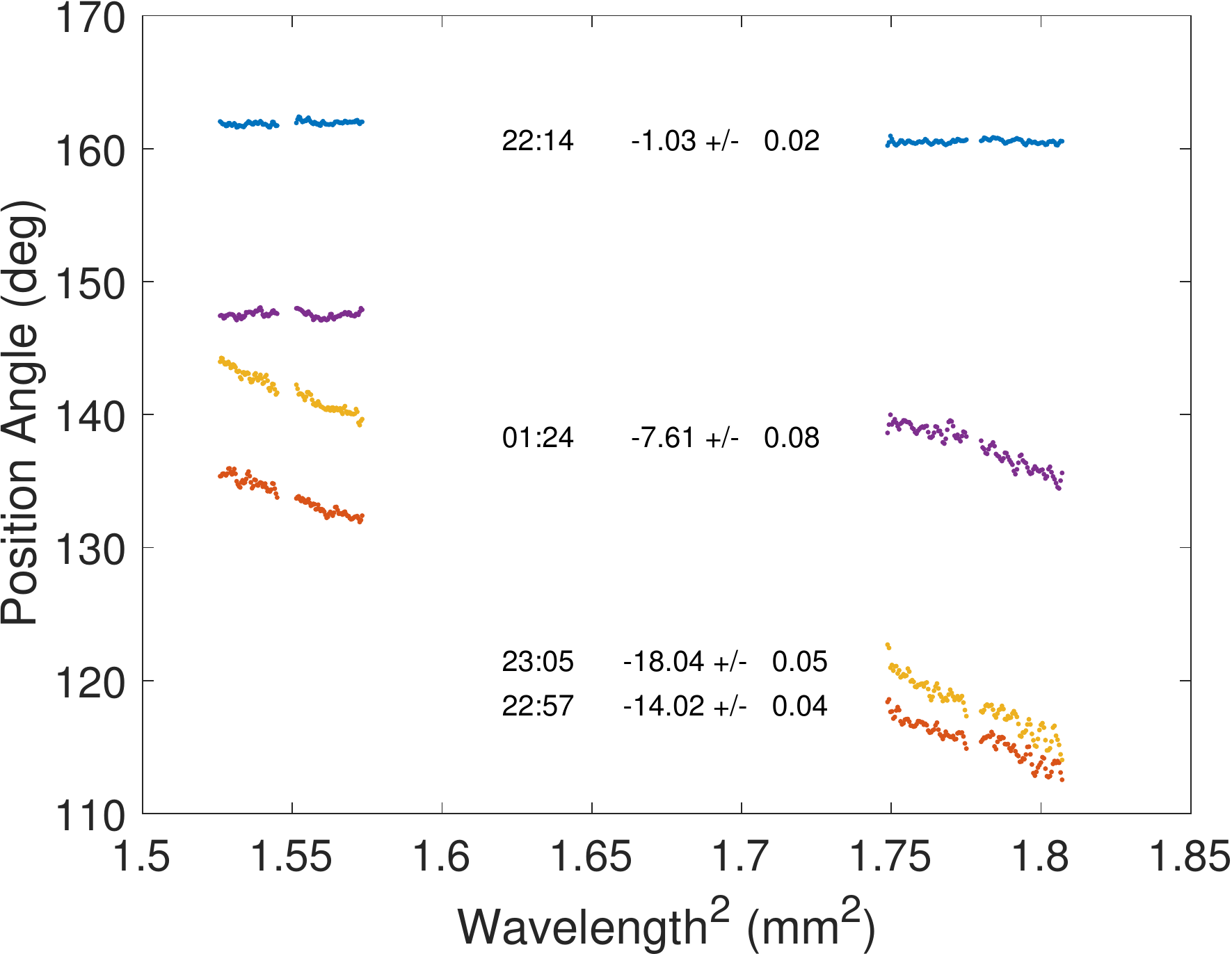}
\caption{Polarization angle versus wavelength squared for a selected
set of scans from epoch 3.  The labels give the time of the scan and
the fitted RM in units of $10^5 \rdm$ and are aligned with the longer
wavelength points to their right.
\label{fig:polchantime3}
}
\end{figure}

\begin{figure}[t!]
\includegraphics[height=0.5\textheight]{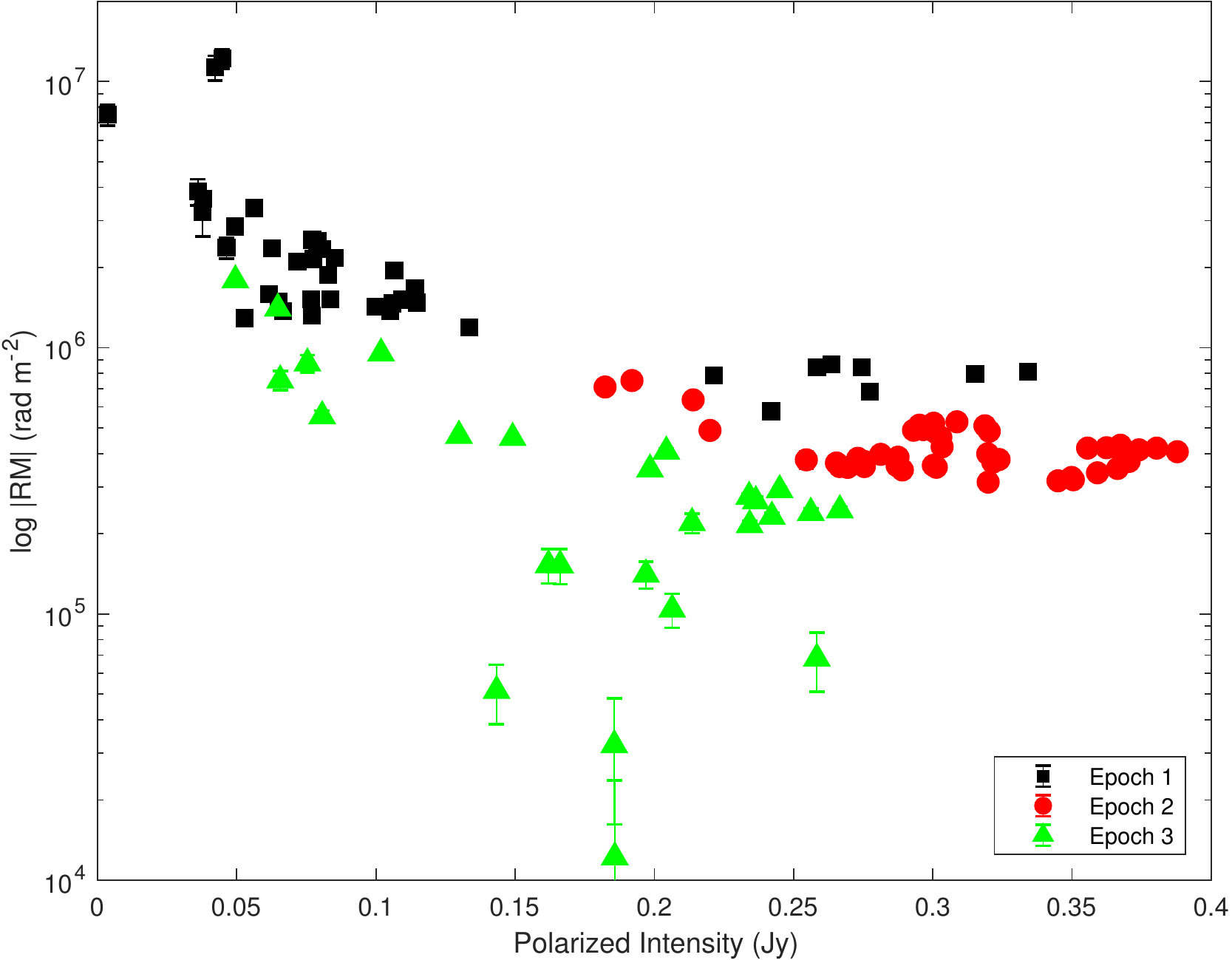}
\includegraphics[height=0.5\textheight]{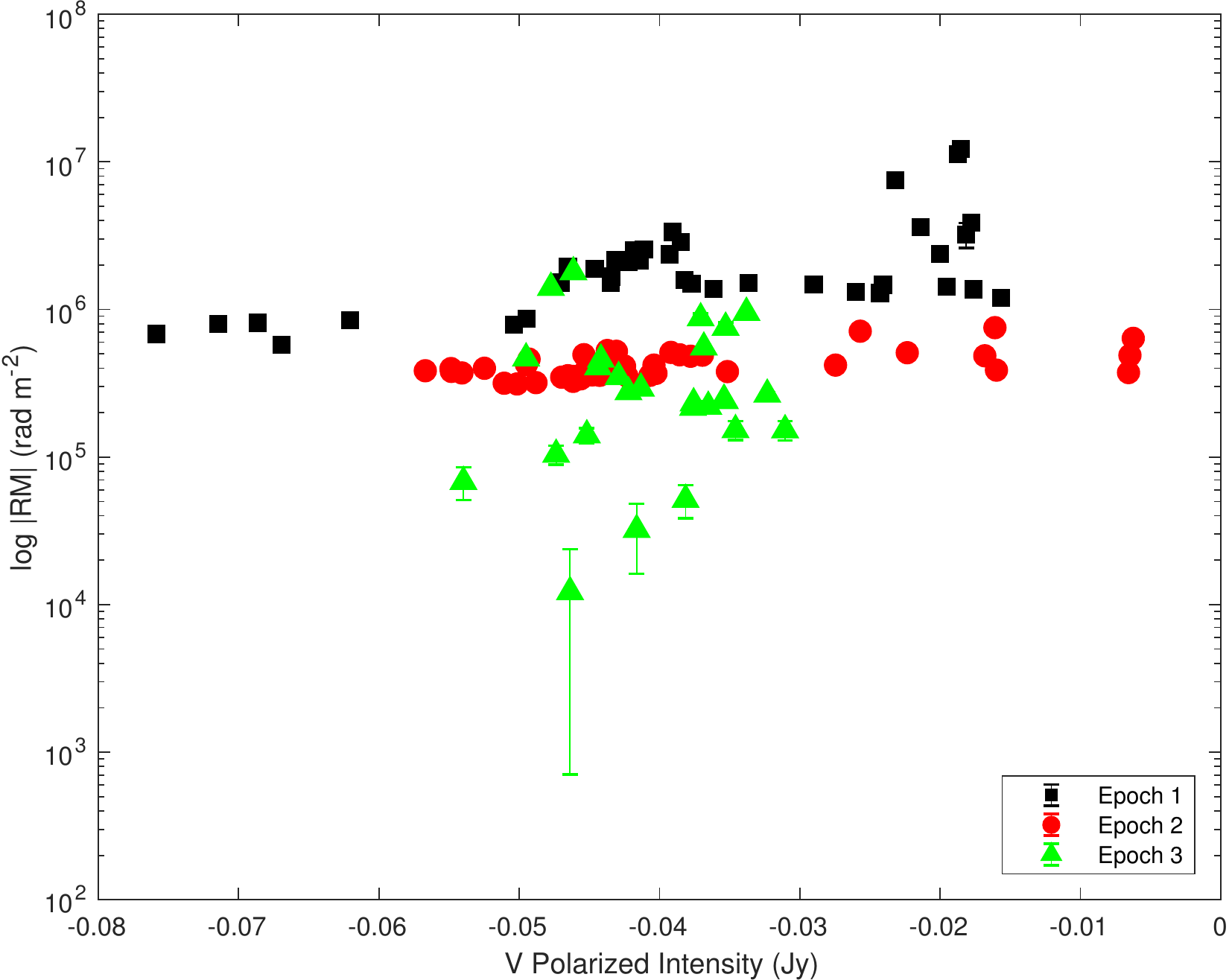}
\caption{Linearly (top) and circularly (bottom) polarized intensity versus $|{\rm RM}|$ for Sgr A* in 3 epochs.
\label{fig:polrm}}
\end{figure}

We show in Figure~\ref{fig:polrm} that variations in the RM appear
to be coupled with the LP flux density.  The lower the polarization
flux density, the higher the absolute value of the RM.  The relation appears
to have an inflection point near a polarized flux density of 100 mJy.  
Below this point, we see the $|{\rm RM}|$ become substantially larger
than $10^6 \rdm$.  At high values of the polarization flux density, the
RM asymptotes to a value near $-5 \times 10^5 \rdm$.  The large
variations in RM do not contribute significantly to the epoch average 
because they are weighted by the polarization fraction.  We find
no correlation between variations in the CP fraction
and the RM, as might be expected for the case where the 
CP is generated through conversion.

\subsection{Non-Faraday Variations}

We have focused on a Faraday interpretation for variations in the polarization
angle with frequency and time.  But it is also possible that 
intrinsic emission processes can produce similar effects.  In particular,
we know from mm VLBI polarimetry that the polarization structure of
\sgra\ is not simple, i.e., not produced in a homogeneous structure
with a single polarization structure \citep{2015Sci...350.1242J}.  While the
VLBI observations cannot be uniquely translated into a map, they clearly
require structure in the polarization vector field on scales smaller than
the source size.  Further, the fact that the total intensity spectrum
peaks at mm wavelengths, implies that some or all of the emission regions
are near the optical depth unity surface \citep{2015ApJ...802...69B}.  Changes 
in individual regions in optical depth or polarization vector magnitude
or orientation can lead to changes in the integrated polarization 
as a function of wavelength.  This could lead to destructive interference
of the polarized signature, a change in the apparent RM, or non-$\lambda^2$
effects.

One of the best tests of intrinsic rather than Faraday origins for
polarization variability is to search for deviations from the $\lambda^2$
law.  
In Figures~\ref{fig:sgratime1},~\ref{fig:sgratime2}, and~\ref{fig:sgratime3},
we show position angle residuals 
after fitting an RM to each integration for \sgra.  For epochs 2 and 3,
the data are all consistent with a Faraday interpretation.  
But for
the end of epoch 1, we see large residuals,
suggesting that a Faraday interpretation is not a good fit.  The largest
deviations from a Faraday interpretation occur for apparent RM $> {\rm few} \times
10^6 \rdm$, or polarized intensity $<100$ mJy.  These differences are also apparent
in Figures~\ref{fig:polchantime1},\,~\ref{fig:polchantime2}, and~\ref{fig:polchantime3}.

The interpretation of these complex wavelength-dependent polarization effects
will likely require comparison with numerical modeling efforts
\citep{2012ApJ...755..133S,2017ApJ...837..180G,2018MNRAS.475...43M,2018MNRAS.478.1875J}.

\section{Conclusions}

These new ALMA results suggest a physical model of long-term stability
coupled with short-term variability in the polarization properties.  The 
long-term stability in the Faraday rotation 
suggests a stable magnetic field configuration and an origin for
much of the Faraday rotation at large radii from \sgra.  These support
the hypothesis that the average RM is a useful constraint on the mean
accretion flow properties.  The variations in the RM on time scales of
months are a potentially useful diagnostic of the scale of turbulent
or secular fluctuations in the accretion flow properties.  Although 
we did not see any effects clearly related to passing of the G2 cloud,
the sensitivity of these measurements confirms that future interactions
may produce detectable Faraday signatures.  

The consistency of the sign of CP over a wide range of wavelengths
and decades in time suggests a stable
magnetic field configuration on scales from a few $R_S$ to hundreds of $R_S$.
This stability is probably best achieved if the CP is arising 
through Faraday conversion in emission regions with poloidal
magnetic fields.  In this model, we exclude an edge-on geometry.  Alternatively,
Doppler boosting in a toroidal magnetic field configuration could lead
to a persistent asymmetry that produces a consistent sign.

The short-term variations observed suggest a complex scenario on scales
of a few to $\sim 10 R_S$, in which both emission and propagation 
effects are important.  Linear and circular polarization are variable
on time scales of hours, comparable to the Keplerian time scale at
these small radii.  The apparent relationship between changes in
LP and wavelength-dependent effects suggests that 
the mildly relativistic electrons that are responsible for the synchrotron
emission also contribute to propagation effects.  It is unclear whether
variations that are modeled as the RM are truly propagation effects
or are the result of a complex, partially optically thick surface
from which the emission originates.  This picture is consistent with 
EHT polarimetric models of \sgra\ that reveal polarimetric structure
on scales smaller than the total intensity region.

We have been conservative in our analysis of systematic errors based
on calibrator observations and do not claim detection of Faraday 
rotation towards these sources.  But it is possible that these calibrators
do in fact reveal large RMs that are indicative of dense, magnetized plasma
in the inner regions of these sources.

Separating intrinsic and propagation effects in \sgra\ can be achieved
through several approaches.  Longer-term monitoring of \sgra\ with
intervals of days to years can provide a more accurate and complete
picture of the scale on which variations originate.  Our data are 
undersampled for establishing the nature of variations on
time scales shorter than years.  Simultaneous measurements at a wider
range of wavelengths, especially at wavelengths longer than 1.3 mm where
Faraday effects are stronger, will be an important probe of 
non-$\lambda^2$ effects that are predicted for models of mixed
emission and propagation effects.  Shorter wavelength observations may provide
more direct probes of the intrinsic emission process as Faraday effects
weaken and the emission region shrinks.

The EHT will obtain polarimetric images of \sgra\ with sensitivity
and fidelity that is substantially improved over past results.  
We expect that images could reveal polarimetric structures 
with independent intrinsic and Faraday characteristics.  Localized
circular polarization signatures can give insights into the
conversion mechanism.  Analysis
of EHT data must be carried out in a domain that is time-dependent,
frequency-dependent, and adaptable to complex
Faraday mechanisms.

\acknowledgements
This paper makes use of the following ALMA data: ADS/JAO.ALMA\#2013.1.00764.S. ALMA is a partnership of ESO (representing its member states), NSF (USA) and NINS (Japan), together with NRC (Canada), MOST and ASIAA (Taiwan), and KASI (Republic of Korea), in cooperation with the Republic of Chile. The Joint ALMA Observatory is operated by ESO, AUI/NRAO and NAOJ.
The National Radio Astronomy Observatory is a facility of the National Science Foundation operated under cooperative agreement by Associated Universities, Inc.
H.F. acknowledges funding from the European Research Council (ERC) Synergy Grant "BlackHoleCam" (Grant 610058).

\facilities{ALMA}
\software{CASA, MATLAB}


\appendix

\section{Calibrator Results and Estimates of Systematic Errors \label{sec:appendix}}

We use calibrator results in this Section to set limits on systematic
errors on LP fraction, CP fraction,
and rotation measure.  We set these limits for inter-epoch
and intra-epoch comparisons, and find that they are comparable 
(Table~\ref{tab:systematics}). Our CASA analysis was similar to that employed for science verification
data, which found a characteristic error $\delta \chi=0.4$ deg and
errors in fractional linear polarization $< 0.1\%$ 
\citep{2016ApJ...824..132N}.

\begin{figure}[t!]
\includegraphics[width=0.4\textwidth]{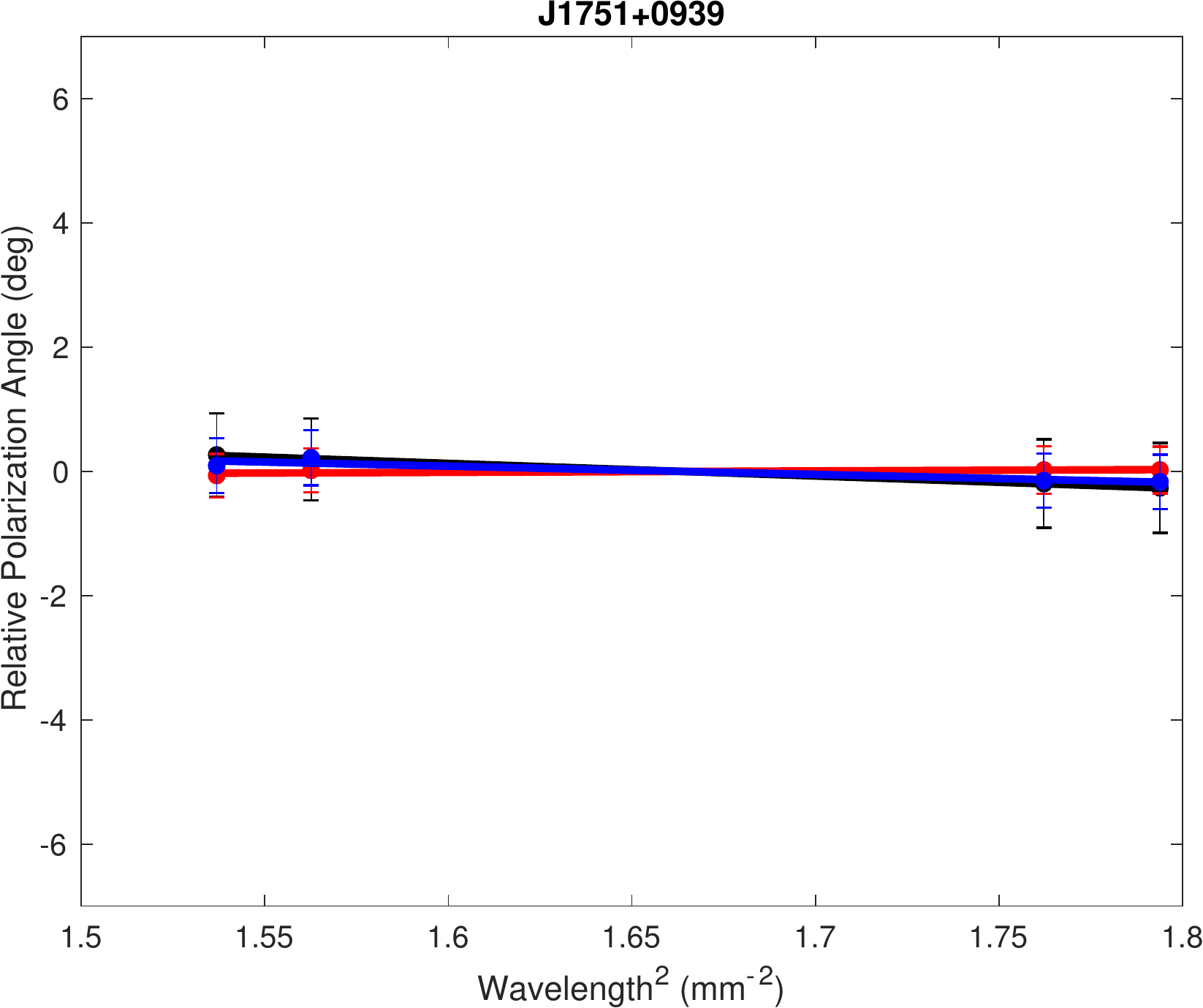}
\includegraphics[width=0.4\textwidth]{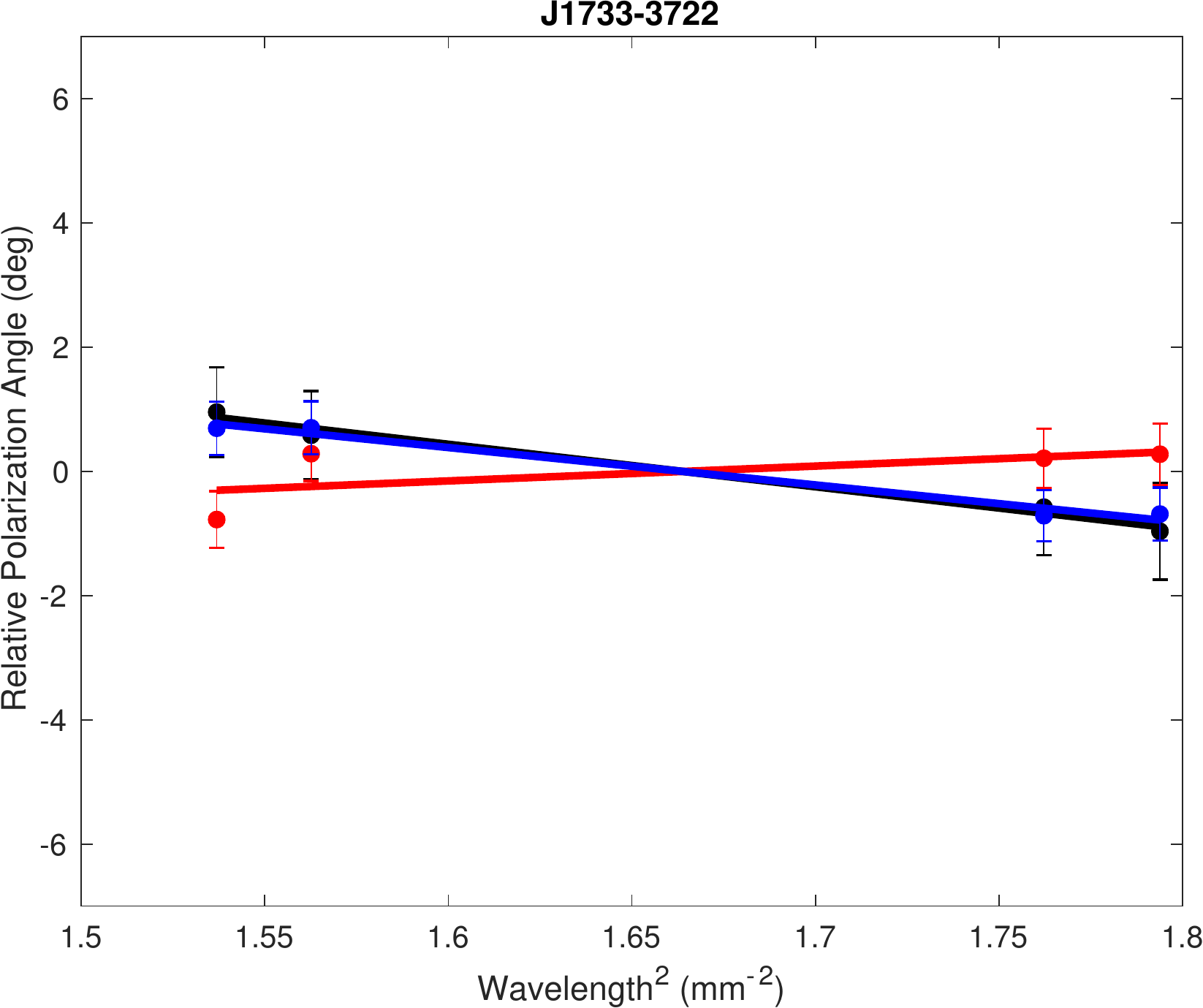}
\includegraphics[width=0.4\textwidth]{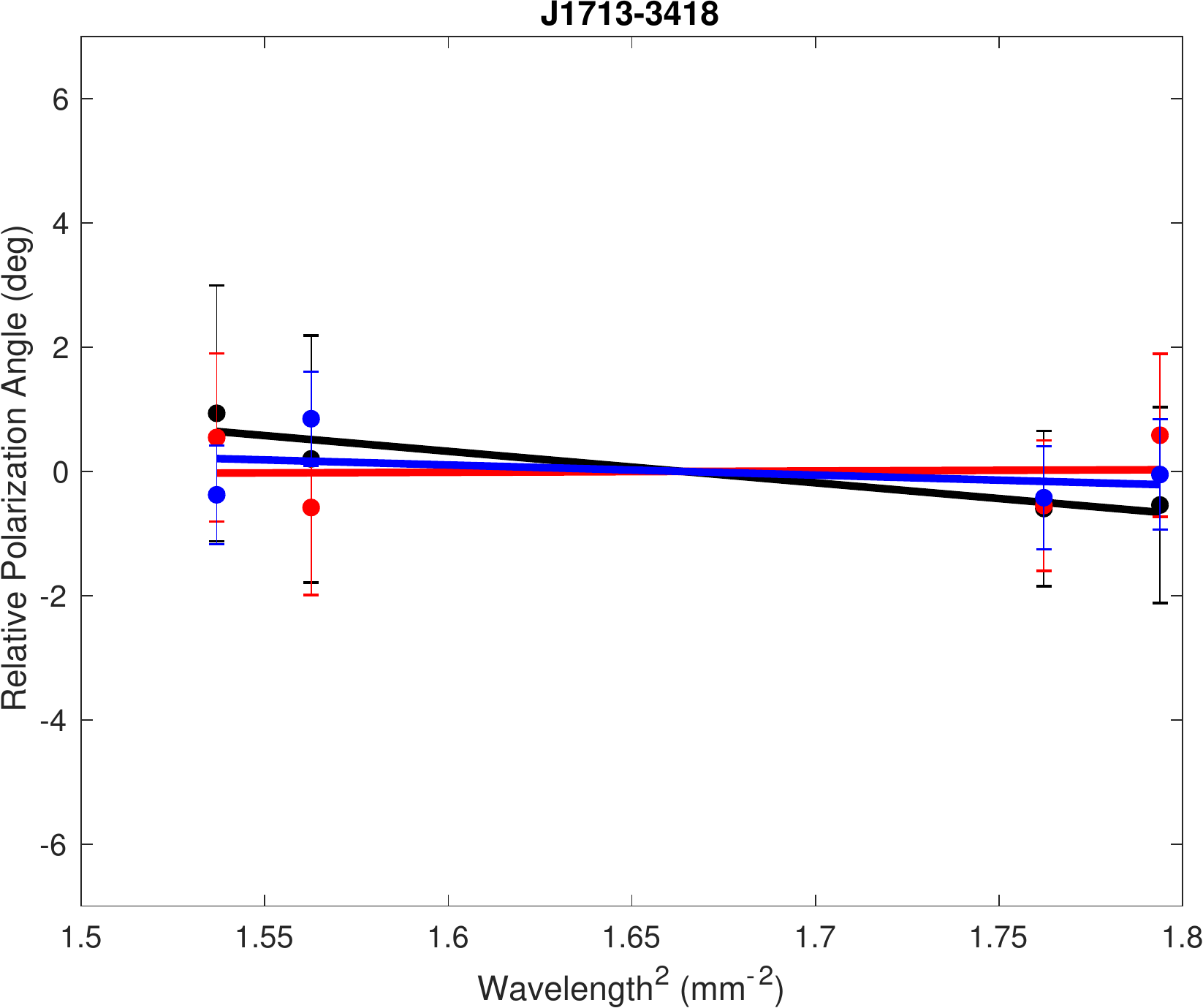}
\includegraphics[width=0.4\textwidth]{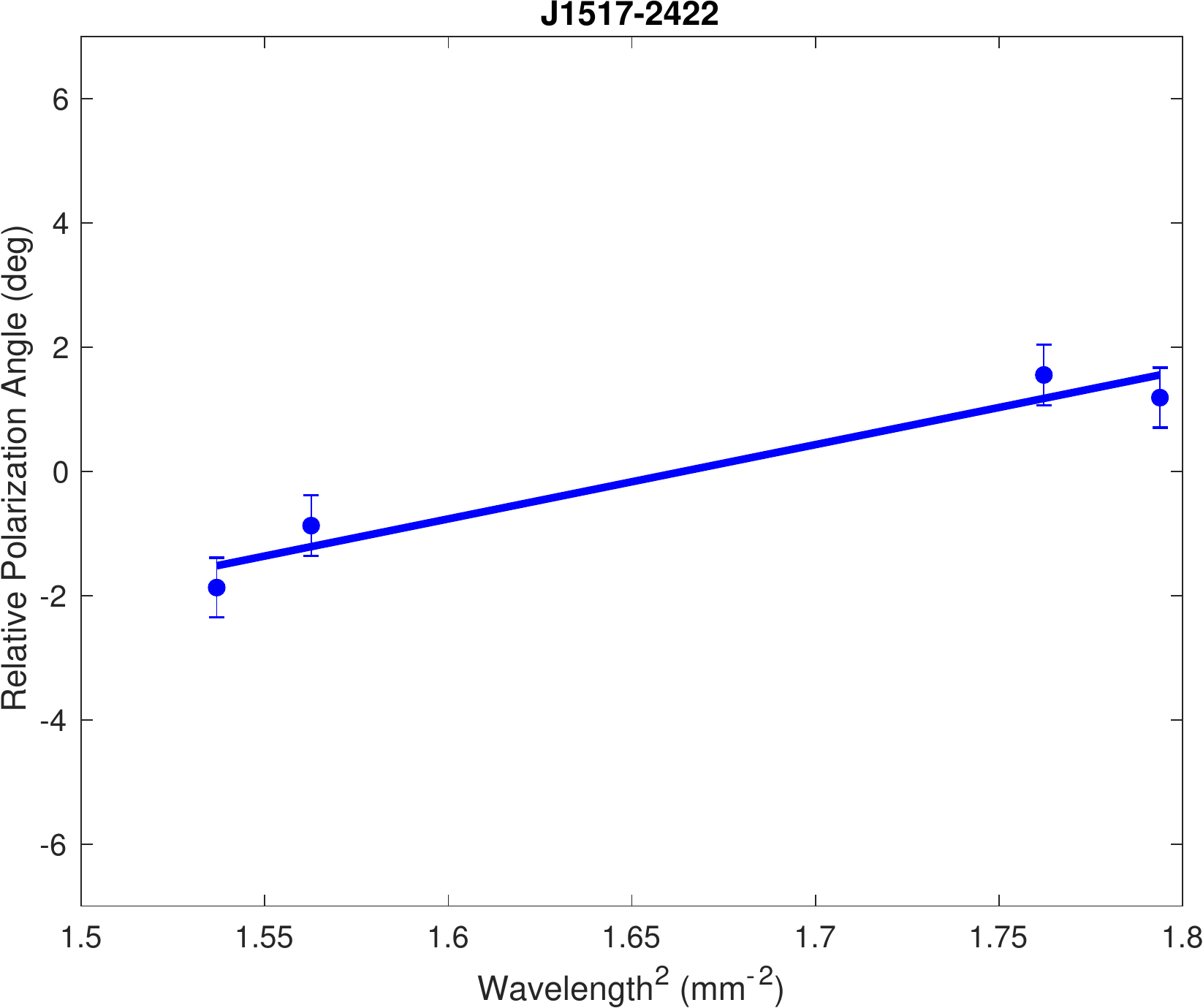}
\includegraphics[width=0.4\textwidth]{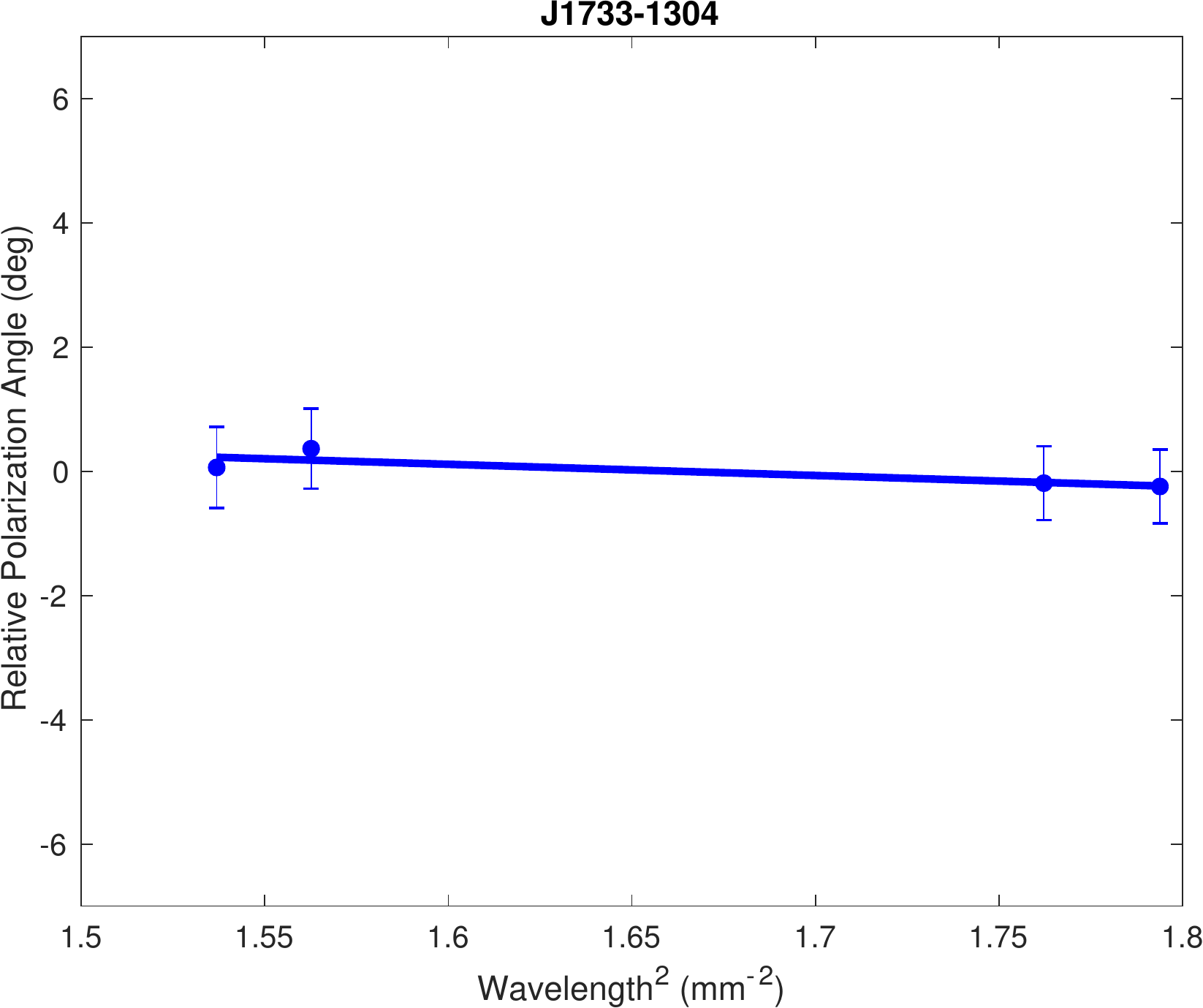}
\includegraphics[width=0.4\textwidth]{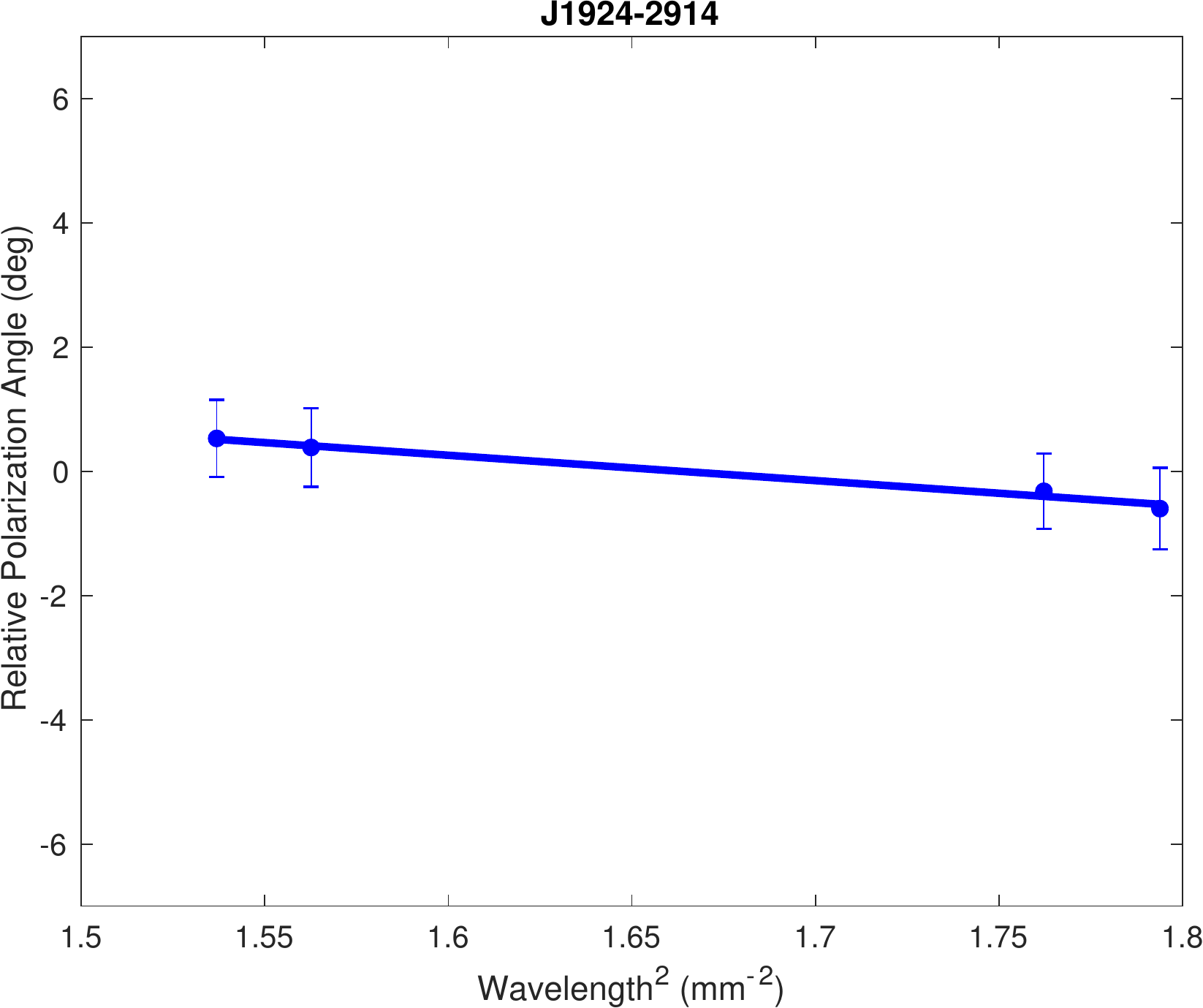}
\caption{Average residual polarization position angle as a function of wavelength-squared for each of the calibrators in epochs 1 (black), 2 (red), and 3 (blue).  We have removed the mean position angle for each source in each epoch to enable clear comparison.  All plots are on the same scale for wavelength-scaled and position angle.
\label{fig:calavg}
}
\end{figure}

\begin{figure}[t!]
\includegraphics[width=0.33\textwidth]{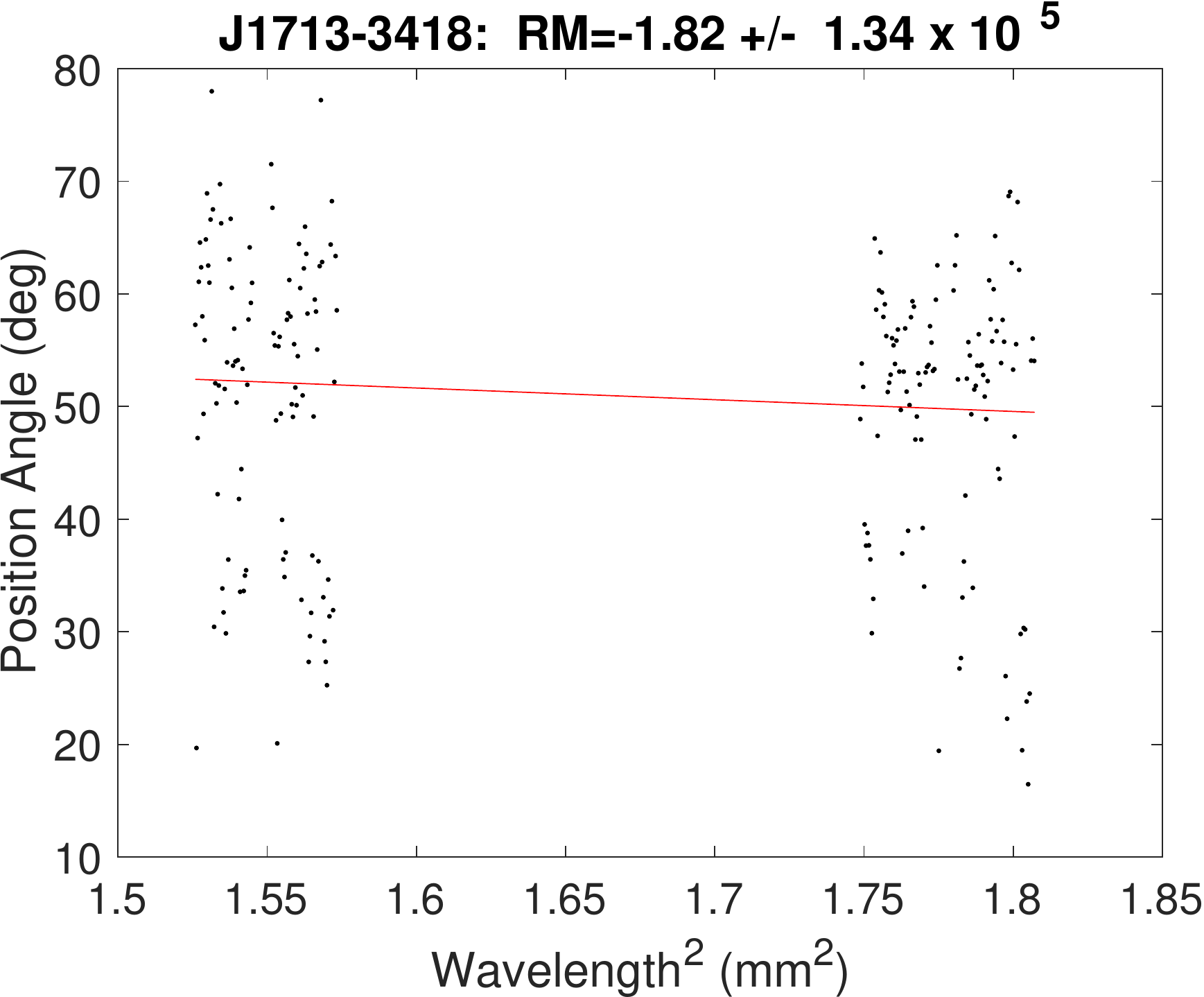}\includegraphics[width=0.33\textwidth]{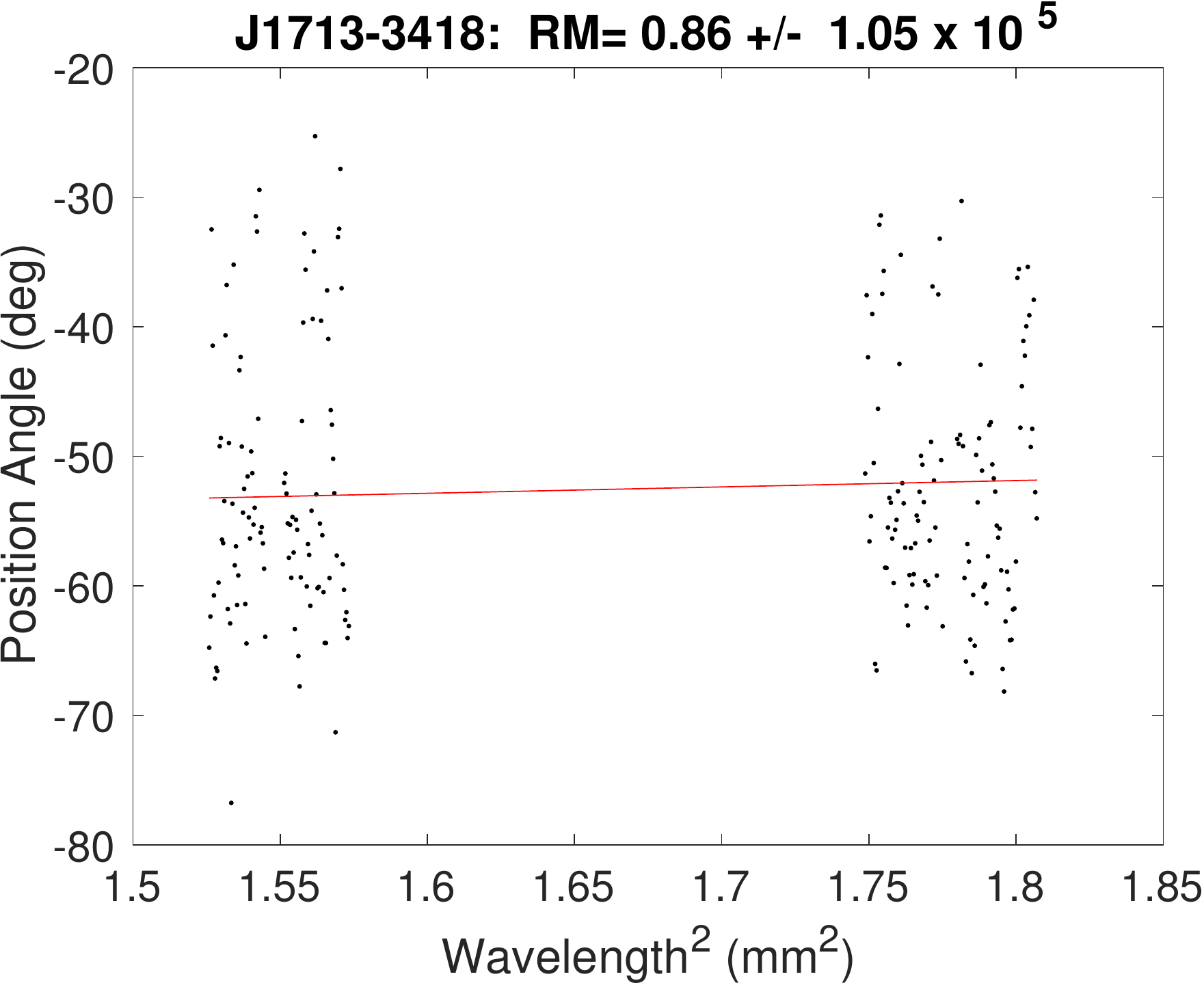}\includegraphics[width=0.33\textwidth]{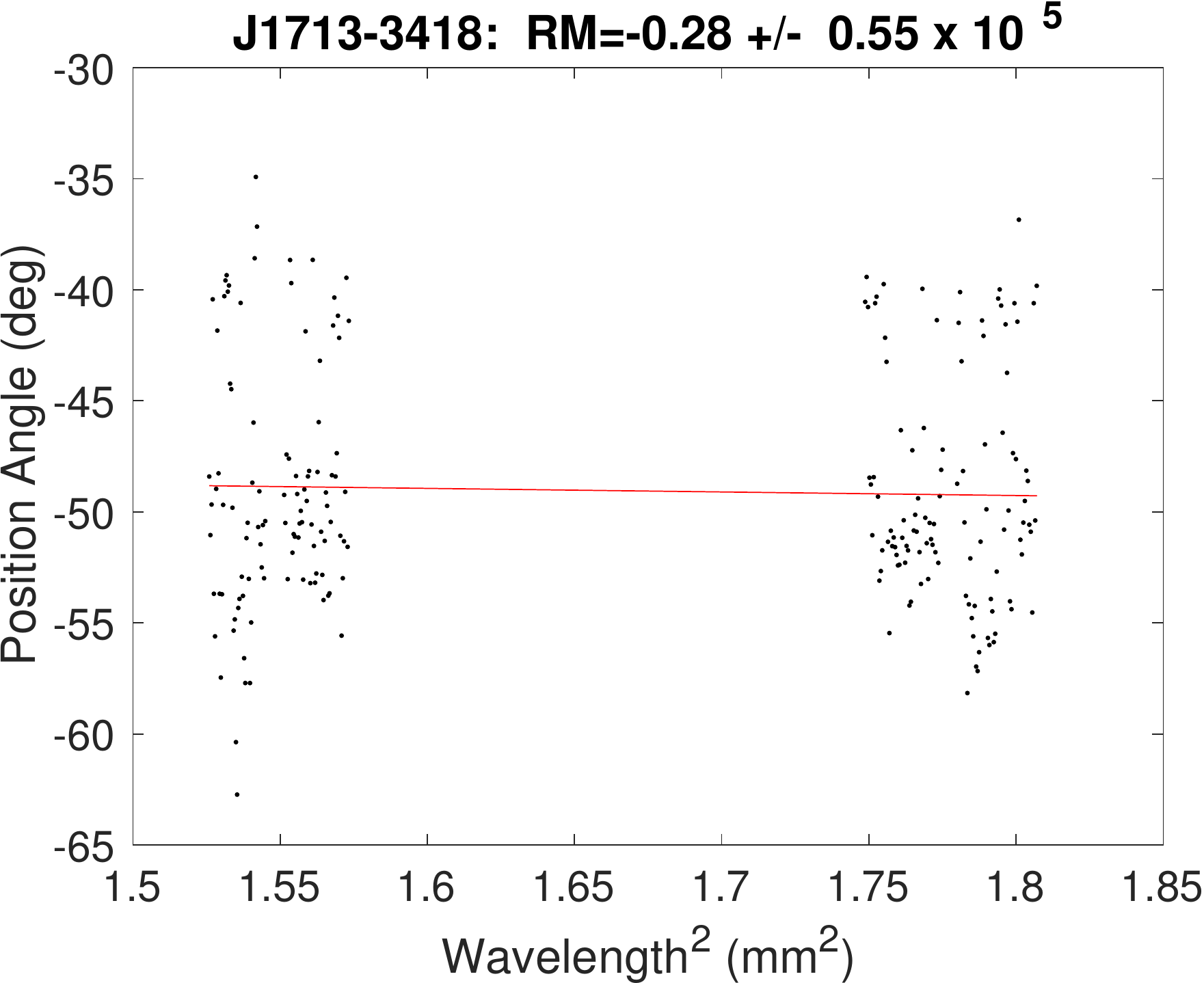}
\includegraphics[width=0.33\textwidth]{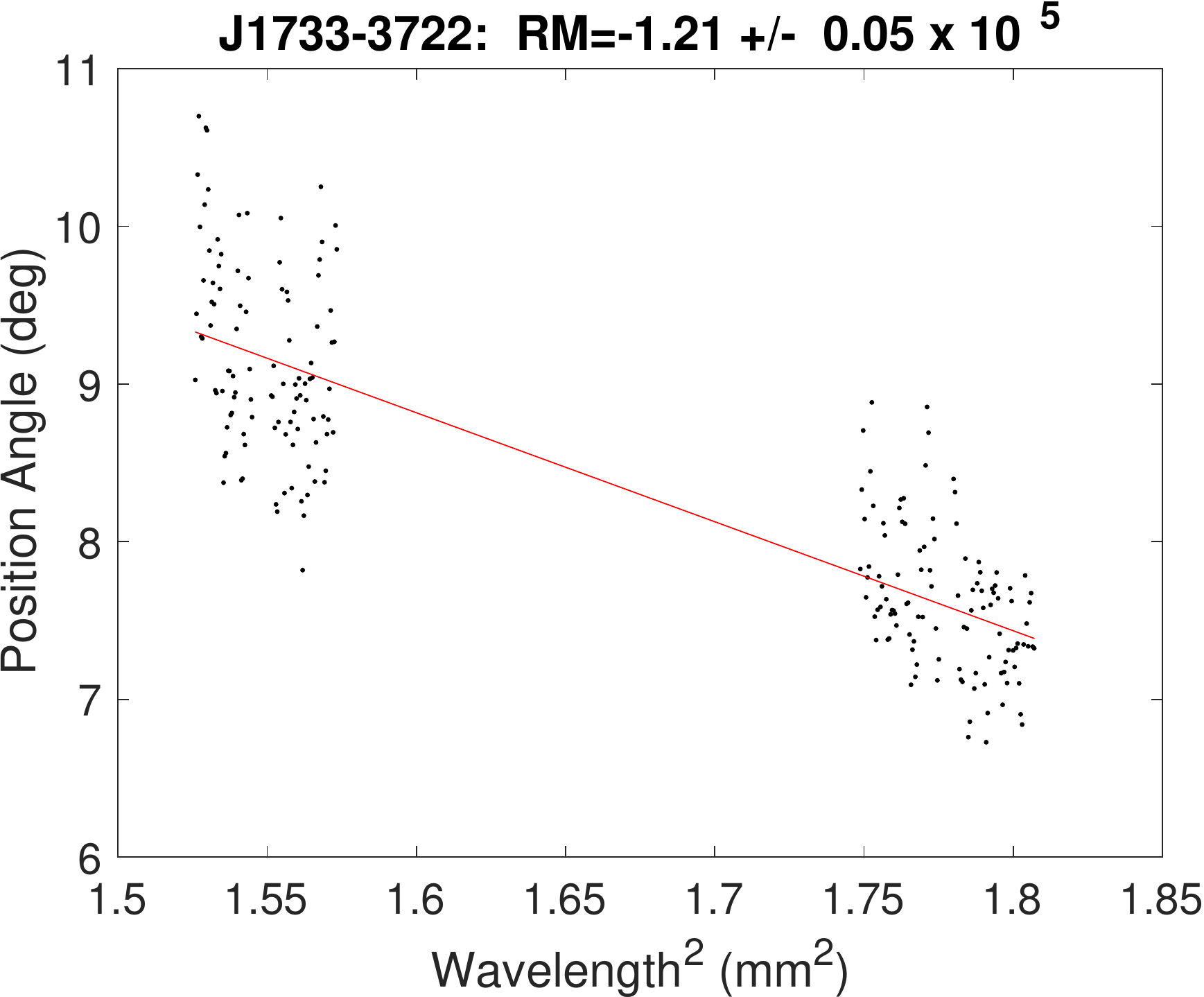}\includegraphics[width=0.33\textwidth]{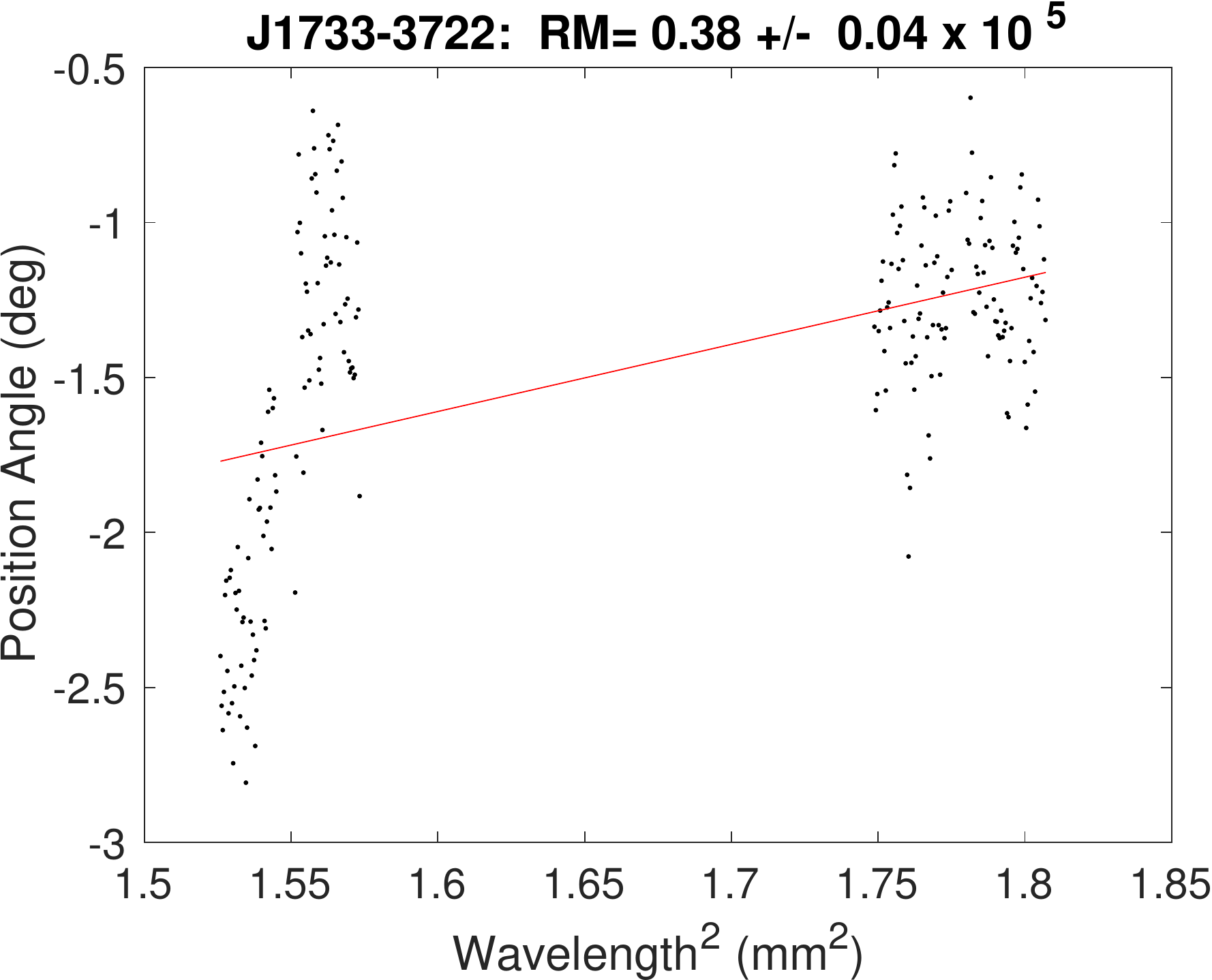}\includegraphics[width=0.33\textwidth]{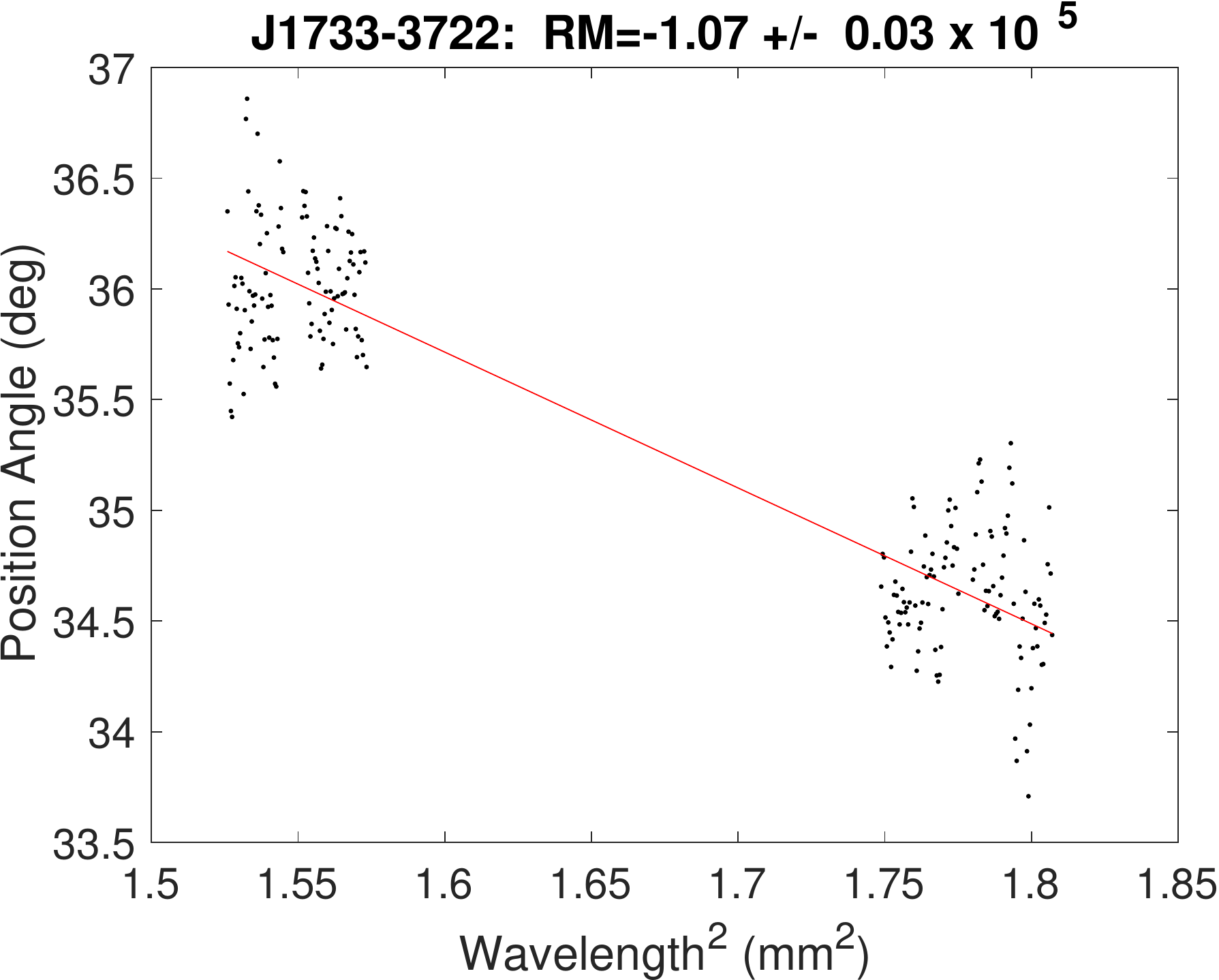}
\includegraphics[width=0.33\textwidth]{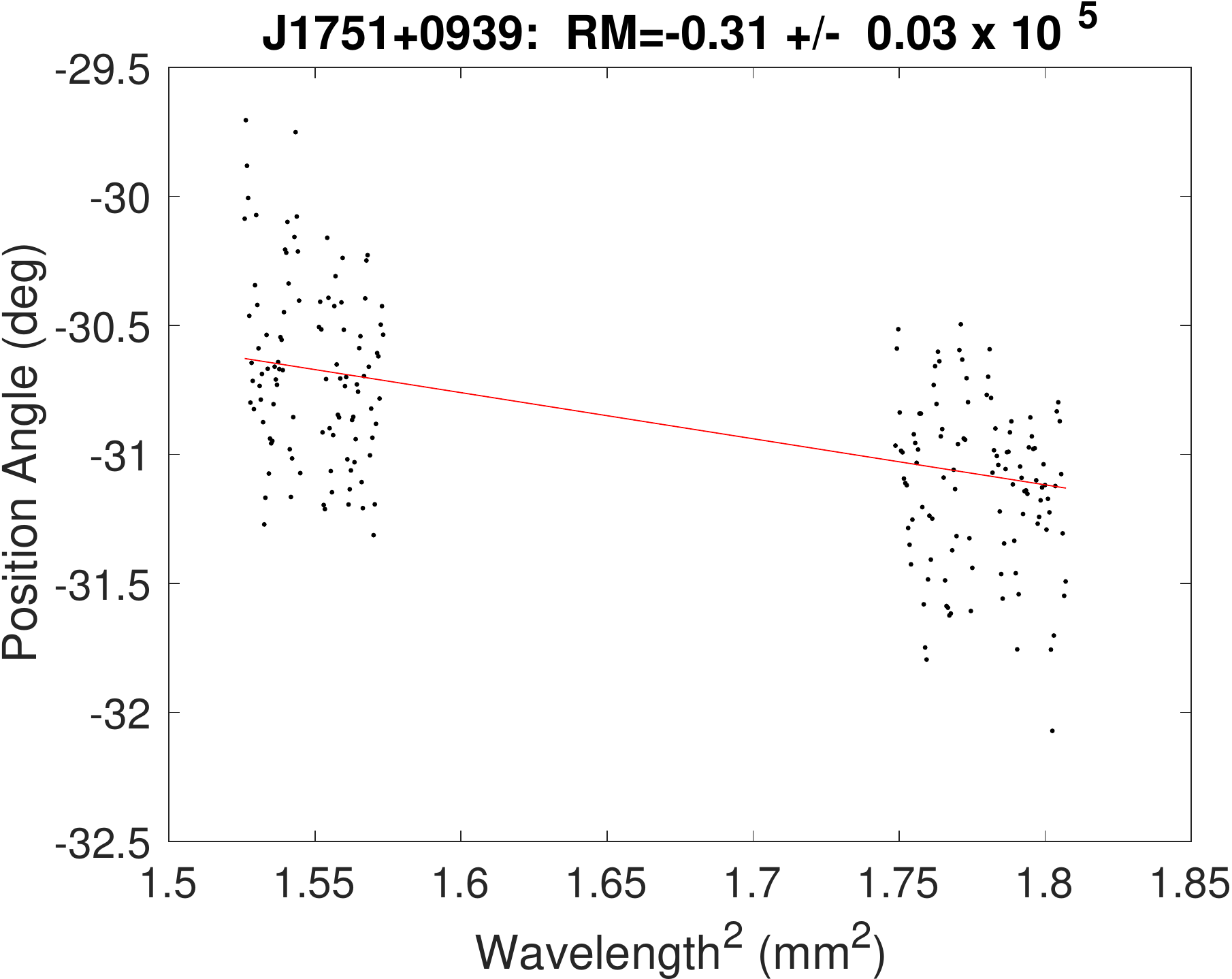}\includegraphics[width=0.33\textwidth]{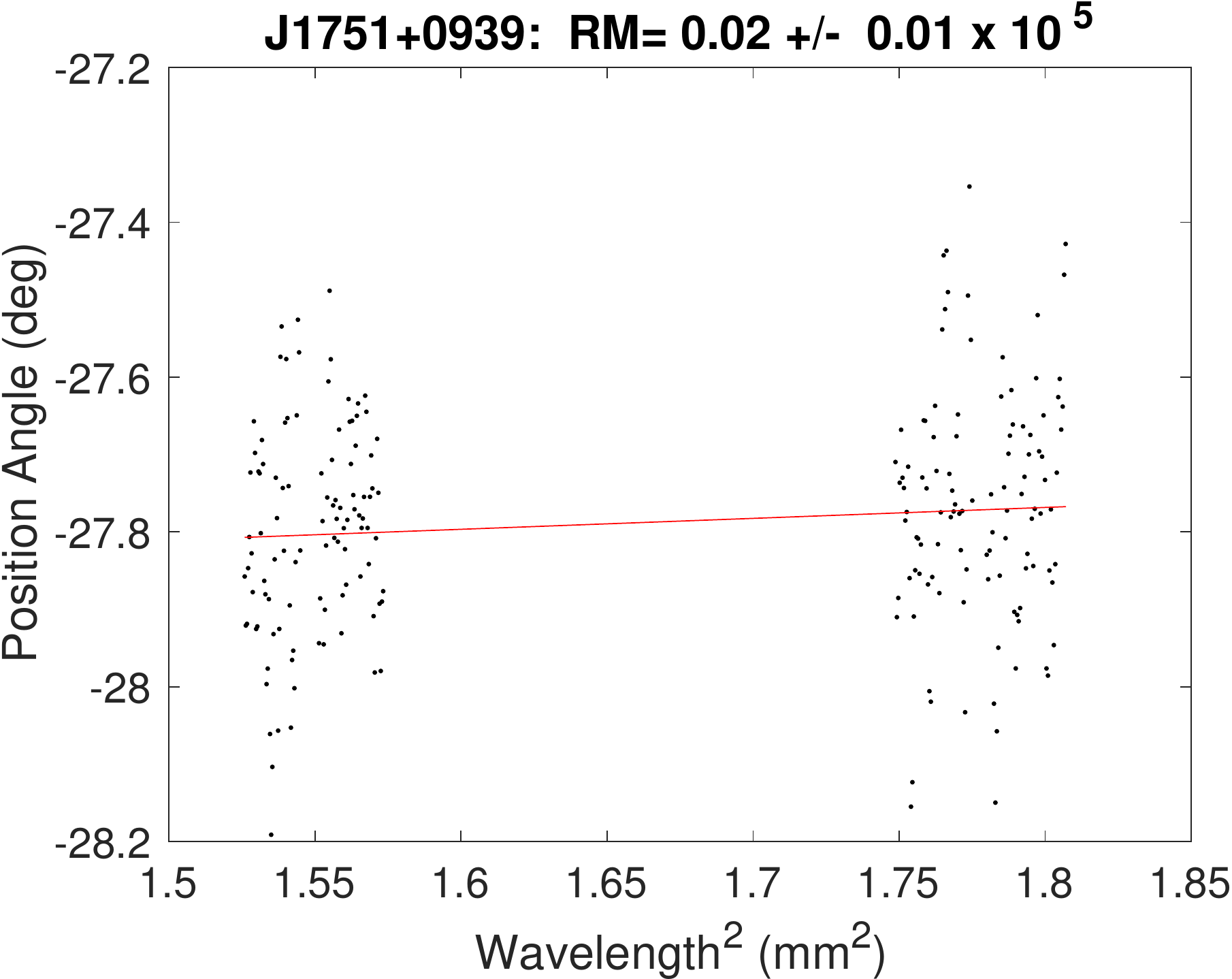}\includegraphics[width=0.33\textwidth]{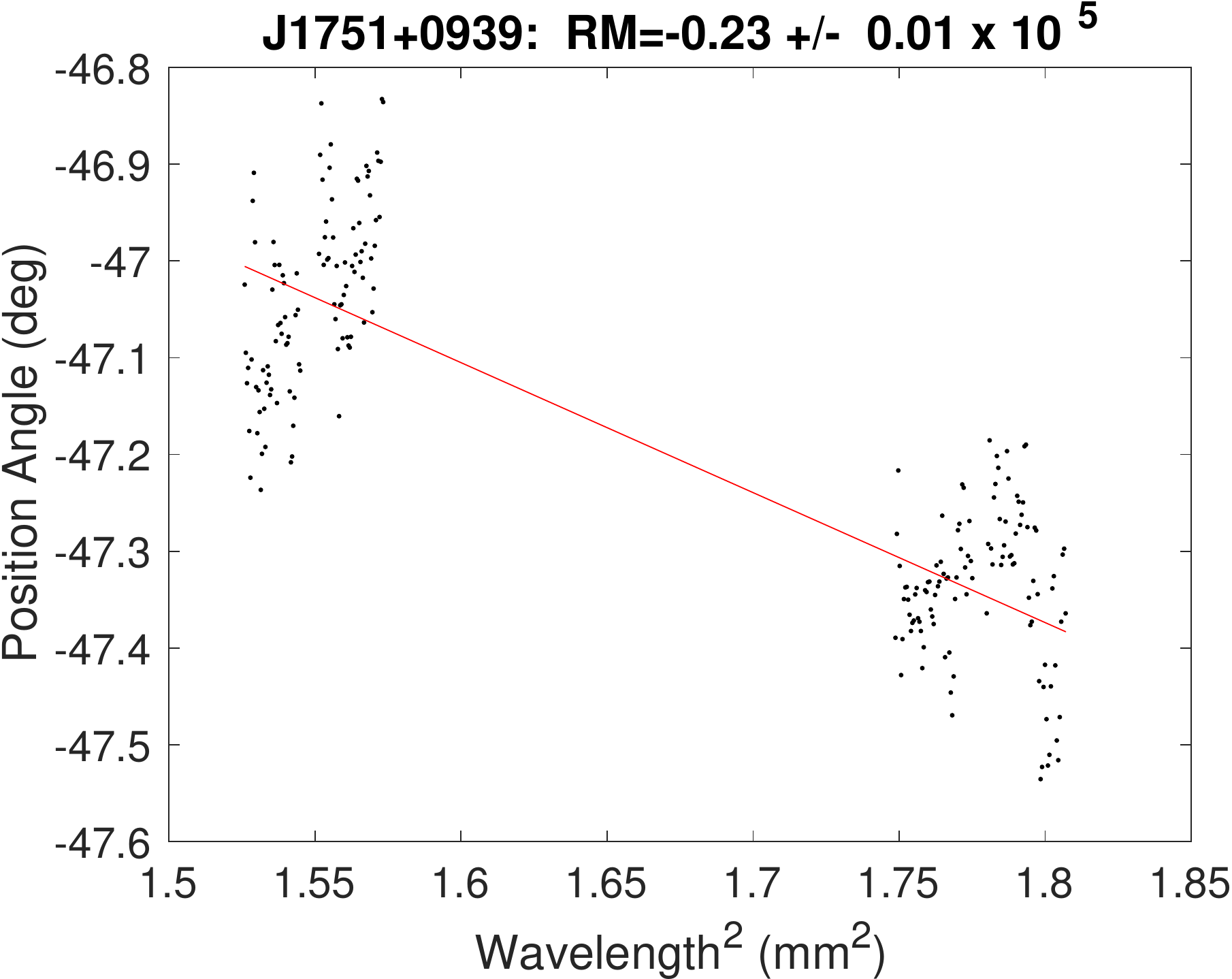}
\includegraphics[width=0.33\textwidth]{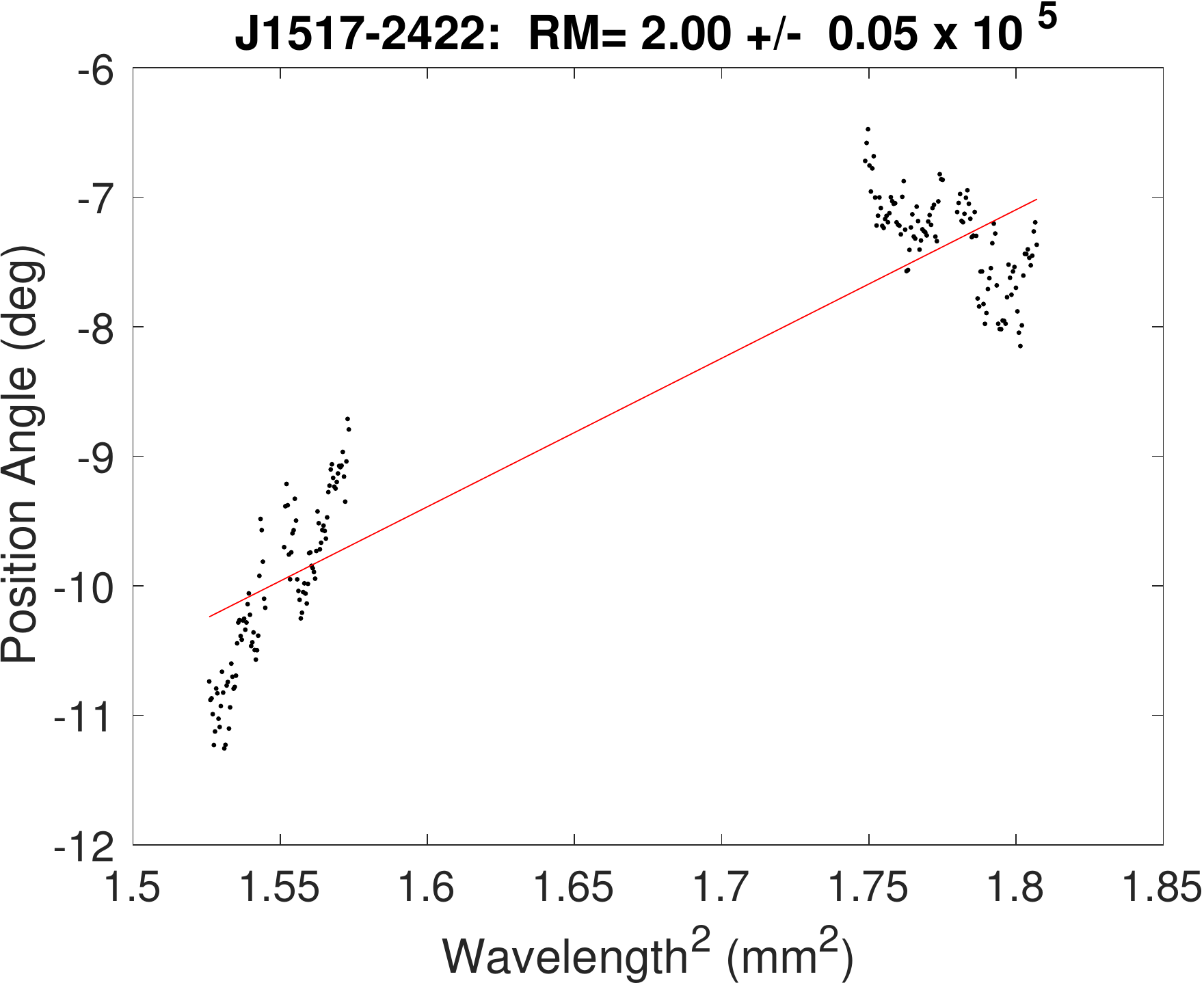}\includegraphics[width=0.33\textwidth]{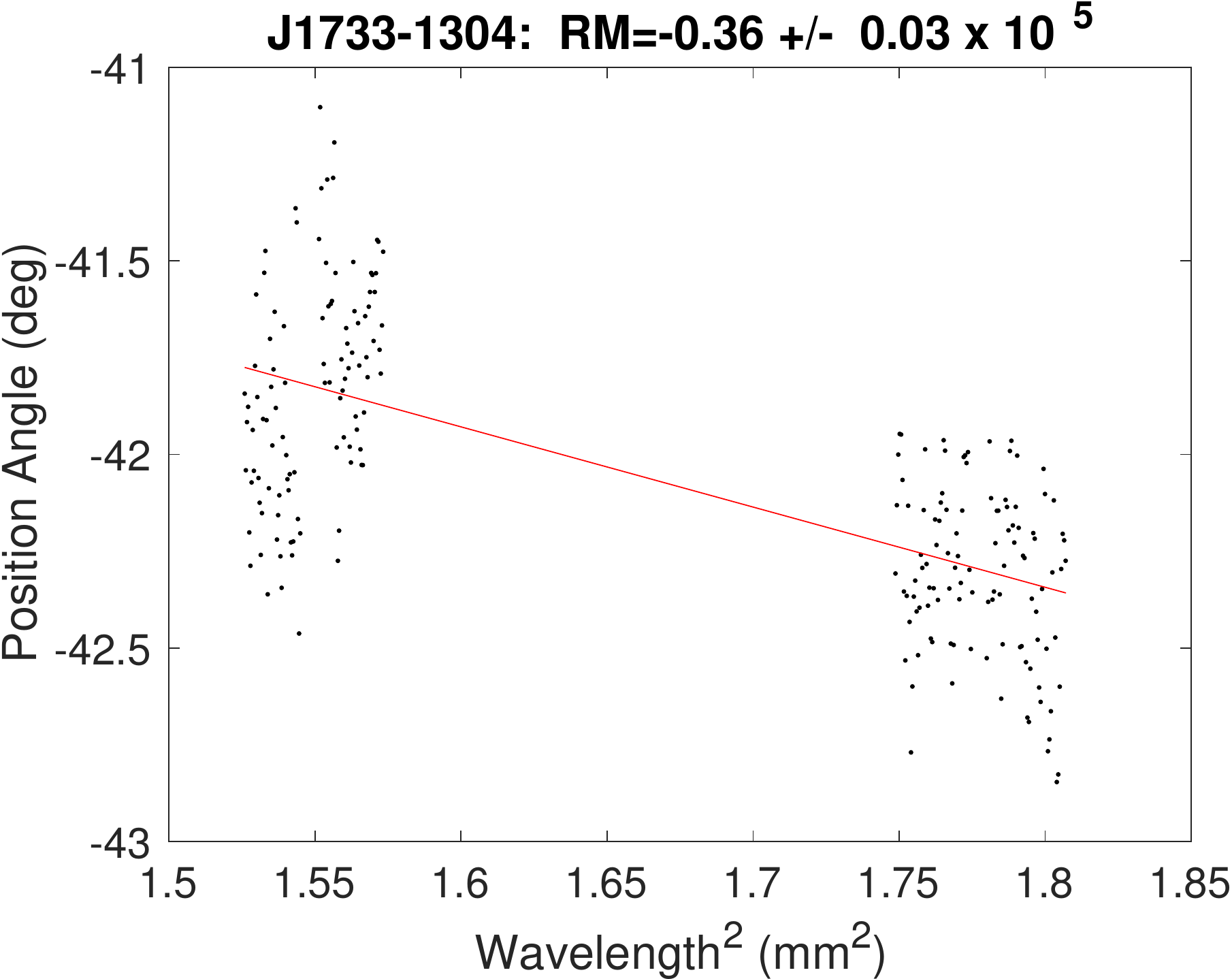}\includegraphics[width=0.33\textwidth]{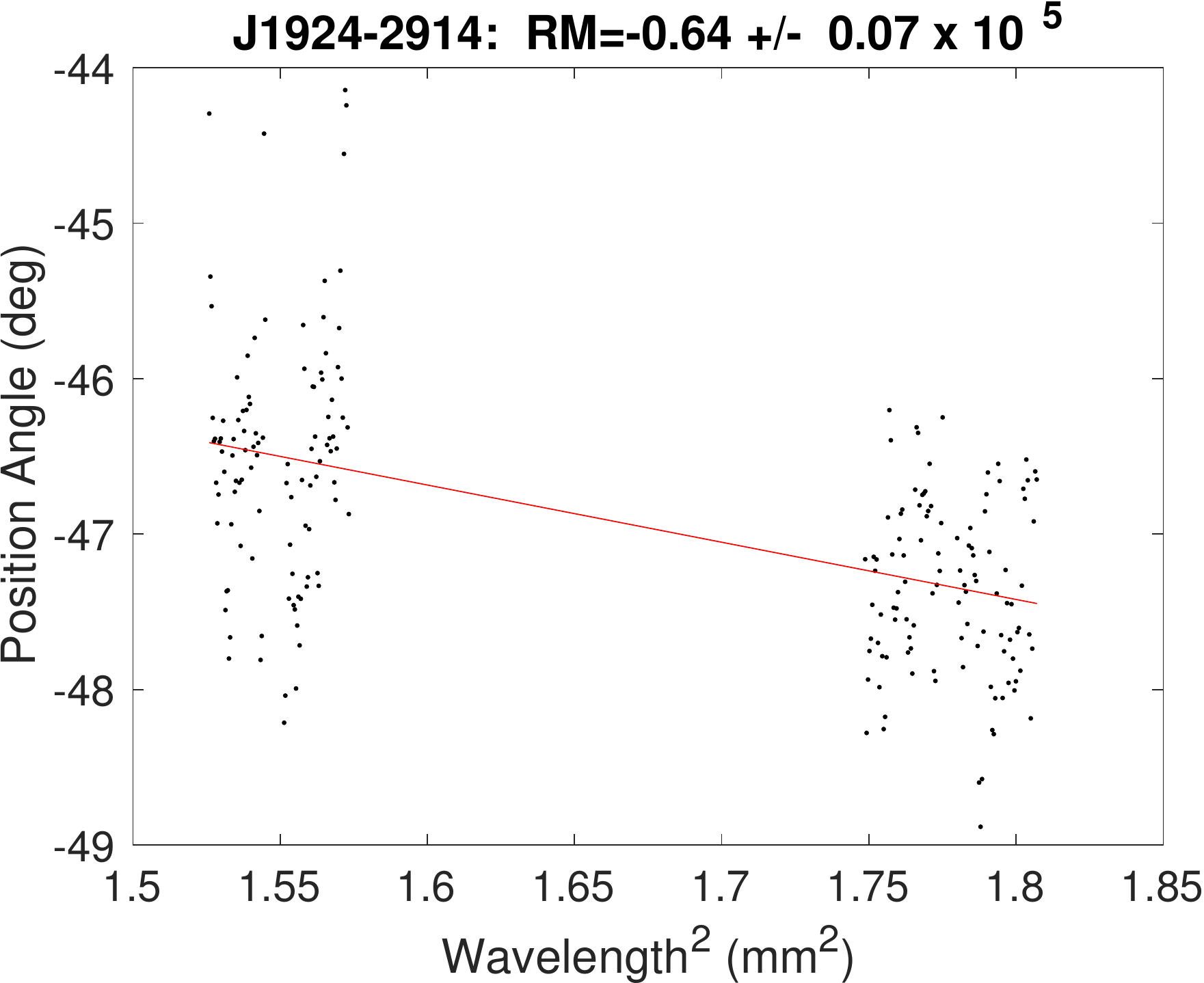}
\caption{Polarization angle as a function of wavelength-squared, presented for each channel and for each epoch for all calibrators.  Epochs 1, 2, and 3
are in left, middle, and right columns except the last row.  In the last row, the three calibrators included only in Epoch 3 are shown.  Note the different scales for position angle.  The title gives the RM units of $\rdm$.
\label{fig:rmchannelcal}
}
\end{figure}

We present similar analysis results for the calibrators as for \sgra\ to
facilitate comparison and demonstration of the reliability of our main
results. Figure~\ref{fig:calavg} shows the SPW-averaged polarization 
angles and fitted RMs for all calibrators.  Figure~\ref{fig:rmchannelcal} shows the channel-averaged
polarization angles for each epoch, as well.  Time-dependent Stokes parameters,
fitted RMs, position angles, and residuals
are shown in Figures~\ref{fig:caltime1} through ~\ref{fig:caltime9}.
In Table~\ref{tab:calavg} we present the epoch-averaged polarization
properties as a function of SPW.  In Table~\ref{tab:calavgrm}, we present 
derived polarization properties for the calibrators in each epoch, 
including RM.  

Treating these calibrator results as limits of calibration accuracy must be given with the caveat that there is evidence of large RMs towards some AGN at millimeter wavelengths. In particular, 3C 84 shows an RM$\approx 10^6 \rdm$ \citep{2014ApJ...797...66P} and M87 shows an RM$\approx 10^5 \rdm$ \citep{2014ApJ...783L..33K}.  ALMA observations of 3C 273 reveal RM$=(3.6 \pm 0.3) \times 10^5 \rdm$ \citep{2018arXiv180309982H}.  \citet{2012A&A...540A..74T} reports RMs as large as $10^5 \rdm$ for AGN at wavelengths near millimeter wavelength. These large RMs are interpreted as originating from the accretion flow, relativistic jet, or the dense gas of the nuclear region \citep[e.g.,][]{2017MNRAS.468.2214M}. Therefore, there is possibly a significant intrinsic contribution to these observed RMs although it cannot be quantified for these sources from these data.  

\begin{deluxetable}{rrr}
\tablecaption{Systematic Limits on Polarization Properties
\label{tab:systematics}}
\tablehead{ \colhead{LP fraction} & \colhead{CP fraction} & \colhead{RM} }
\startdata
 0.1\% & 0.2\% & $10^5 \rdm$ \\
\enddata
\end{deluxetable}

\subsection{Inter-Epoch Properties}

The most optimistic limits on polarization calibration are set 
by analysis of the polarization calibrator, J1751+0939.  The polarization calibration assumes $V=0$, which is achieved with an accuracy of $\lsim 0.1$\% for the calibrator, J1751+039.    The mean polarization angle for J1751+0939 within an epoch is determined to an accuracy of $< 0.1$ deg.  The RM for each epoch is constrained to less than a few $\times 10^4 \rdm$.  

The results obtained for the other calibrators provide a more realistic estimate of accuracy of polarization calibration and gain solution transfer.  In particular, the phase calibrator, J1733-3722, and the check source, J1713-3418, are observed with similar cycles to Sgr A*.  J1713-3418 is consistent with a mean RM$=-0.68 \times 10^5 \pm 0.59 \times 10^5 \rdm$. J1733-3722 is inconsistent with a constant RM. Consistency can be forced by adding in quadrature an error of $0.6 \times 10^5 \rdm$.  The upper bound on RM found in any epoch for these sources is epoch 1 for J1733-3722 with a value of $-1.19 \times 10^5 \pm 0.10 \times 10^5 \rdm$.  We find similar limits for J1733-1304 and J1924-2914 but a somewhat large value for a one-epoch limit for J1517-2422 ($2.09 \times 10^5 \pm 0.38 \times 10^5 \rdm$).  The limits for these sources are less reliable because they are the result of only single snapshot observations, whereas J1733-3722 and J1713-3418 were observed $>10$ and $>5$ times per epoch, respectively, over a wide range of parallactic angles.

We use the calibrators and estimates of intrinsic RM to establish a systematic limit of $10^5 \rdm$ for changes between sources and epochs.  This limit corresponds to a change in the position angle across the band of 1.3 deg.  As discussed below, we estimate that systematic errors within an epoch for a given source are less than this value.

Calibration of Stokes V for polarization and gain assumes $V=0$.  As a result, we can only set systematic limits on Stokes V with the check source J1713-3418.  
The gain calibration for J1713-3418 is shared with that of Sgr A*.  Stokes V is detected for this check source with statistical significance in only
epoch 2 at a level $V \lsim 0.2$\%, which we adopt as our systematic
threshold for detection and change between sources and epochs.  
We note in its proposal materials for Cycle 6, ALMA suggests a circular polarization systematic error of 0.6\%.   Detailed analysis for ALMA observations of 
3C 273 finds $V=0.2\%$ \citep{2018arXiv180309982H}.

\subsection{Intra-Epoch Properties}

\begin{figure}
\includegraphics[width=\textwidth]{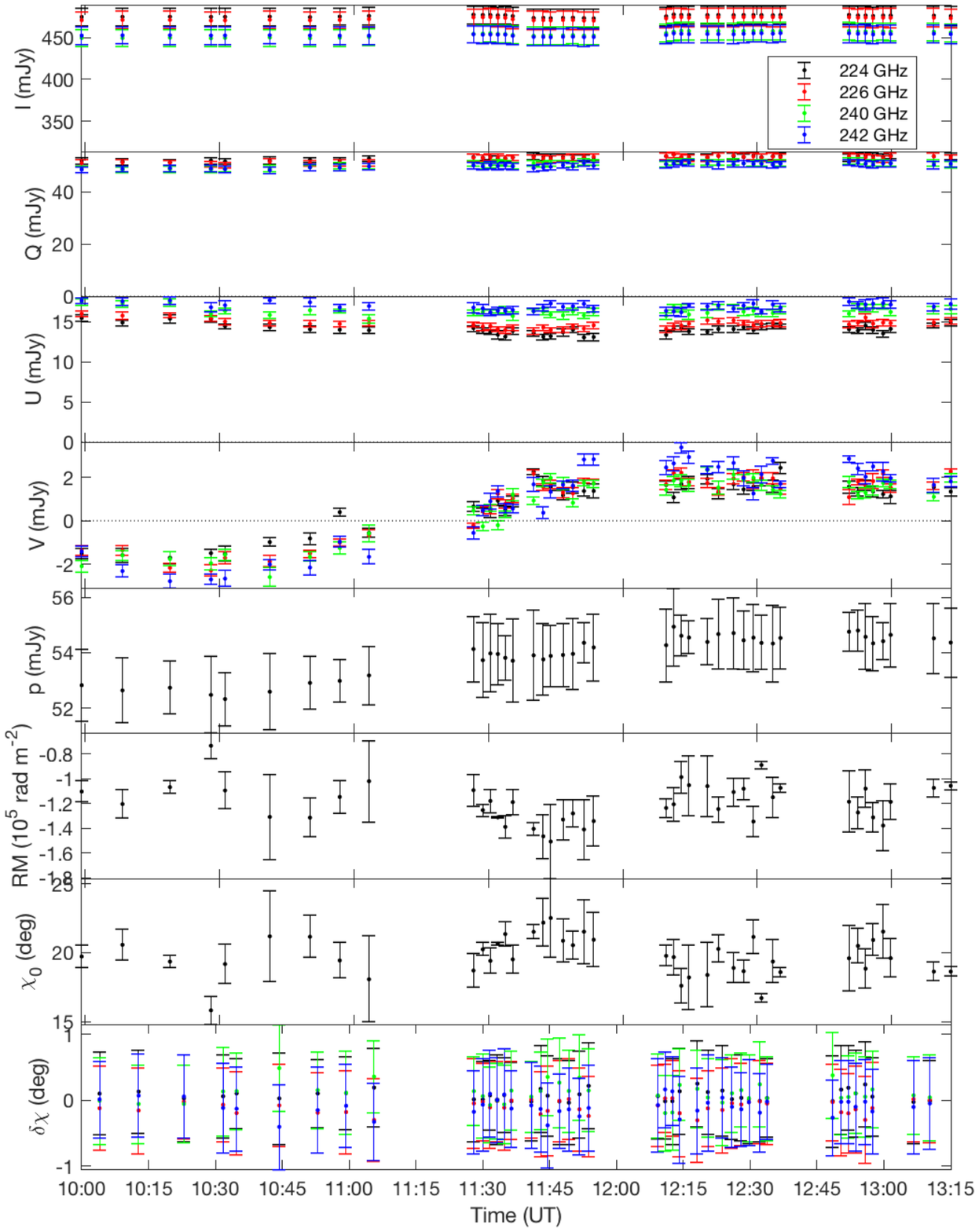}
\caption{Time series plot for J1733-3722 in epoch 1 following
Figure~\ref{fig:sgratime1}.
\label{fig:caltime1}
}
\end{figure}

\begin{figure}
\includegraphics[width=\textwidth]{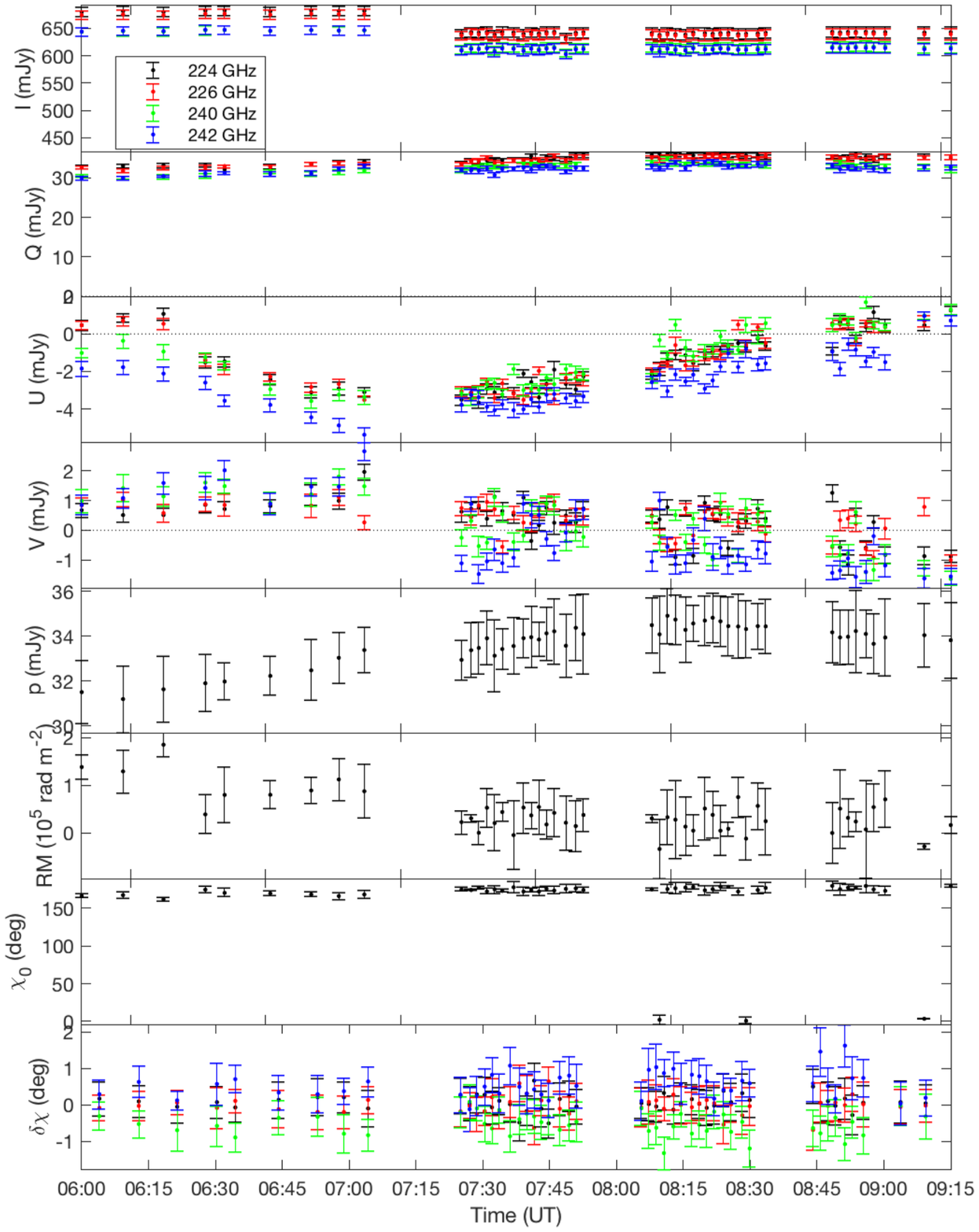}
\caption{Time series plot for J1733-3722 in epoch 2 following
Figure~\ref{fig:sgratime1}.
\label{fig:caltime2}
}
\end{figure}

\begin{figure}
\includegraphics[width=\textwidth]{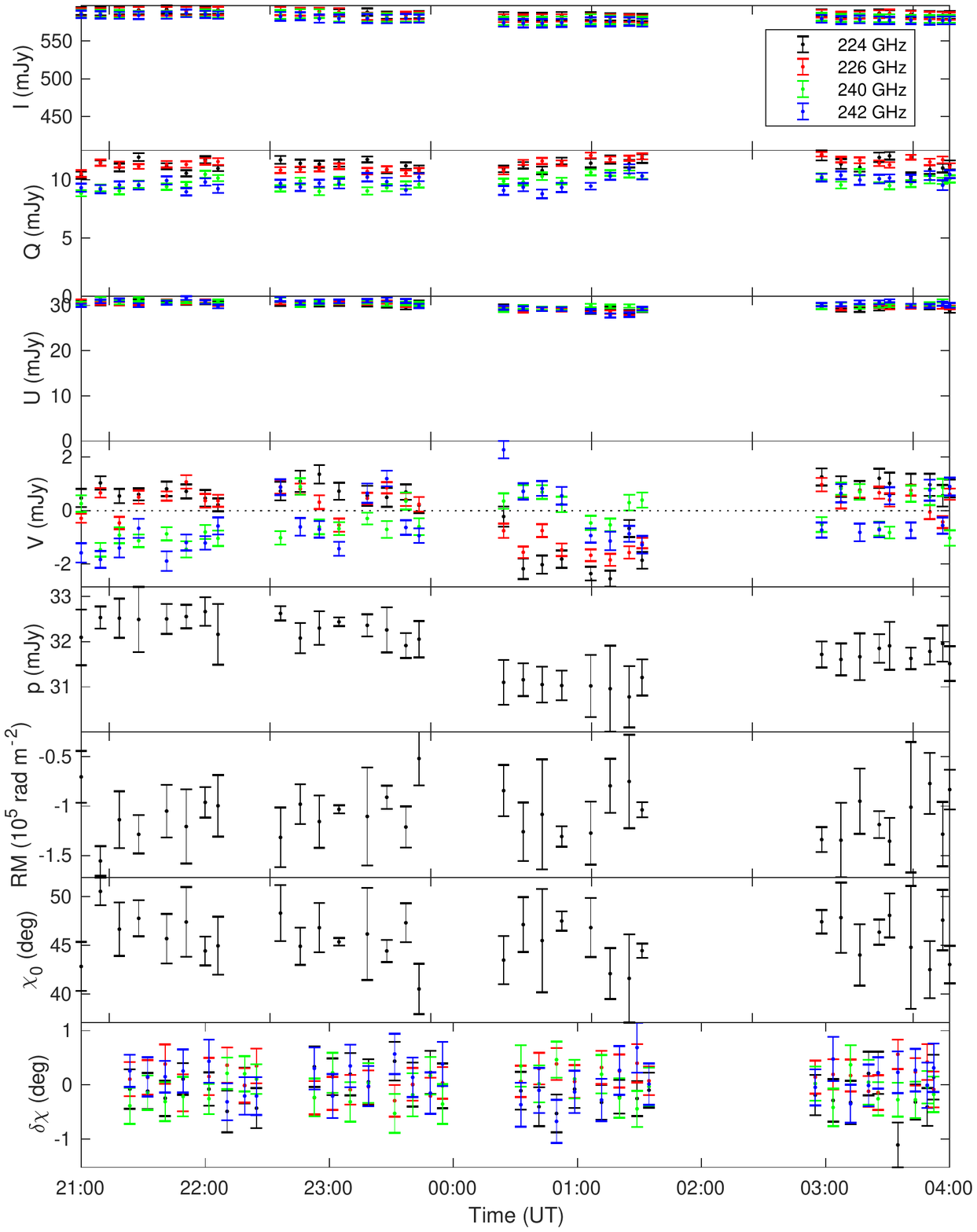}
\caption{Time series plot for J1733-3722 in epoch 3 following
Figure~\ref{fig:sgratime1}.
\label{fig:caltime3}
}
\end{figure}

\begin{figure}
\includegraphics[width=\textwidth]{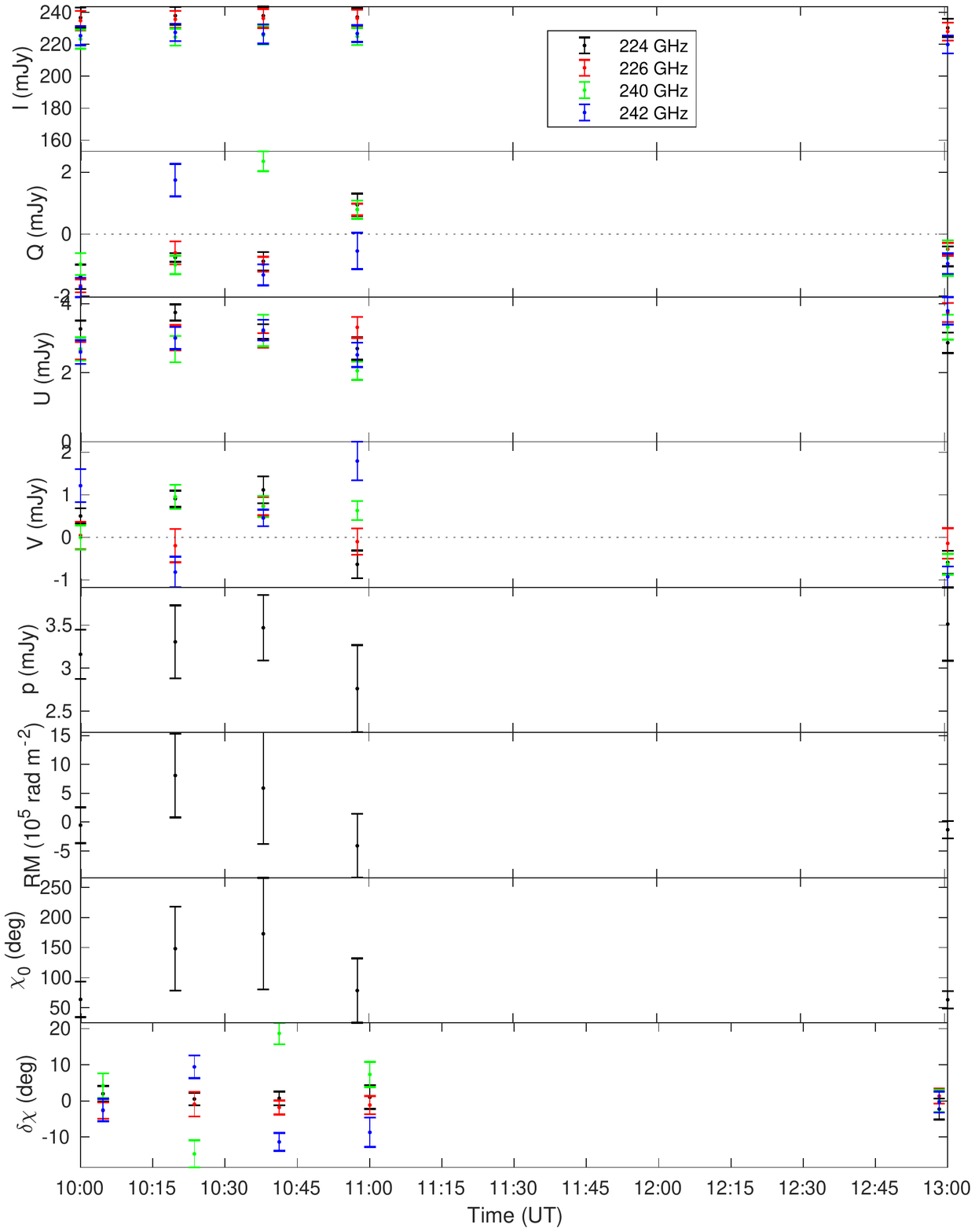}
\caption{Time series plot for J1713-3418 in epoch 1 following
Figure~\ref{fig:sgratime1}.
\label{fig:caltime4}
}
\end{figure}

\begin{figure}
\includegraphics[width=\textwidth]{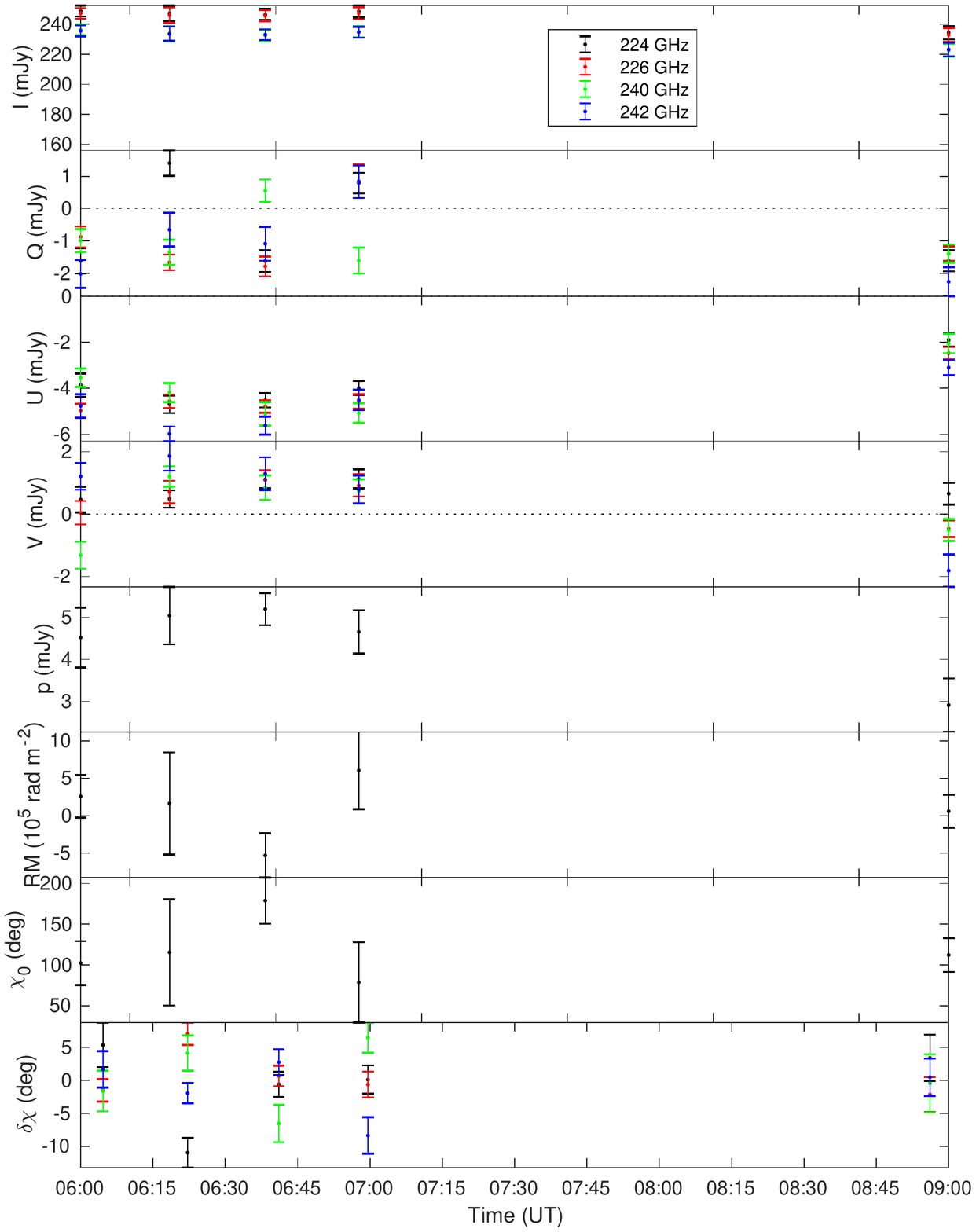}
\caption{Time series plot for J1713-3418 in epoch 2 following
Figure~\ref{fig:sgratime1}.
\label{fig:caltime5}
}
\end{figure}

\begin{figure}
\includegraphics[width=\textwidth]{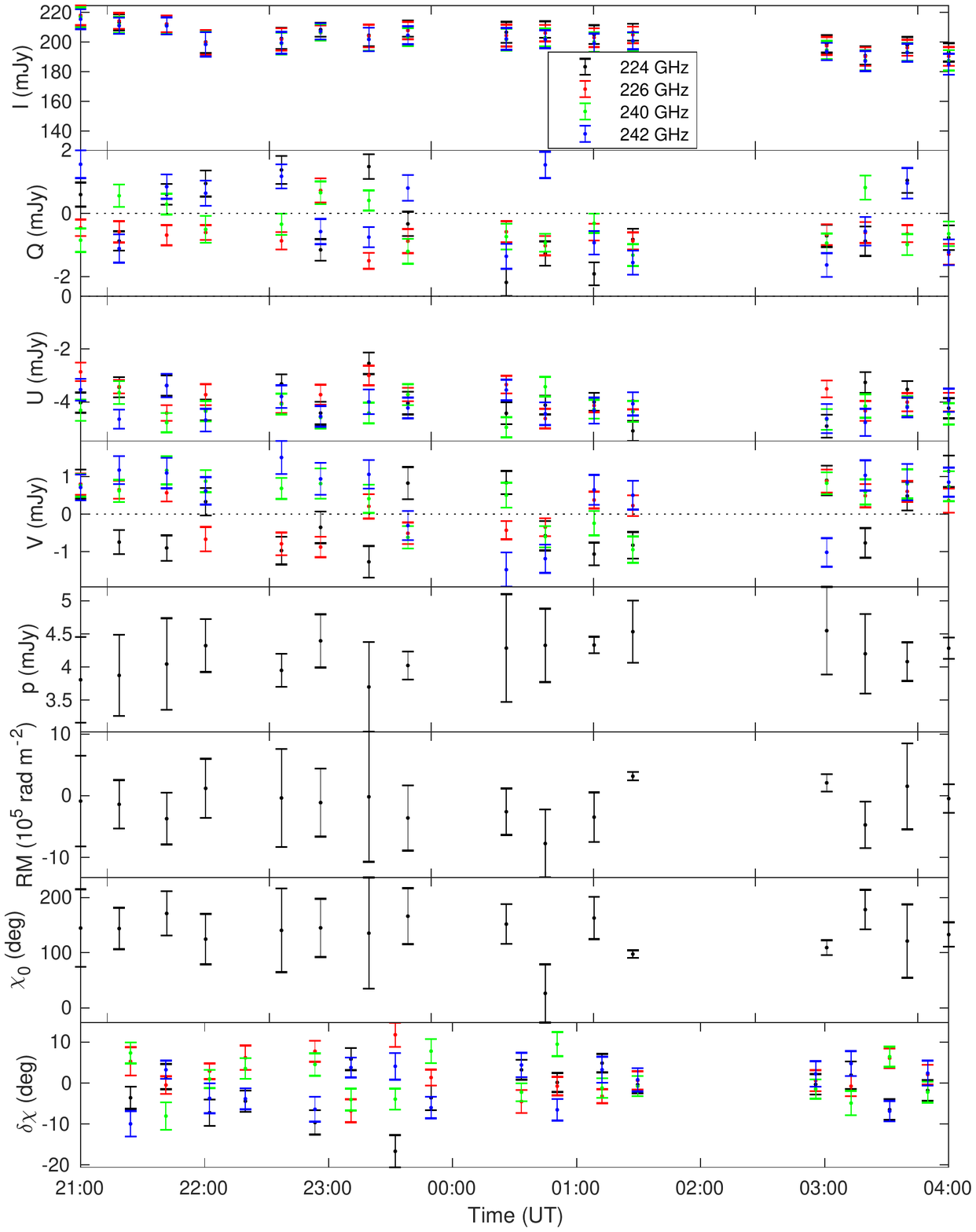}
\caption{Time series plot for J1713-3418 in epoch 3 following
Figure~\ref{fig:sgratime1}.
\label{fig:caltime6}
}
\end{figure}

\begin{figure}
\includegraphics[width=\textwidth]{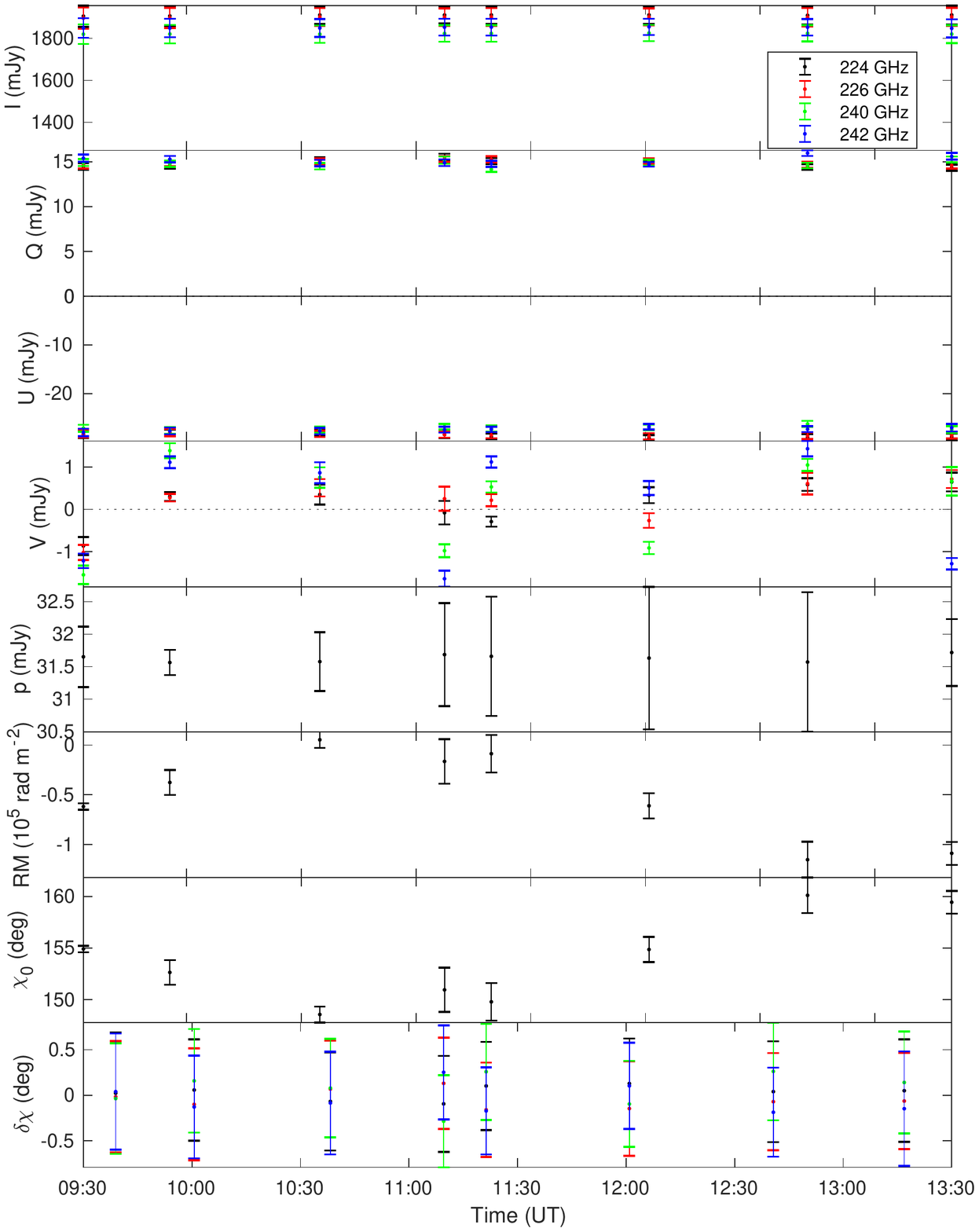}
\caption{Time series plot for J1751+0939 in epoch 1 following
Figure~\ref{fig:sgratime1}.
\label{fig:caltime7}
}
\end{figure}

\begin{figure}
\includegraphics[width=\textwidth]{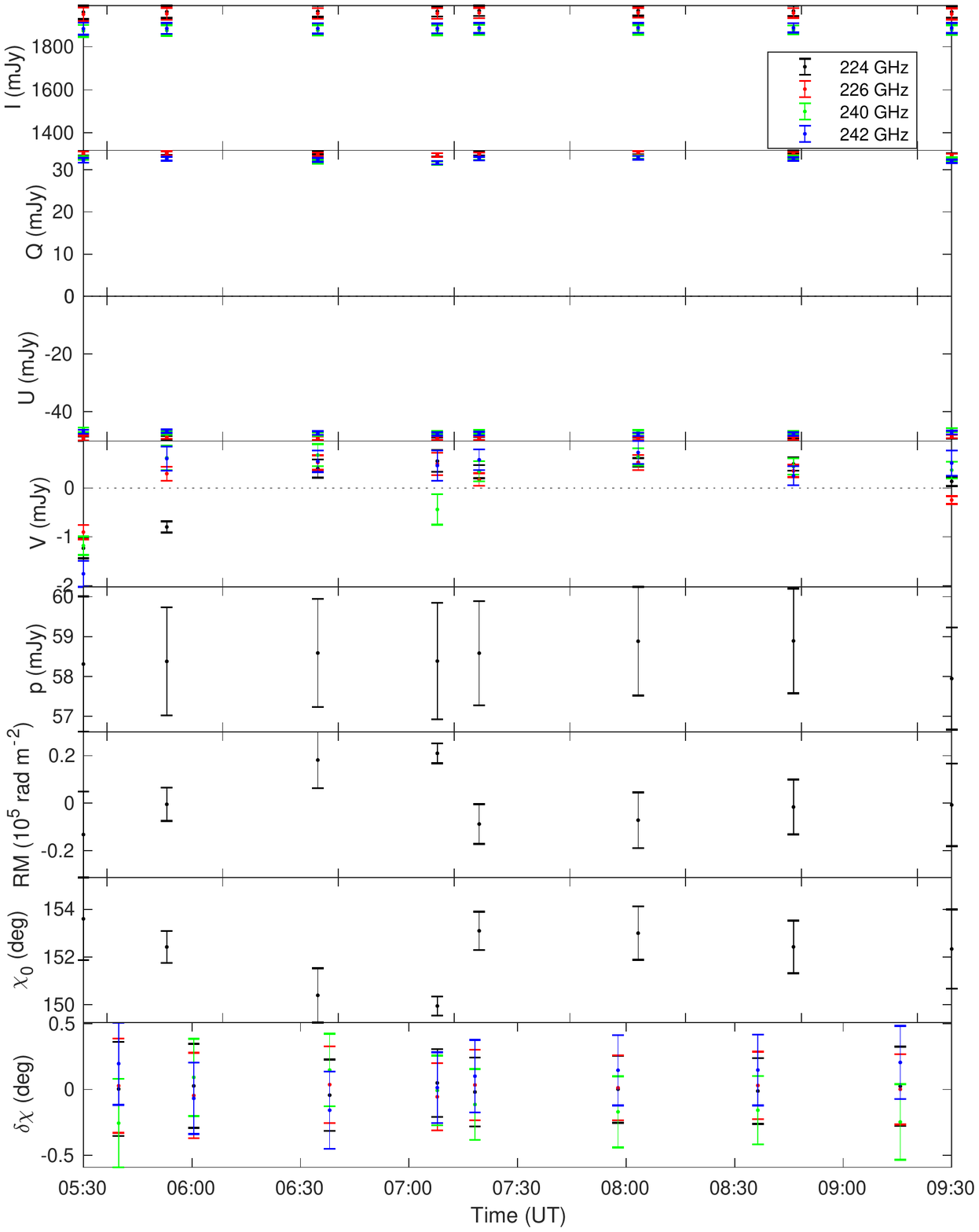}
\caption{Time series plot for J1751+0939 in epoch 2 following
Figure~\ref{fig:sgratime1}.
\label{fig:caltime8}
}
\end{figure}

\begin{figure}
\includegraphics[width=\textwidth]{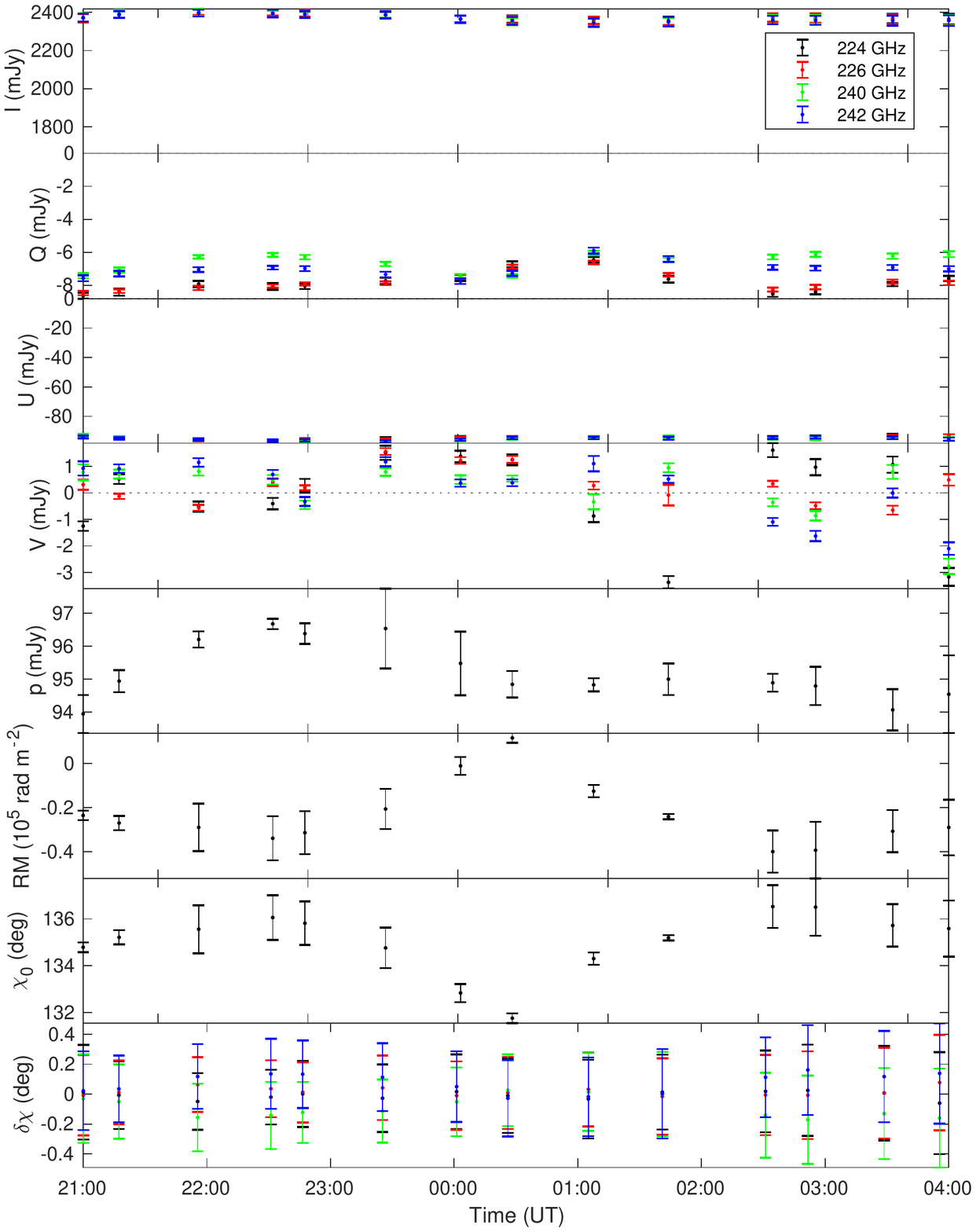}
\caption{Time series plot for J1751+0939 in epoch 3 following
Figure~\ref{fig:sgratime1}.
\label{fig:caltime9}
}
\end{figure}

As with the epoch-averaged results, the results for the polarization
calibrator, J1751+0939, present the limits of calibration accuracy.  In all three epochs, we see very stable measurements in each Stokes parameter as a function of time.  Variations in the fitted RM are $\lsim 3 \times 10^4 \rdm$ and in the fitted $\bar{\chi} \lsim 0.5$ deg.  Results for the phase calibrator J1733-3722 and for the check source are also stable over the course of each track in the Stokes parameters.  The stability of the calibrator results are best seen in Figures~\ref{fig:caltime1} through~\ref{fig:caltime9}.  For J1733-3722, we see a maximum variation in the fitted position angle $\bar{\chi}$ of $\lsim 4$ deg and rms variations of $\lsim 1$ deg. The RM has an rms variation $<10^4 \rdm$. For J1713-3418, the variations are slightly larger due to the lower source flux but consistent with no change in either RM or $\bar{\chi}$.  RMS variations in $\bar{\chi}$ are at most 6 deg and RM variations are $\lsim 10^5 \rdm$.


\vfill




\startlongtable
\begin{deluxetable}{llrrrrr}
\tablecaption{Calibrator Average Polarization Properties \label{tab:calavg}}
\tablehead{
\colhead{Source} & \colhead{Epoch} & \colhead{SPW} & \colhead{$I$} & \colhead{$Q$} & \colhead{$U$} & \colhead{$V$} \\
                 &                 &               & \colhead{(mJy)} & \colhead{(mJy)} & \colhead{(mJy)} & \colhead{(mJy)} 
} 
\startdata
J1713-3418 &     1 & 0 &  $   234 \pm      7 $ & $   -0.763 \pm     0.157 $ & $    3.069 \pm    0.175 $ & $   -0.040 \pm    0.133 $\\ 
     \dots & \dots & 1 &  $   232 \pm      7 $ & $   -0.762 \pm     0.157 $ & $    3.091 \pm    0.138 $ & $    0.030 \pm    0.140 $\\ 
     \dots & \dots & 2 &  $   222 \pm      6 $ & $   -0.753 \pm     0.173 $ & $    2.725 \pm    0.198 $ & $    0.373 \pm    0.159 $\\ 
     \dots & \dots & 3 &  $   224 \pm      6 $ & $   -0.913 \pm     0.235 $ & $    3.002 \pm    0.225 $ & $   -0.412 \pm    0.220 $\\ 
J1713-3418 &     2 & 0 &  $   243 \pm      4 $ & $   -0.906 \pm     0.246 $ & $   -3.697 \pm    0.169 $ & $    0.251 \pm    0.152 $\\ 
     \dots & \dots & 1 &  $   242 \pm      4 $ & $   -1.208 \pm     0.152 $ & $   -4.203 \pm    0.161 $ & $    0.377 \pm    0.172 $\\ 
     \dots & \dots & 2 &  $   230 \pm      4 $ & $   -1.120 \pm     0.172 $ & $   -3.881 \pm    0.201 $ & $    0.426 \pm    0.134 $\\ 
     \dots & \dots & 3 &  $   230 \pm      4 $ & $   -1.169 \pm     0.219 $ & $   -4.744 \pm    0.232 $ & $    0.620 \pm    0.210 $\\ 
J1713-3418 &     3 & 0 &  $   196 \pm      4 $ & $   -0.667 \pm     0.133 $ & $   -3.920 \pm    0.123 $ & $   -0.336 \pm    0.112 $\\ 
     \dots & \dots & 1 &  $   195 \pm      4 $ & $   -0.691 \pm     0.077 $ & $   -3.764 \pm    0.112 $ & $   -0.213 \pm    0.084 $\\ 
     \dots & \dots & 2 &  $   193 \pm      4 $ & $   -0.576 \pm     0.091 $ & $   -4.173 \pm    0.112 $ & $    0.206 \pm    0.081 $\\ 
     \dots & \dots & 3 &  $   193 \pm      4 $ & $   -0.753 \pm     0.099 $ & $   -4.140 \pm    0.117 $ & $    0.242 \pm    0.102 $\\ 
J1733-3722 &     1 & 0 &  $   475 \pm     14 $ & $   53.035 \pm     1.535 $ & $   14.043 \pm    0.397 $ & $    0.817 \pm    0.061 $\\ 
     \dots & \dots & 1 &  $   472 \pm     14 $ & $   52.687 \pm     1.524 $ & $   14.717 \pm    0.422 $ & $    0.882 \pm    0.064 $\\ 
     \dots & \dots & 2 &  $   454 \pm     12 $ & $   50.531 \pm     1.382 $ & $   16.346 \pm    0.451 $ & $    0.739 \pm    0.064 $\\ 
     \dots & \dots & 3 &  $   453 \pm     12 $ & $   50.036 \pm     1.398 $ & $   16.908 \pm    0.467 $ & $    0.999 \pm    0.060 $\\ 
J1733-3722 &     2 & 0 &  $   648 \pm     11 $ & $   34.860 \pm     0.601 $ & $   -1.448 \pm    0.060 $ & $    0.191 \pm    0.048 $\\ 
     \dots & \dots & 1 &  $   645 \pm     11 $ & $   34.557 \pm     0.573 $ & $   -1.515 \pm    0.042 $ & $    0.146 \pm    0.065 $\\ 
     \dots & \dots & 2 &  $   620 \pm     10 $ & $   32.813 \pm     0.506 $ & $   -1.354 \pm    0.055 $ & $   -0.159 \pm    0.051 $\\ 
     \dots & \dots & 3 &  $   618 \pm     10 $ & $   32.421 \pm     0.521 $ & $   -2.542 \pm    0.062 $ & $   -0.355 \pm    0.056 $\\ 
J1733-3722 &     3 & 0 &  $   570 \pm      9 $ & $   11.000 \pm     0.194 $ & $   29.049 \pm    0.487 $ & $    0.266 \pm    0.105 $\\ 
     \dots & \dots & 1 &  $   569 \pm      9 $ & $   11.050 \pm     0.194 $ & $   29.101 \pm    0.475 $ & $   -0.226 \pm    0.048 $\\ 
     \dots & \dots & 2 &  $   565 \pm      9 $ & $    9.529 \pm     0.169 $ & $   29.396 \pm    0.482 $ & $   -0.441 \pm    0.052 $\\ 
     \dots & \dots & 3 &  $   563 \pm      9 $ & $    9.507 \pm     0.169 $ & $   29.304 \pm    0.483 $ & $   -0.448 \pm    0.068 $\\ 
J1751+0939 &     1 & 0 &  $  1905 \pm     59 $ & $   14.873 \pm     0.459 $ & $  -28.325 \pm    0.882 $ & $   -0.097 \pm    0.068 $\\ 
     \dots & \dots & 1 &  $  1896 \pm     58 $ & $   14.936 \pm     0.436 $ & $  -28.282 \pm    0.870 $ & $   -0.015 \pm    0.067 $\\ 
     \dots & \dots & 2 &  $  1820 \pm     52 $ & $   14.795 \pm     0.416 $ & $  -27.112 \pm    0.778 $ & $   -0.033 \pm    0.083 $\\ 
     \dots & \dots & 3 &  $  1846 \pm     53 $ & $   15.082 \pm     0.415 $ & $  -27.476 \pm    0.808 $ & $   -0.196 \pm    0.073 $\\ 
J1751+0939 &     2 & 0 &  $  1964 \pm     35 $ & $   33.880 \pm     0.577 $ & $  -49.329 \pm    0.865 $ & $   -0.244 \pm    0.059 $\\ 
     \dots & \dots & 1 &  $  1954 \pm     34 $ & $   33.720 \pm     0.598 $ & $  -49.098 \pm    0.874 $ & $   -0.208 \pm    0.051 $\\ 
     \dots & \dots & 2 &  $  1876 \pm     31 $ & $   32.353 \pm     0.525 $ & $  -47.128 \pm    0.776 $ & $   -0.290 \pm    0.083 $\\ 
     \dots & \dots & 3 &  $  1886 \pm     31 $ & $   32.294 \pm     0.533 $ & $  -47.337 \pm    0.765 $ & $   -0.352 \pm    0.074 $\\ 
J1751+0939 &     3 & 0 &  $  2342 \pm     36 $ & $   -7.714 \pm     0.139 $ & $  -93.228 \pm    1.435 $ & $   -0.197 \pm    0.106 $\\ 
     \dots & \dots & 1 &  $  2340 \pm     36 $ & $   -7.668 \pm     0.113 $ & $  -93.420 \pm    1.426 $ & $   -0.074 \pm    0.050 $\\ 
     \dots & \dots & 2 &  $  2327 \pm     36 $ & $   -6.473 \pm     0.113 $ & $  -93.514 \pm    1.454 $ & $    0.204 \pm    0.055 $\\ 
     \dots & \dots & 3 &  $  2325 \pm     36 $ & $   -6.883 \pm     0.118 $ & $  -93.509 \pm    1.457 $ & $    0.142 \pm    0.058 $\\ 
J1517-2422 &     3 & 0 &  $  2963 \pm     53 $ & $   69.603 \pm     1.255 $ & $  -18.583 \pm    0.367 $ & $   -0.185 \pm    0.171 $\\ 
     \dots & \dots & 1 &  $  2971 \pm     53 $ & $   70.258 \pm     1.274 $ & $  -17.801 \pm    0.334 $ & $    0.275 \pm    0.106 $\\ 
     \dots & \dots & 2 &  $  2966 \pm     56 $ & $   77.661 \pm     1.468 $ & $  -26.849 \pm    0.516 $ & $    1.677 \pm    0.144 $\\ 
     \dots & \dots & 3 &  $  2968 \pm     56 $ & $   77.915 \pm     1.482 $ & $  -30.013 \pm    0.571 $ & $    2.605 \pm    0.180 $\\ 
J1733-1304 &     3 & 0 &  $  2191 \pm     46 $ & $   10.036 \pm     0.283 $ & $ -106.769 \pm    2.242 $ & $    0.600 \pm    0.294 $\\ 
     \dots & \dots & 1 &  $  2188 \pm     46 $ & $   10.301 \pm     0.253 $ & $ -107.436 \pm    2.256 $ & $    0.273 \pm    0.158 $\\ 
     \dots & \dots & 2 &  $  2156 \pm     49 $ & $   12.350 \pm     0.323 $ & $ -106.954 \pm    2.437 $ & $    3.158 \pm    0.157 $\\ 
     \dots & \dots & 3 &  $  2150 \pm     49 $ & $   11.136 \pm     0.344 $ & $ -106.260 \pm    2.456 $ & $    3.572 \pm    0.178 $\\ 
J1924-2914 &     3 & 0 &  $  3784 \pm     79 $ & $   -3.980 \pm     0.206 $ & $  -44.584 \pm    1.026 $ & $   -5.749 \pm    0.514 $\\ 
     \dots & \dots & 1 &  $  3769 \pm     80 $ & $   -3.408 \pm     0.169 $ & $  -42.889 \pm    0.914 $ & $   -3.519 \pm    0.429 $\\ 
     \dots & \dots & 2 &  $  3706 \pm     79 $ & $   -2.062 \pm     0.176 $ & $  -37.685 \pm    0.835 $ & $   -1.293 \pm    0.341 $\\ 
     \dots & \dots & 3 &  $  3686 \pm     79 $ & $   -1.899 \pm     0.217 $ & $  -38.272 \pm    0.832 $ & $    0.510 \pm    0.336 $\\ 
\enddata
\end{deluxetable}

\movetabledown=1in
\begin{rotatetable}
\begin{deluxetable}{llrrrrrrrr}
\tabletypesize{\scriptsize}
\tablecaption{Calibrator Frequency-Averaged Polarization and Rotation Measures \label{tab:calavgrm}}
\tablehead{
\colhead{Source} & \colhead{Epoch} & 
\colhead{$I$} & \colhead{$\delta I$} &
\colhead{$P$} & \colhead{$\delta P$} &
\colhead{$V$} & \colhead{$\delta V$} &
\colhead{RM} & \colhead{$\chi_0$} 
\\
                 &                 & 
\colhead{(mJy)} & \colhead{(mJy GHz$^{-1}$)} &
\colhead{(mJy)} & \colhead{(mJy GHz$^{-1}$)} &
\colhead{(mJy)} & \colhead{(mJy GHz$^{-1}$)} &
\colhead{($10^5 \rdm$)} & \colhead{(deg)} 
}
\startdata
J1713-3418 & 1 &  $    228 \pm      1 $ & $  -0.6 \pm   0.8 $ & $   3.07 \pm   0.07 $ & $  -0.01 \pm   0.01 $ & $    0.045 \pm    0.169 $ & $ 0.005 \pm 0.021 $ & $ -0.88 \pm  0.24 $ & $  61.0 \pm   2.3 $ \\ 
     \dots & 2 &  $    237 \pm      0 $ & $  -0.8 \pm   0.4 $ & $   4.27 \pm   0.23 $ & $   0.02 \pm   0.03 $ & $    0.397 \pm    0.046 $ & $ 0.012 \pm 0.006 $ & $  0.04 \pm  0.61 $ & $ -52.8 \pm   5.8 $ \\ 
     \dots & 3 &  $    194 \pm      0 $ & $  -0.2 \pm   0.1 $ & $   4.05 \pm   0.05 $ & $   0.02 \pm   0.01 $ & $   -0.020 \pm    0.016 $ & $ 0.031 \pm 0.002 $ & $ -0.28 \pm  0.51 $ & $ -47.1 \pm   4.9 $ \\ 
J1733-3722 & 1 &  $    463 \pm      0 $ & $  -1.3 \pm   0.4 $ & $  53.87 \pm   0.02 $ & $  -0.11 \pm   0.00 $ & $    0.863 \pm    0.066 $ & $ 0.003 \pm 0.008 $ & $ -1.19 \pm  0.10 $ & $  19.8 \pm   1.0 $ \\ 
     \dots & 2 &  $    633 \pm      0 $ & $  -1.7 \pm   0.4 $ & $  33.71 \pm   0.02 $ & $  -0.13 \pm   0.00 $ & $   -0.041 \pm    0.036 $ & $ -0.027 \pm 0.004 $ & $  0.42 \pm  0.38 $ & $  -5.4 \pm   3.6 $ \\ 
     \dots & 3 &  $    567 \pm      0 $ & $  -0.4 \pm   0.3 $ & $  30.98 \pm   0.03 $ & $  -0.01 \pm   0.00 $ & $   -0.288 \pm    0.084 $ & $ -0.021 \pm 0.011 $ & $ -1.05 \pm  0.10 $ & $  45.3 \pm   1.0 $ \\ 
J1751+0939 & 1 &  $   1866 \pm      9 $ & $  -4.1 \pm   9.2 $ & $  31.54 \pm   0.15 $ & $  -0.05 \pm   0.02 $ & $   -0.090 \pm    0.043 $ & $ -0.004 \pm 0.005 $ & $ -0.35 \pm  0.01 $ & $ -27.5 \pm   0.1 $ \\ 
     \dots & 2 &  $   1920 \pm      5 $ & $  -4.8 \pm   5.1 $ & $  58.47 \pm   0.12 $ & $  -0.15 \pm   0.01 $ & $   -0.273 \pm    0.018 $ & $ -0.006 \pm 0.002 $ & $  0.04 \pm  0.03 $ & $ -28.2 \pm   0.3 $ \\ 
     \dots & 3 &  $   2334 \pm      0 $ & $  -0.9 \pm   0.1 $ & $  93.70 \pm   0.04 $ & $   0.01 \pm   0.01 $ & $    0.034 \pm    0.031 $ & $ 0.017 \pm 0.004 $ & $ -0.23 \pm  0.06 $ & $ -45.0 \pm   0.6 $ \\ 
J1517-2422 & 3 &  $   2967 \pm      2 $ & $   0.1 \pm   2.2 $ & $  77.55 \pm   0.24 $ & $   0.66 \pm   0.03 $ & $    1.084 \pm    0.158 $ & $ 0.127 \pm 0.021 $ & $  2.09 \pm  0.38 $ & $ -28.6 \pm   3.7 $ \\ 
J1733-1304 & 3 &  $   2171 \pm      0 $ & $  -2.3 \pm   0.4 $ & $ 107.42 \pm   0.27 $ & $  -0.02 \pm   0.03 $ & $    1.804 \pm    0.139 $ & $ 0.194 \pm 0.018 $ & $ -0.31 \pm  0.13 $ & $ -39.1 \pm   1.3 $ \\ 
J1924-2914 & 3 &  $   3736 \pm      3 $ & $  -5.1 \pm   2.6 $ & $  40.93 \pm   0.42 $ & $  -0.36 \pm   0.05 $ & $   -2.462 \pm    0.533 $ & $ 0.268 \pm 0.067 $ & $ -0.71 \pm  0.06 $ & $ -40.2 \pm   0.6 $ \\ 
\enddata
\end{deluxetable}
\end{rotatetable}

\end{document}